%% file: T22.tex
\documentclass[fleqn,usenatbib]{mnras}

\usepackage{newtxtext,newtxmath}
\usepackage{caption}
\usepackage{adjustbox}
\usepackage[T1]{fontenc}

\DeclareRobustCommand{\VAN}[3]{#2}
\let\VANthebibliography\thebibliography
\def\thebibliography{\DeclareRobustCommand{\VAN}[3]{##3}\VANthebibliography}

\usepackage{graphicx}	
\usepackage{amsmath}	

\newcommand{\gaia}{\textit{Gaia}}

\providecommand{\grp}{\ensuremath{G_\mathrm{RP}}}
\providecommand{\gbp}{\ensuremath{G_\mathrm{BP}}}

\title[C-MetaLL Survey: II]{Cepheid Metallicity in the Leavitt Law (C- MetaLL) survey: II. High-resolution spectroscopy of the most metal poor Galactic Cepheids\thanks{Based on observations European Southern Observatory programs P105.20MX.001/P106.2129.001}}

\author[E. Trentin et al.]{
E. Trentin,$^{1,2,3}$\thanks{E-mail: etrentin@aip.de}
V. Ripepi,$^{3}$
G. Catanzaro,$^{4}$
J. Storm,$^{1}$
M. Marconi,$^{3}$
G. De Somma,$^{3,5}$
V. Testa,$^{6}$
I. Musella$^{3}$
\\
$^{1}$Leibniz-Institut f\"ur Astrophysik Potsdam (AIP), An der Sternwarte 16, D-14482 Potsdam, Germany \\
$^{2}$Institut für Physik und Astronomie, Universität Potsdam, Haus 28, Karl-Liebknecht-Str. 24/25, D-14476 Golm (Potsdam), Germany\\
$^{3}$INAF-Osservatorio Astronomico di Capodimonte, Salita Moiariello 16, 80131, Naples, Italy\\
$^{4}$INAF-Osservatorio Astrofisico di Catania, via Santa Sofia 78, 95125, Catania, Italy\\
$^{5}$Istituto Nazionale di Fisica Nucleare (INFN)-Sez. di Napoli, Via Cinthia, 80126 Napoli, Italy \\
$^{6}$INAF – Osservatorio Astronomico di Roma, via Frascati 33, I-00078 Monte Porzio Catone, Italy 
}

\date{Accepted XXX. Received YYY; in original form ZZZ}

\pubyear{2015}

\begin{document}
\label{firstpage}
\pagerange{\pageref{firstpage}--\pageref{lastpage}}
\maketitle

\begin{abstract}
Classical Cepheids (DCEPs) are the first fundamental step in the calibration of the cosmological distance ladder. Furthermore, they represent powerful tracers in the context of Galactic studies. We have collected high-resolution spectroscopy with UVES@VLT for a sample of 65 DCEPs.  The majority of them are the faintest DCEPs ever observed in the Milky Way. For each target, we derived accurate atmospheric parameters, radial velocities, and abundances for 24 different species. The resulting iron abundances range between +0.3 and $-$1.1 dex with the bulk of stars at [Fe/H]$\sim-0.5$ dex. Our sample includes the most metal-poor DCEPs observed so far with high-resolution spectroscopy.  We complement our sample with literature data obtaining a complete sample of 637 DCEPs and use Gaia Early Data Release 3 (EDR3) photometry to determine the distance of the DCEPs from the Period-Wesenheit-Metallicity relation.  
Our more external data trace the Outer arm (at Galactocentric radius ($R_{GC})\sim$16--18 kpc) which appears significantly warped.  
We investigate the metallicity gradient of the Galactic disc using this large sample, finding a slope of $-0.060 \pm 0.002$ dex kpc$^{-1}$, in very good agreement with previous results based both on DCEPs and open clusters. We also report a possible break in the gradient at $R_{GC}$=9.25 kpc with slopes of $-0.063 \pm 0.007$ and $-0.079 \pm 0.003$ dex kpc$^{-1}$ for the inner and outer sample, respectively. The two slopes differ by more than 1 $\sigma$. A more homogeneous and extended DCEPs sample is needed to further test the plausibility of such a break.

\end{abstract}

\begin{keywords}
Stars: distances –- Stars: variables: Cepheids –-  Stars: abundances -- Stars: fundamental parameters -- Galaxy: disc 
\end{keywords}



\section{Introduction}

Classical Cepheids (DCEPs) are fundamental astrophysical objects. Since their discovery, the period-luminosity (PL) and period-Wesenheit\footnote{The Weseheit magnitudes are quantities that are reddening free by construction, once the reddening law is known \citep[see][]{Madore1982}} (PW) relations that hold for these pulsators, represent the base for the cosmic distance scale \citep[e.g.][]{Leavitt1912,Madore1982,Caputo2000,Riess2016}. 
Once calibrated by means of geometric methods such as trigonometric parallaxes, eclipsing binaries and water masers, these relations can be used to calibrate secondary distance indicators like Type Ia Supernovae (SNe), which are sufficiently powerful to reach the unperturbed Hubble flow, allowing us to measure the Hubble constant \citep[$H_0$, see eg.][and references therein]{Sandage2006,Freedman2012,Riess2016,Riess2019,Riess2021}.

In recent years this topic has been at the center of a heated debate as the values of $H_0$ based on the cosmic distance scale \citep[e.g.][and references therein]{verde2019,Riess2021b} are in significant disagreement with those estimated from the Planck Cosmic Microwave Background (CMB) measurements under the flat $\Lambda$ Cold Dark Matter ($\Lambda$CDM) model. The latest estimate of $H_0$ from the cosmic distance scale is $H_0$=73.04$\pm$1.04 km s$^{-1}$ Mpc$^{-1}$ \citep{Riess2021}, which is 5$\sigma$ different from the value estimated by Planck+$\Lambda$CMB, namely $H_0$=67.4$\pm$0.5 km s$^{-1}$ Mpc$^{-1}$ \citep{Planck2020}. This discrepancy, known as the $H_0$ tension, has still unknown origins, in spite of many observational and theoretical efforts to study the possible causes as well as residual systematics\citep[see e.g.][and reference therein]{Dainotti2021,Freedman2021,Riess2021b}. 

In this context, it is crucial to identify any residual systematic effect which could in principle affect the cosmic distance scale path to the measure of $H_0$. One of the most controversial candidates is the metallicity dependence of the DCEP $PL$ and $PW$ relations. Several recent results based on \gaia\ mission Data Release 2 \citep[DR2][]{Gaia2016,Gaia2018} and Early Data Release 3 \citep[EDR3][]{Gaia2021} parallaxes are indeed in disagreement one each other \citep[see e.g.][]{Groenewegen2018,Ripepi2019,Ripepi2020,Riess2021,Breuval2021,Ripepi2021a,Ripepi2022}. To overcome these difficulties and to establish accurately the metallicity dependence of the DCEPs $PL$ and $PW$ relations and its impact on the measure of $H_0$, we undertook a project dubbed C-MetaLL (Cepheid - Metallicity in the Leavitt Law) which is fully described in \citet{Ripepi2021a}. One of the immediate objectives of the project is to measure the chemical abundance of a sample of 250-300 Galactic DCEPs through high-resolution spectroscopy, specifically aiming at enlarging as much as possible the metallicity range of the targets towards the metal-poor regime ([Fe/H]$<-$0.4 dex), where only a few objects are present in the literature. 

Here we present the spectroscopic observations of a considerable sample (i.e. 65 objects) of recently discovered faint DCEPs whose Galactocentric distance, $R_{GC}>15$ kpc, makes them excellent metal-poor candidates (see~Section \ref{sect:selectionTerget} for details).    

The DCEP sample discussed in this work has not only relevance for the cosmic distance scale, but also for Galactic studies. Indeed, the accurate distances which can be measured for DCEPs allow one to carry out the chemical tagging of the disc and in particular to study how the abundance of the different chemical elements varies with $R_{GC}$, from the central regions of the MW towards the outer regions of the disc. Many studies exist already in the literature facing this problem with a variety of tracers other than DCEPs, such as OCs \citep[e.g.][just to mention the most recent ones]{cunha2016chemical,netopil2016metallicity,casamiquela2019occaso,donor2020open,spina2021galah}. Focusing on the most recent works using DCEPs, we notice that their majority \citep[][]{Luck2011,Genovali2014,Luck2018,lemasle2018milky,Ripepi2022,da2022new} agree on measured iron gradients in the interval 0.05-0.06 dex kpc$^{-1}$. However, all these investigations followed the radial gradient of the MW disc only up to about $R_{GC}\sim$13-14 kpc, not exploring the outer and possibly most metal-poor regions of the disc, with the exception of \citet{Minniti2020}, who obtained a metallicity estimate  for three DCEPs with 15$<R_{GC}<$22 kpc, beyond the Galactic bulge. Our sample will allow us increase significantly the number of DCEPs with $R_{GC}>$15 kpc and get new insights on the metallicity gradient of the MW disc. The application of our sample to the determination of the metallicity dependence of the DCEPs PL/PW relations will be the subject of a separate paper.

\section{Data}

\subsection{Selection of the targets}
\label{sect:selectionTerget}

Our targets have different characteristics due to the displacement of the MW disc during the different seasons. During the southern winter we can observe in the direction of the Galactic centre, while the opposite is true for the summer season. Given the Galactic gradient, the southern winter and summer seasons allow us to observe DCEPs with solar or super-solar and sub-solar metallicity, respectively. The targets towards the Galactic centre were selected among the list published by \citet{Skowron2019}, requiring that the relative error on the parallax published in the {\it Gaia} DR2 were better than 20\% and that $G<13$ mag (note that EDR3 was not available at the time of proposal submission). This limit ensures that the target were observable in less than 1 h (the maximum allowed duration of an "Observing Block" at ESO) in any weather condition (so called ESO "filler program"). To choose the winter sample, we used again the \citet{Skowron2019} catalog, and by relying on the distances therein, we selected the targets with Galactocentric distance larger than 15 kpc. In this way, thanks to the metallicity gradient of the Galactic disc \citep[see e.g.][and references therein]{Luck2011,kovtyukh2000precise,Ripepi2022}, we could reasonably expect that the large majority of these DCEPs had [Fe/H] values lower than $-$0.4 dex, which is one of our main aims.    

\subsection{Observations and data reduction}

The spectra analysed in this work have been obtained in the context of two ESO (European Southern Observatory) proposals, namely P105.20MX.001 and P106.2129.001. The observations foreseen for P105 were postponed due to pandemic issues and were collected in service mode between January, 2nd and September, 13th 2021. For P106 the observations were carried out in virtual visitor mode in the period December 12--19, 2020 under very good sky conditions.     
We used the UVES (Ultraviolet and Visual Echelle Spectrograph\footnote{https://www.eso.org/sci/facilities/paranal/instruments/uves.html}) instrument, attached at the UT2 (Unit Telescope 2) of VLT (Very Large Telescope), placed at Paranal (Chile). We used the red arm equipped with the grism CD\#3, covering the wavelength interval 4760--6840 {\AA}, and the central wavelength at 5800 {\AA}. We adopted the 1 arcsec slit, which provides a dispersion of R$\sim$47,000. The exposure times and the signal to noise (S/N) for each target are listed in Table~\ref{tab:radec}, together with the identification of the targets, coordinates, period, $G$ magnitude and mode of pulsation.

The data reduction was carried out using the standard UVES pipeline version 6.1.6. which provides fully calibrated spectra. The spectra were normalized using the {\it IRAF} task \emph{continuum} after dividing each spectrum into intervals (usually 3/4) in order to have a better continuum detection in each interval. 
\input{T22_ra_dec}

\begin{figure}
	\includegraphics[width=\columnwidth]{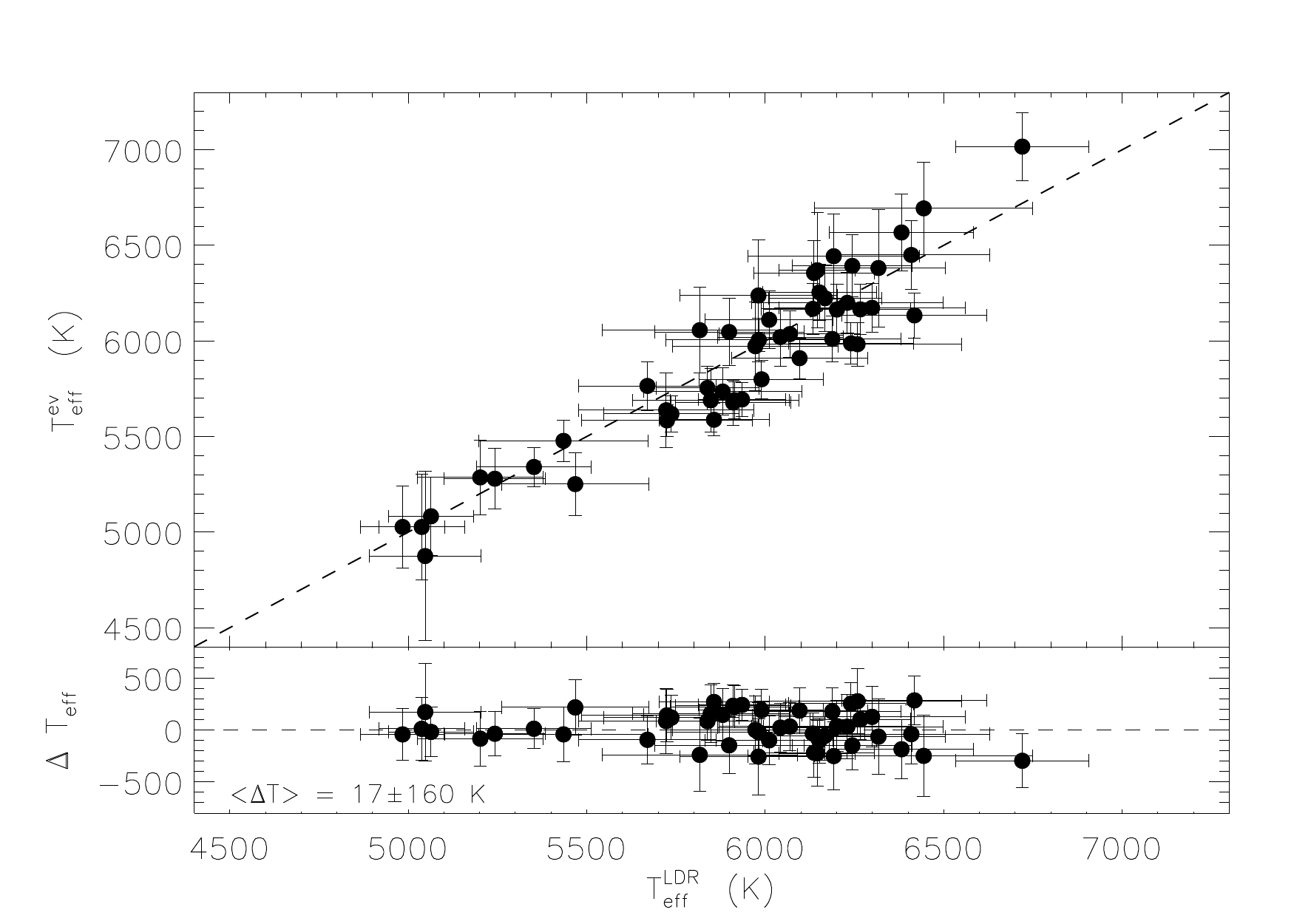}
    \caption{Effective temperatures derived by excitation potential of neutral iron lines versus those derived by LDR. The upper panel shows the direct comparison between the two estimates, while the lower panel displays their differences.}
    \label{fig:comparison}
\end{figure}

\section{Abundance analysis}
\subsection{Stellar parameters}
\begin{figure*}
	\includegraphics[width=18cm,bb=0 80 504 360]{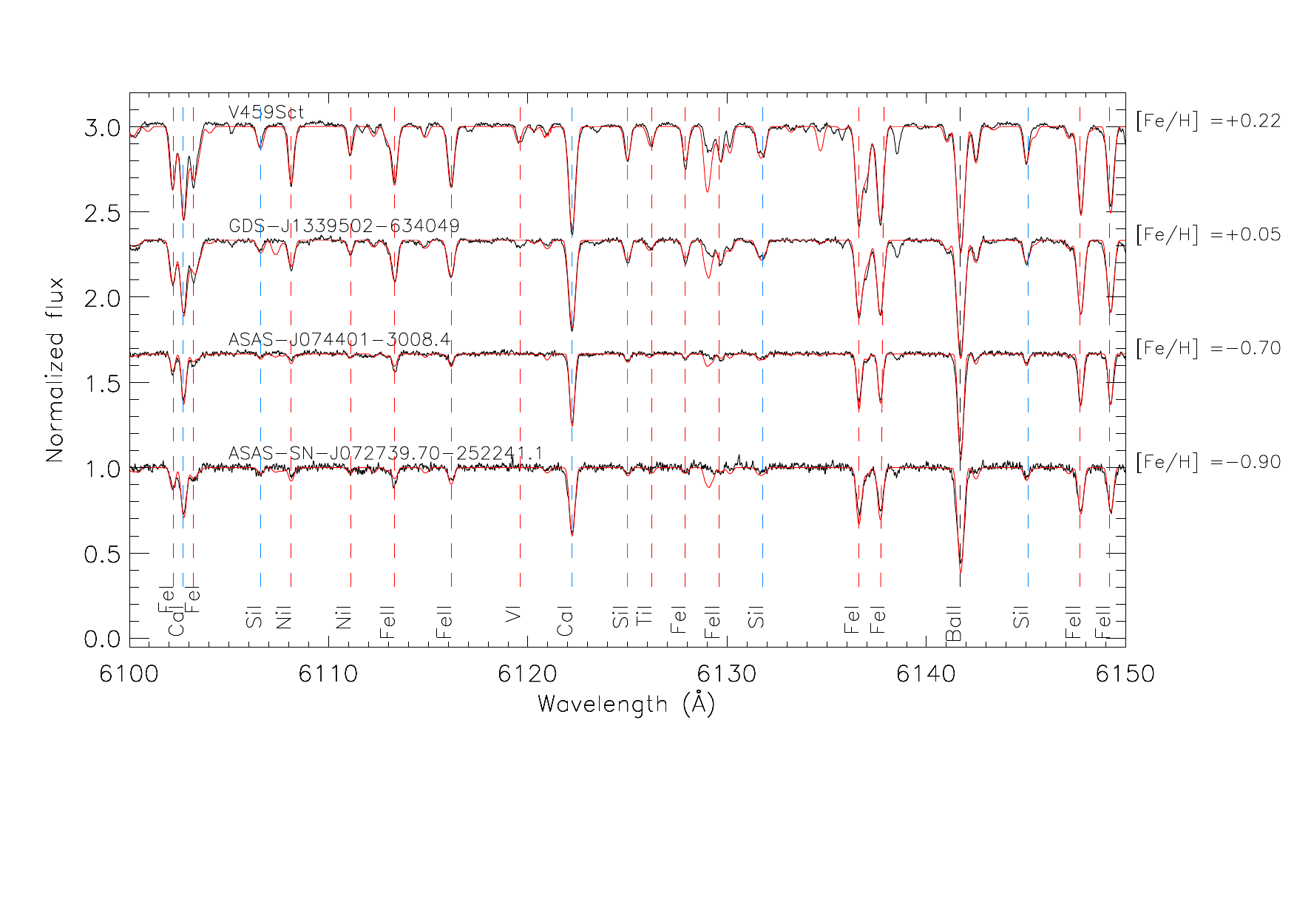}
    \caption{Example of high-resolution spectra in the range $\lambda\lambda$ 6100-6150 {\AA}. Stars are ordered from top to bottom with decreasing metallicity (see Sec.~\ref{Abundances} for its estimation). The vertical dashed lines identify the position of atomic lines, the meaning of the colors are: red for lines of vanadium, iron, and nickel, blue for lines of silicon and calcium, and black for the line of barium.}
    \label{fig:spectra}
\end{figure*}

The first step when measuring chemical abundances is the estimation of the main atmospheric parameters, namely the effective temperature (T$_{\rm eff}$), surface gravity (log {\itshape g}), microturbulent velocity ($\xi$) and line broadening parameter (v$_{\rm br}$), i.e. the combined effects of macroturbulence
and rotational velocity (in the case of DCEPs, macroturbulence often represents the dominating factor).

A useful tool extensively adopted in the literature to evaluate the effective temperature, is the line depth ratios (LDR) method \citep{kovtyukh2000precise}. This method has the advantage of being sensitive to temperature variations, but not to abundances and interstellar reddening. Typically in our targets we measured about 32 LDRs for each spectrum.

For some targets, because of their low metallicity, the number of couples of spectral lines useful for the LDR method is not enough to guarantee accurate results. For these stars the effective temperature can be determined by imposing excitation equilibrium, that is, by imposing that there is no residual correlation between the iron abundance and the excitation potential of the neutral iron lines \citep[see e.g.][]{mucciarelli2020facing}. 
To check the consistency of both methods, we computed the effective temperatures by using 
excitation potentials for all the targets for which LDRs gave accurate results. In total we
selected 55 targets for which we calculated T$_{\rm eff}$ with both methods. The average difference among all
these values is $\Delta$T\,=\,17\,$\pm$\,160~K, thus we can conclude that both methods are consistent,
The results of this comparison are shown in Fig.~\ref{fig:comparison}. 

For the other parameters ($\xi$ and log {\itshape g}) an iterative approach was adopted: microturbulences were estimated by demanding the iron abundances do not depend on the equivalent widths (EWs), that is, the slope of the [Fe/H] as a function of EWs is null. For this purpose, we first measured the equivalent widths of a sample of 145 \ion{Fe}{I} lines using a python3 semi-automatic custom routine. The lines sample was extracted from the line list published by \citet{romaniello2008influence} and the routine minimizes errors in the continuum estimation on the wings of the spectral lines. The conversion of the EWs in abundances has been performed through the WIDTH9 code \citep{kurucz1981solar} applied to the corresponding atmospheric model calculated by using ATLAS9 \citep{kurucz1993new}. In this calculation we did not consider the influence of $\log$ {\itshape g} since neutral iron lines are insensitive to it. Then, the surface gravities were estimated with a similar iterative procedure imposing the ionization equilibrium between \ion{Fe}{I} and \ion{Fe}{II}. The adopted list of 24 \ion{Fe}{II} lines was extracted from \citet{romaniello2008influence}.

The atmospheric parameters estimated, and summarized in  Table~\ref{tab:atm_err}, were used as input values for the abundance analysis presented in the next section.

\input{T22_atm_err}
\subsection{Abundances}\label{Abundances}
In order to avoid problems from spectral line blending caused by line broadening, a spectral synthesis technique was applied to our spectra. Synthetic spectra were generated in three steps: i) plane parallel local thermodynamic equilibrium (LTE) atmosphere models were computed using the ATLAS9 code \citet{kurucz1993new}, using the stellar parameters in Table~\ref{tab:atm_err}; ii) stellar spectra were synthesized by using SYNTHE \citet{kurucz1981solar}; 
iii) the synthetic spectra were convoluted for instrumental and line broadening. This was evaluated by matching the synthetic line profiles to a selected set of the observed metal lines.

For a total of 24 different chemical elements it was possible to detect spectral lines used for the estimation of the abundances. For all elements we performed the following analysis: we divided the observed spectra into intervals, 25 Å or 50 Å wide, and derived the abundances in each interval by performing a $\chi^2$ minimization of the differences between the observed and synthetic spectra. The minimization algorithm was written in {\it IDL}\footnote{IDL (Interactive Data Language) is a registered trademark of L3HARRIS Geospatial} language, using the {\it amoeba} routine. 

We considered several sources of uncertainties in our abundances. First, we evaluated the expected errors caused by variations in the fundamental stellar parameters of $\delta T_{\mbox{\scriptsize eff}} = \pm 150$, $\delta \log${\itshape g}$= \pm$\,0.2~dex, and $\delta\xi \pm$\,0.3~km\,s$^{-1}$. 
According to our simulations, those errors contribute $\approx \pm$\,0.1~dex to the total error budget. Total errors were evaluated by summing in quadrature this value to the standard deviations obtained from the average abundances.

The adopted lists of spectral lines and atomic parameters are from \citet{castelli2004spectroscopic}, who have updated the original parameters of \citet{kurucz1995kurucz}. When necessary we also checked the NIST database \citep{ralchenko2019nist}.

An example of four spectra in the range $\lambda\lambda$ 6100-6150 {\AA} is displayed in Fig.~\ref{fig:spectra}. 
With the aim of comparing the differences in the spectral line depth, each spectrum reports, as a side label, with the metallicity derived in our analysis (see Sec.~\ref{Abundances}).

\begin{figure*}
	\includegraphics[width=15cm]{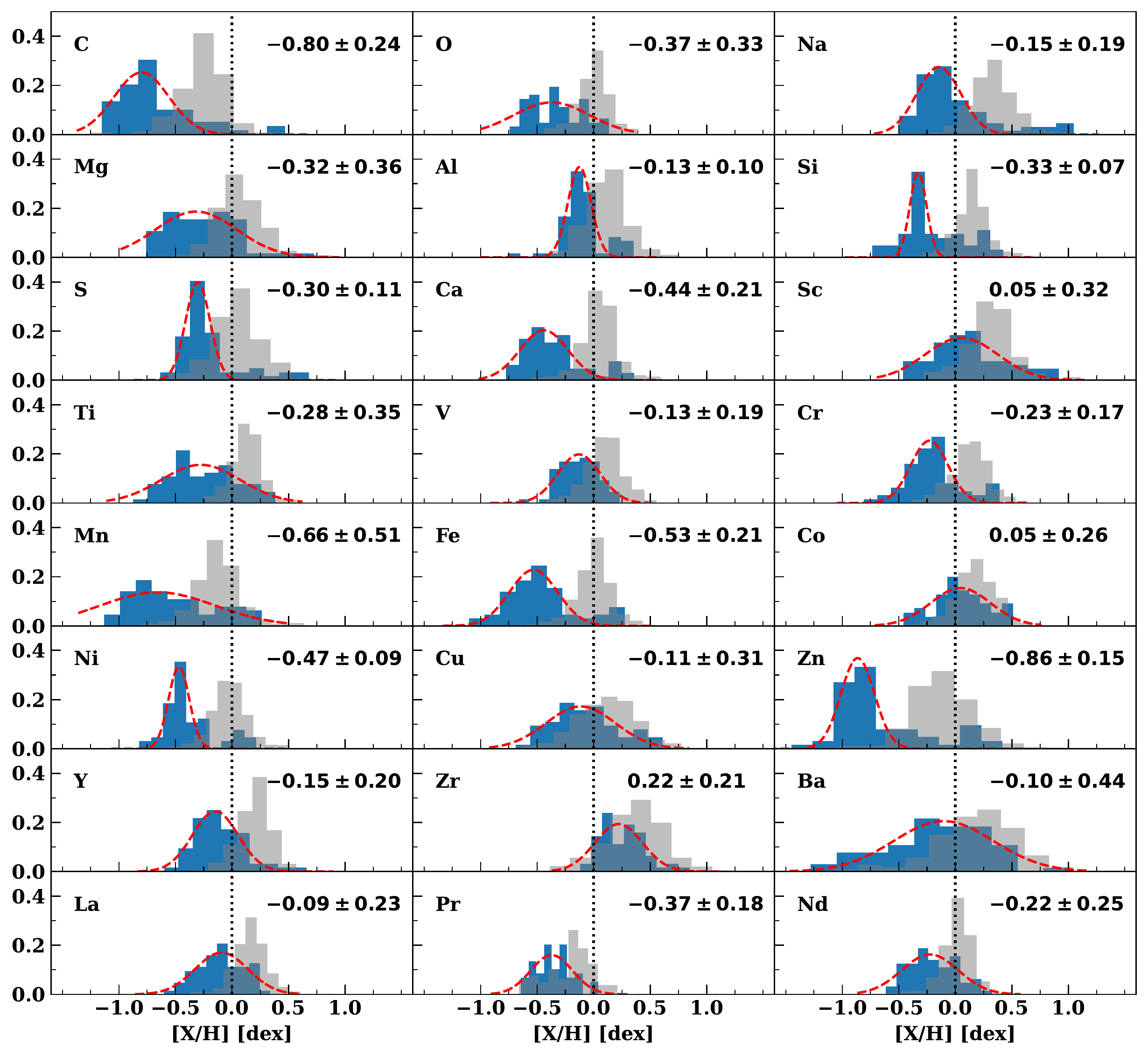}
    \caption{Histograms of the distribution of the chemical abundances derived in this study (blue histograms) compared with literature data as defined in Sect.~\ref{literatueSample} (grey histograms). Gaussian fits have been overimposed in our data (red dashed curves) and the respective mean value and FWHM reported in each box.}
    \label{fig:histograms}
\end{figure*}

In Fig.~\ref{fig:histograms} we show the distribution of the elements derived in this analysis in the form of histograms, where bins have been fixed to 0.15 dex, in order to be representative of the errors. The literature sample is also shown for comparison (see Sect.~\ref{literature}). For each element, The Gaussian fit has been over-imposed with the respective mean value and its FWHM is reported in the upper right corner of each box.

\begin{itemize}
    \item Carbon, Oxygen: the following spectral lines were used for carbon: $\lambda\lambda$ 4932.049, 5052.144 and 5380.0325 Å. The Gaussian fit reports a mean abundance of 0.80 dex below the solar one, with an asymmetrical tail towards higher values. For oxygen, the forbidden line \ion{O}{I} at 6300.304 Å and the \ion{O}{I} triplet at 6155-8 Å, obtaining a broader distribution centered at $-0.37$ dex. No Nitrogen lines could be detected in our observed spectral range.
    \item Sodium: six neutral lines have been found for this element: $\lambda\lambda$ 5682.647, 5688.217, 5889.930, 5895.910, 6154.230 and 6160.753 Å. The mean value is slightly under abundant, at $-0.15$ dex. 
    \item Aluminum: Two aluminum neutral lines have been detected for our spectra, $\lambda\lambda$ 6696.018 and 6698.667 Å, obtaining a distribution slightly under abundant.
    \item $\alpha$-elements: Magnesium (four \ion{Mg}{I} spectral lines: $\lambda\lambda$ 5167.320, 5172.680, 5183.600 and 5528.405 Å), Silicon (four \ion{Si}{II} lines at $\lambda\lambda$ 5041.024, 5055.984, 6347.11 and 6371.370 Å) and Calcium (via \ion{Ca}{I} lines, $\lambda\lambda$ 5598.480, 6162.170, 6166.433, 6169.042, 6169.583, 6462.567, 6471.662, 6493.781, 6499.65 Å) result under-abundant, while Sulfur (three \ion{S}{II} at $\lambda\lambda$ 6743.585, 6748.153, 6757.153 Å) and Titanium (several \ion{Ti}{I} and \ion{Ti}{II} all over the spectrum) are consistent with about $-0.30$ dex under solar abundances although the latter has a wider distribution.
    \item Scandium: Five \ion{Sc}{II} lines were measured at $\lambda\lambda$ 5031.021, 5237.813, 5526.818, 5684.000 and 6245.621 Å. The resulting distribution is centered on about solar value.
    \item Iron peak elements: V, Cr, Mn, Fe, Co, Ni, Co and Zn: four neutral vanadium lines ($\lambda\lambda$  5698.482, 5703.569, 6090.194 and 6243.088 Å) distributed around $-0.13$ dex. Whilst Cobalt is consistent with solar abundances (measured using three \ion{Co}{I} lines at $\lambda\lambda$ 5266.490, 5352.000 and 6450,067 Å), manganese has a spread distribution around -0.66 dex (five \ion{Mn}{I} lines at $\lambda\lambda$ 4823.515, 5341.057, 6013.458, 6016.638 and 6021.790 Å) and Chromium is slightly under abundant (with five \ion{Cr}{II} spectral lines: $\lambda\lambda$ 4824.131, 4876.339, 5204.506, 5206.037, 5208.419 Å), Iron and Nickel (\ion{Ni}{I}: $\lambda\lambda$ 5080.533, 5081.110, 5084.011, 5094.411, 5099.931 Å) are significantly under abundant, with distributions centered at $-0.53$ and $-0,47$ dex respectively. We used two \ion{Cu}{I} neutral lines at $\lambda\lambda$ 5105.500 and 5218.201 Å, with abundances consistent with the solar ones. For zinc only one neutral line could be measured at $\lambda$ 6362.338 Å. The mean value of the distribution is centered on $-0.86$ dex.
    \item Yttrium and Zirconium: with four \ion{Y}{II} lines ($\lambda\lambda$  4883.682, 4900.120, 5087.418, 5402.774 Å) and one \ion{Zr}{II} line ($\lambda$ 6114.853 Å) we obtained two distributions centered at $-0.15$ and 0.22 dex respectively.
    \item Barium: as expected, using four \ion{Ba}{II} spectral lines at $\lambda\lambda$ 4934.100, 5853.625, 6141.713, 6496.897 Å, we estimated abundances centered on $-0.10$ dex but distributed over a wide range, from $\approx$ -1.5 to $\approx$ 1.0 dex.
    \item Rare earth elements: we obtained distributions centered at $-0.09$, $-0.37$ and $-0.22$ dex, respectively for Lanthanum (\ion{La}{II} lines at $\lambda\lambda$ 4921.776, 5114.559 and 5290.818 Å), Praseodymium (one \ion{Pr}{II} line at $\lambda$ 5219.045 Å ) and Neodymium (five \ion{Nd}{II} lines at $\lambda\lambda$ 4959.119, 4989.950, 5092.794, 5293.163, 5688.518 Å).
\end{itemize}
A comprehensive list of all estimated abundances is shown in Table~\ref{tab:abundances}.

The differences between our results and the literature values are largely due to the location of a significant fraction of our targets in the galactic anti-center region, characterized by significantly lower abundances than in regions closer to the galactic center (see discussion in Sect.~\ref{literature}).

\section{Comparison with the literature} \label{literature}



\subsection{Literature sample}
\label{literatueSample}

For the following analysis, we complemented the sample studied in this work with stars from the literature. More in detail, we considered the large compilation of homogenised literature iron abundances for 436 DCEPs presented by \citet[][G18 hereinafter]{Groenewegen2018}, complemented with literature results for a few additional stars by \citet[][GC17 hereinafter]{Gaia2017}. To these data we added the sample of 49 stars presented in our previous works \citep[collectively called R21 hereinafter]{Catanzaro2020,Ripepi2021a,Ripepi2021b} and the recently published large sample of 104 stars by \citet[][K22 hereinafter]{Kovtyukh2022}. There is almost no overlap between the G18/GC17 and the R21 sample, apart from the stars V5567 Sgr and X Sct \citep[see][]{Ripepi2021a}. On the contrary, there is some overlap between K22 and the other samples, including the dataset presented in this paper. More precisely, K22 has 13, 2, 14 and 2 stars in common with G18, GC17, R21 and this paper, respectively. 
In order to use all the data together, we seek for any systematic difference between the K22 and the other datasets. Albeit there is a large scatter, it can be seen that K22 on average tends to overestimate the [Fe/H] value by 0.06$\pm$0.02 dex (the error is the standard deviation of the mean). In order to obtain a literature homogenised sample, we decided to remove from the K22 sample the stars in common with the other datasets and to add -0.06 dex to the [Fe/H] of the remaining 73 stars. Therefore, the total sample of DCEPs with [Fe/H] from high-resolution spectroscopy is therefore composed of 637 objects.  


Concerning the elements other than iron that are not present in G18 or GC17, we adopted the homogeneous and extensive list published by \citet[][L18 hereinafter]{Luck2018} for 435 Galactic DCEPs and merged it with that of K22, as performed above for iron. There were 9 stars in common between L18 and K22 which were removed from the merged list.

\subsection{Element by element comparison}\label{elementbyelement}

\begin{figure*}
	\includegraphics[width=15cm]{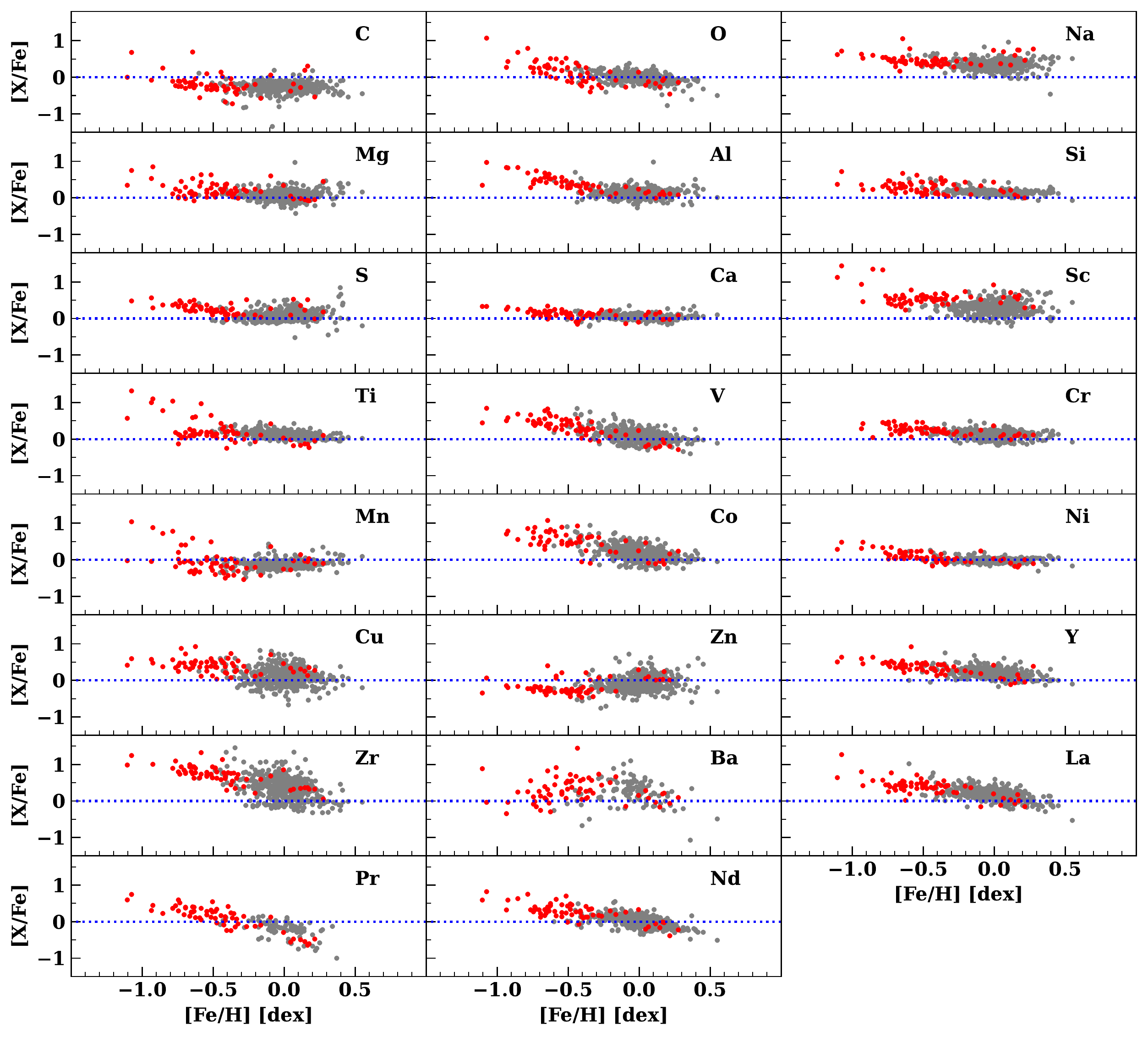}
    \caption{Chemical element abundances (expressed as [X/Fe]) as a function of the iron content for both literature and our sample of stars. Red and gray points are the observed and literature abundances, respectively. The blue dotted horizontal lines point out the solar reference. The corresponding element of each plot is written on the top right.}
    \label{fig:elem_vs_iron}
\end{figure*}

In Fig.~\ref{fig:elem_vs_iron} we plot both our abundances and those from the literature, expressed in the form [X/Fe], as a function of metallicity, expressed as [Fe/H]. 
It is worth mentioning that, as can be seen also in Fig.~\ref{fig:histograms}, most of our stars cover the metallicity region between $-$0.8/$-$0.4 dex, previously less populated. 



In the following, we describe the behavior of each element with respect to the iron content, dividing each group of elements in a similar way as listed in Sect.\ref{Abundances}.

\begin{figure}
	\includegraphics[width=8.5cm]{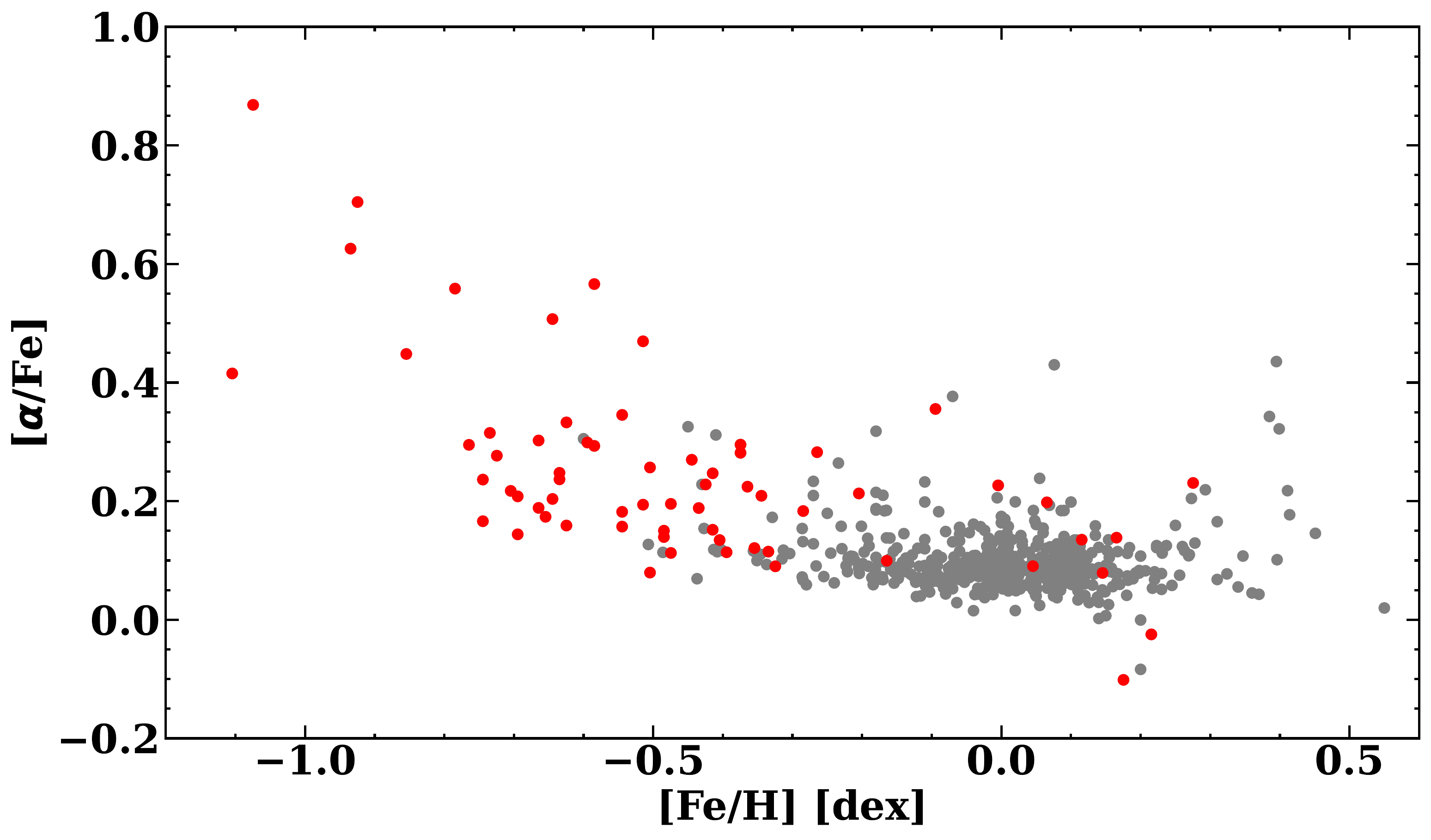}
    \caption{As in Fig.~\ref{fig:elem_vs_iron} but for $\alpha$-elements abundance.}
    \label{fig:alpha_vs_iron}
\end{figure}

\begin{itemize}
    \item Carbon results from the first dredge-up \citet{iben1967stellar}, where the incomplete CN-process outcomes mix to the surface. As for Oxygen, it is mostly released by type II supernovae (SNe II), whose progenitors are young massive stars. 
    Due to their short lifetime, they pollute the interstellar medium essentially instantaneously with respect to the timescale of Galactic evolution, so we expect higher abundances for lower metallicities, with decreasing values as the metallicity increases and other slower mechanisms start to have a major impact. What we find is a negative trend for oxygen, with metal-poor stars having over solar abundances, and an almost flat distribution for carbon. 
    \item both Sodium and Aluminum are odd-z elements that should be produced by core-collapse SNe. Nevertheless, we find a relatively flat trend for sodium and an average overabundance, interpreted by \cite{sasselov1986normal} as the result of the Na-Ne cycle for luminous stars. On the other hand, we find for Al an almost flat trend down to [Fe/H]\,$\approx$\,-0.5 and a rise of abundance toward lower metallicities.
    \item $\alpha$-elements (Mg, Si, S, Ca, Ti): as for Oxygen, the $\alpha$-elements production is due to core collapse supernovae, so we should expect a similar trend as Oxygen. However, in Fig.~\ref{fig:alpha_vs_iron} we plotted  $\alpha$-elements vs. metallicity and we observe a general increase of their abundance toward lower metallicities. 
    \item Scandium can be seen as an intermediate element between the $\alpha$ elements and the iron peak group. Similar to $\alpha$ elements, Scandium is produced in the innermost ejected layers of core-collapse SNe (type II) as reviewed by \cite{romano2010sc}, while the contribution from type Ia SNe seems to be negligible \cite{clay03}.
    For this element we find a quite flat distribution over the metallicities, with the exception of the most metal poor of our sample, that seem to have an overabundance of scandium.
    \item Iron peak elements (V, Cr, Mn, Co, Ni, Cu, Zn) are those elements close to Iron, produced especially by Type Ia supernovae (SN Ia), the final stage of white dwarfs that undergo a binary merging. In general we observe a increase of abundances toward lower metallicities for elements like vanadium, chromium, cobalt and nickel, an almost flat trend for elements such as manganese and copper, while only zinc seems to increase with metallicities.
    \item For heavier elements, starting from Y, there are two main processes both based on neutron-capture.
    The so-called rapid neutron-process (r-process) whose possible channels refer to neutron star mergers, supernova explosions and electron capture SN (see \citet{argast2004neutron, cowan2021origin, surman2008r, korobkin2012astrophysical} and are responsible for the production of Y, Zr, La, Pr and Nd, for which we have found negative trend with metallicities. The slow neutron-process (or s-process) which occurs in low and intermediate mass stars during the AGB phase (\cite{karakas2016stellar}), is the primary source for barium formation. The trend of barium with respect to iron seems to have a maximum close to [Fe/H]\,$\approx$\,-0.2~dex and decrease both toward higher and lower metallicities, but given the high scatter of the measurements no conclusions can be drawn. The cause of this high scatter of the measurements has been noticed in previous papers (see e.g. \citet{simmerer2004,andy2013}) and ascribed to the variation of microturbulent velocities from star to star. 
    
\end{itemize}

The behaviour of each element with respect to the iron abundance is in general agreement with recent results based on high-resolution spectroscopy of static stars belonging to the Galactic disc \citep[see for example][and references therein]{silvaguirre2018,spitoni2019,hayden2017}. A detailed discussion of the evolution of the chemical elements is beyond the scope of this paper. Here we only mention that, in general, the time-delay model \citep[see][for a review]{matteucci2021} represents a robust interpretation for the observed abundance ratio versus metallicity ([Fe/H]) diagrams.

\section{The Galactic radial gradient}

\subsection{Distance estimate}

To estimate the distances $D$ to each  DCEP in our sample, we used the distance modulus definition $w-W$=$-5+5\log_{10} D$, where $w$ and $W$ represent the apparent and absolute Wesenheit magnitudes, respectively. These magnitudes are reddening-free by construction \citep[see][]{Madore1982}, assuming that we know the extinction law. We adopted the Wesenheit magnitude in the \gaia\ bands which was empirically defined by \citet{Ripepi2019} as $w=G-1.90 \times (G_{BP}-G_{RP})$, where $G$, \gbp\ and \grp\ are the magnitudes in the \gaia\ bands. The Wesenheit coefficient was estimated on the basis of the synthetic \gaia\ photometry by \citet[][]{Jordi2010} and then fine-tuned to minimize the spread in the PW relation in the LMC. However, the same coefficient can b safely used in the MW \citep[see][]{Gordon2003}. The absolute Wesenheit magnitude $W$ was calculated according to the period-Wesenheit-metallicity ($PWZ$) relation published by \citet{Ripepi2022}: 

\begin{equation}
\begin{aligned}
W=(-5.988\pm0.018)-(3.176\pm0.044)(\log P-1.0)\\-(0.520\pm0.090){\rm [Fe/H]}
\label{eq:pwz}
\end{aligned}
\end{equation}

\noindent
where P and [Fe/H] are the period and the iron abundance of each DCEP, listed in Table~\ref{tab:radec} and Table~\ref{tab:abundances}. To calculate the $w$ values, we retrieved the \gaia\ magnitudes for our DCEP sample from the EDR3 catalog \citep[][]{Gaia2021}. In principle, we should adopt intensity-averaged magnitudes, that is, calculated over the light curves after transforming the magnitudes into intensities and then converting the resulting mean intensity into magnitude \citep[e.g.][]{Caputo1999,Clementini2016}. However, these magnitudes will be available for all our targets only after the publication of \gaia\ data release 3. Luckily enough, we can safely use the mean magnitudes in the {\it Gaia} EDR3 catalogue because it has been demonstrated that the difference between the Wesenheit magnitudes calculated in the two different fashions is very small \citep[see a detailed discussion on this point in][]{Poggio2021,Ripepi2022}.
The distances can thus be derived from the values of $w$ and $W$. To estimate the uncertainty on the distance, we adopted the  equation $\sigma_D=0.4605 \sigma_\mu D$, where $\sigma_\mu$ is the error on the distance modulus, estimated by adding  in quadrature the errors on $w$ and $W$. The former was estimated by propagating the errors on the {\it Gaia} magnitudes taking also into account the uncertainty of $\pm$0.05 in the Wesenheit coefficient \citep[][]{Ripepi2019}, while for the latter we propagated the errors in the coefficients of Eq.~\ref{eq:pwz}, as well as that on the iron abundance\footnote{The error on the periods is negligible.}. Distances and relative errors are listed in Table~\ref{tab:radec}. Having estimated the distances, we transformed the equatorial coordinates listed in Table~\ref{tab:radec} into the Galactic coordinate system and adopted standard transformations to calculate Galactocentric Cartesian coordinates \citep[see e.g. eq. 4 in][]{Ripepi2019}, using a distance of 8.277 kpc for the Galactic centre \citep[][]{Gravity2022}. From the Cartesian coordinates it was straightforward to calculate the Galactocentric radius $R_{GC}$ for each target.

\begin{table*}
\caption{Results of the Galactic metallicity gradient derived in this work. Literature estimates are also shown for comparison. The radial gradient has the form: ${\rm [Fe/H]}=\alpha \times R_{GC}+\beta$. Columns 1 and 2 list the values of the slope ($\alpha$) and zero point ($\beta$) of the radial gradient; column 3 (when possible) the rms; column 4 reports the number of objects used; column 5 and 6 includes the literature source and the notes, respectively.}
\label{tab:radialGradient} 
\footnotesize\setlength{\tabcolsep}{3pt}
\centering          
\begin{tabular}{cccccc} 
\hline\hline             
$\alpha$ & $\beta$ & rms & n.obj. & Source & Notes\\
(dex kpc$^{-1}$) & (dex)  & (dex) &  & & \\ 
\hline
\multicolumn{6}{c}{\bf This work} \\
\hline
 $-0.060 \pm 0.002$ & $0.573\pm0.017$ & 0.12 & 637   & This Work & $R_{GC}>4$ kpc\\ 
 $-0.064\pm0.007$ & $0.588\pm0.053$  & 0.11 & 333   & This Work & $4$ kpc $\leq R_{GC} \leq$ 9.25 kpc\\
 $-0.080 \pm 0.003$ & $0.829\pm0.038$ & 0.13 & 304   & This Work & $R_{GC}>9.25$ kpc \\ 
 $-0.066 \pm 0.009$ & $0.64\pm0.11$ & 0.07 & 12   & This Work & Bin sample, $R_{GC}>4$ kpc \\ 
 $-0.068 \pm 0.004$ & $0.62\pm0.27$ & 0.04 & 4   & This Work & Bin sample, $4$ kpc $\leq R_{GC} \leq$ 9.25 kpc \\ 
 $-0.078 \pm 0.002$ & $0.81\pm0.26$ & 0.08 & 8   & This Work & Bin sample, $R_{GC}>9.25$ kpc \\ 
 
\hline
\multicolumn{6}{c}{\bf Literature (using Cepheids)} \\
\hline 
 $-0.062 \pm 0.002$ & $0.605 \pm 0.021$ & & 313   & \citet{Luck2011} (L11)& \\ 
 $-0.051\pm0.003$ & $0.49 \pm 0.03$ & & 128 & \citet{Genovali2014} (G14)& UVES and FEROS only\\
 $-0.060\pm0.002$ & $0.57 \pm 0.02$ & & 450 & \citet{Genovali2014} (G14)& Whole sample \\
 $-0.051 \pm 0.002$ &  & & 411   & \citet{Luck2018} (L18)& \\ 
 $-0.045 \pm 0.007$& & &25 & \citet{lemasle2018milky} (Lem18)& Mixed F/1O Cepheids\\
 $-0.054 \pm 0.008$ & $0.52 \pm 0.08$ & & 30 & \citet{Minniti2020} (M20) & Far side of the MW \\
 $-0.062 \pm 0.013$ & $0.59 \pm 0.13$ & & 28 & \citet{Minniti2020} (M20) & $R_{GC}\leq 17 kpc$ \\
 $-0.0527\pm0.0022$ & $0.511 \pm 0.022$ & 0.11 & 489 & \citet{Ripepi2022} (R22)\\
 $-0.055 \pm 0.003$ & $0.43 \pm 0.03$ &  &  & \citet{da2022new} (Da22)\\
 \hline                                   
 \multicolumn{6}{c}{\bf Literature (using Open Clusters)}\\
 \hline
 $-0.035 \pm 0.007$&&&29&\citet{cunha2016chemical} (C16)\\
 $-0.068 \pm 0.017$&&&29&\citet{cunha2016chemical} (C16) & Two-fit line $R_{GC}<12$ kpc\\
 $-0.030 \pm 0.009$&&&&\citet{cunha2016chemical} (C16)& $R_{GC}>12$ kpc\\
 $-0.025 \pm 0.017$&&&7&\citet{cunha2016chemical} (C16)& age < 0.5 Gyr\\
 $-0.086 \pm 0.008$& $0.72 \pm 0.08$&&88&\citet{netopil2016metallicity} (N16)& $R_{GC}<12$ kpc\\
 $-0.066 \pm 0.007$& $0.54 \pm 0.07$&&82&\citet{netopil2016metallicity} (N16)& Excluding outliers\\
 $-0.016 \pm 0.007$& $-0.04 \pm 0.12$&&12&\citet{netopil2016metallicity} (N16)& $R_{GC}>12$ kpc\\
 $-0.079 \pm 0.015$& $0.62 \pm 0.12$&&35&\citet{netopil2016metallicity} (N16)& Age $\leq$ 0.5 Gyr\\
 $-0.06 \pm 0.01$ & & & 11 & \citet{casamiquela2019occaso} (C19)& Age < 1.5 Gyr\\
 $-0.068 \pm 0.004$& & & 128 & \citet{donor2020open} (D20)& Whole sample\\
 $-0.068 \pm 0.004$& & & 71 &\citet{donor2020open} (D20)& Two-fit line, $R_{GC}<13.9$ kpc\\
 $-0.009 \pm 0.011$& & & & \citet{donor2020open}  (D20)& and $R_{GC}>13.9$ kpc\\
 $-0.050 \pm 0.003$& & & 13 & \citet{donor2020open} (D20)& Age $\leq$ 0.4 Gyr\\
 $-0.073 \pm 0.003$& & & 16 & \citet{donor2020open} (D20)& 0.4 $<$ Age $\leq$ 0.8 Gyr\\
 $-0.076 \pm 0.009$& $0.60 \pm 0.08$ & & 134& \citet{spina2021galah} (S21)\\
 $-0.054 \pm 0.008$ & $0.32 \pm 0.07$& 0.116&503&\citet{Gaia2022gaia} (G22) & \\
 \hline
\end{tabular}
\end{table*}

\begin{figure*}
	\includegraphics[width=18cm]{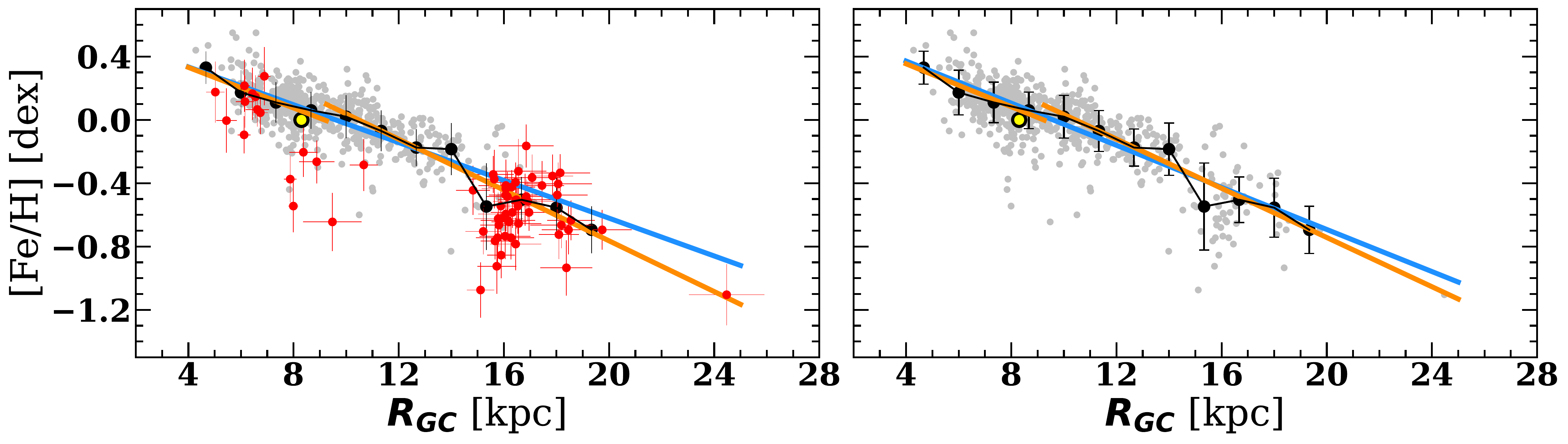}
    \caption{Galactic iron abundance radial gradient. Left: Grey and red dots show the data from the literature and this work, respectively. The big black dots with relative error-bars represent a binning of the data with bins of 1.33 kpc each. The light blue line shows a linear fitting to the data over the entire range of Galactocentric distances, while the two orange lines represent the fit to the data carried out by dividing the sample into two pieces with a break at $R_{GC}$=9.25 kpc. The sun is shown with a yellow-black symbol. Right: Same as left, where all the data are in grey dots, and the lines represent the fit on the binning points.}
    \label{fig:metGradient}
\end{figure*}

    \begin{figure*}
\hbox{
	\includegraphics[width=17cm]{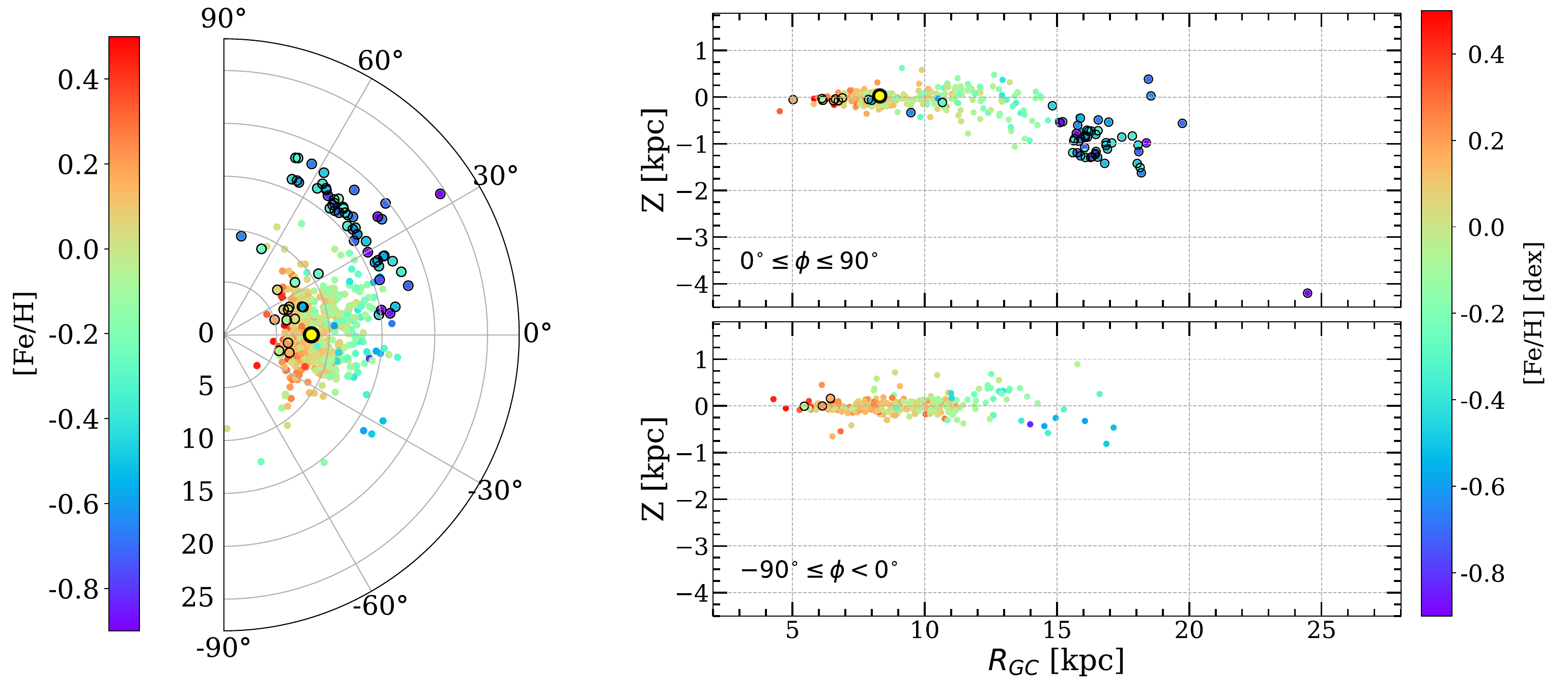}
    }
    \caption{Left panel: polar representation (the radial coordinate is $R_{GC}$) of the 637 Galactic DCEPs with metallicity from high-resolution spectroscopy (see Sect.~\ref{literature}). The colour is coded according to the measured iron abundance. The stars presented in this work are identified by additional black circles. The position of the sun is shown with a yellow-black symbol. Right panel: Cartesian representation of the height above or below the galactic plane as a function of $R_{GC}$ in two opposite directions with respect to the Galactic centre-Sun line (top and bottom panels, respectively). The colour coding and the symbols are as in the left panel.}
    \label{fig:maps}
\end{figure*}


\subsection{Metal radial gradient of the MW disc}\label{metal_gradient}

The relation between the iron abundance (${\rm [Fe/H]}$) and the Galactocentric radius ($R_{GC}$), is displayed in Fig.~\ref{fig:metGradient}, for both our and literature samples. Our data extends significantly the range in $R_{GC}$ over which it is now possible to investigate the metallicity disc gradient, reaching up to 18-20 kpc. It is noticeable the presence of the star ASAS\,J062939-1840.5 with [Fe/H]=$-$1.10$\pm$0.19 dex, which is perfectly placed at the ideal extension of the metallicity disc gradient at 25 kpc. 

We carried out a linear regression adopting the {\tt python LtsFit} package \citep{Cappellari2013}. This software allows one to use weights on both axes and implement a robust outlier removal. We adopted a conservative 3$\sigma$ clipping procedure, obtaining the ${\rm [Fe/H]}$-$R_{GC}$ relation. At a frist glance, it appears that the fitting line does not represent well the data at low values of [Fe/H]. This could be due to the non-uniform sampling of the data. For this reason we operated a binning division of the entire sample in 12 intervals of around 1.33 kpc and subsequently estimated the coefficients of the ${\rm[Fe/H]}$-$R_{GC}$ relation. In the right panel of Fig.~\ref{fig:metGradient} we highlight the binned points and the relative fitting line.
Furthermore, different regimes might be identified in this diagram. Following the suggestion made by \citep{Genovali2014}, it could be possible to distinguish an inner and an outer sample, with a break at about 9.25 kpc. The extension at lower metallicities presented in this work allows us to better verify this early proposal.
Therefore, we decided to quantitatively estimate the metallicity radial gradient in an alternative way, that is dividing the sample into two sub-samples separated at $R_{GC}$=9.25 kpc. We applied this division to both the data and the binning points. 
All the coefficients are listed at the beginning of Table~\ref{tab:radialGradient}. For each relation we estimated the root means square (rms) as a test of the goodness of the fit. The first three relations are over-imposed on the data on the left panel in Fig.~\ref{fig:metGradient}, the last three, corresponding to the fitted relation on the binned points, are plotted on the right panel of the same figure. We point out that in both cases the inner and outer slopes differ at more than 1$\sigma$. The two-samples fits appear to reproduce well both the data and the binned point distributions. This occurrence suggests that the break in the Galactic disc metallicity gradient is plausible, but since the rms are comparable with that of the single line fit, for both the data and the binned points, we can conclude that a single line fit can certainly reproduce the data, but the hypothesis of the break cannot be ruled out on the basis of the present data sample.

It is instructive to inspect the location on the Galactic disc of the sample adopted for the analysis of the metallicity gradient. This is shown in the left panel of Fig.~\ref{fig:maps}, where we show the location of both literature and our samples in polar coordinates. The metal gradient is qualitatively visible also in this representation, and it is noticeable how our outer sample nicely traces a spiral arm of the MW disc, which we identify as the  Outer arm \citep[see e.g.][]{minniti2021using}.
The scatter of iron abundance at this location is of the order of 0.4 dex which is not surprising, as we sampled the Outer arm for more than 15 kpc.
The right panel of Fig.~\ref{fig:maps} shows the height above/below the Galactic disc ($Z$ coordinate of the Galactocentric Cartesian representation) as a function of $R_{GC}$. The sample was divided according to the positive (top panel) or negative (bottom panel) values of the polar angle $\phi$. 
The presence of the disc warp \citep[see e.g.][]{Skowron2019} can be clearly seen, especially in the top panel, where our outer sample is present. Indeed, in the panel many objects have $Z<-0.5$ kpc for $R_{GC}>15$ kpc, while in the bottom panel, in the same region there are very few stars. The lower  visibility of the warp for negative values of $phi$ could be due to the fact that this direction is closer to the line of nodes of the warped disc structure 
\citep[see fig. 2 in][]{Skowron2019}.

Also noticeable is the extreme negative height below the Galactic plane of ASAS\,J062939-1840.5, which appears however in line with the bending of the disc. 
The same figure exhibits a small but noticeable difference in the metallicity distribution between the upper and lower panels, that is at positive and negative azimuthal angles, respectively. Indeed, for $\phi>0$ it seems that the iron abundance decreases more steeply than for $\phi<0$, especially at $R_{GC}>$8--10 kpc which could tentatively be associated with the previously mentioned break in the radial gradient visible in Fig.~\ref{fig:metGradient}.     

\begin{figure*}
	\includegraphics[width=18cm]{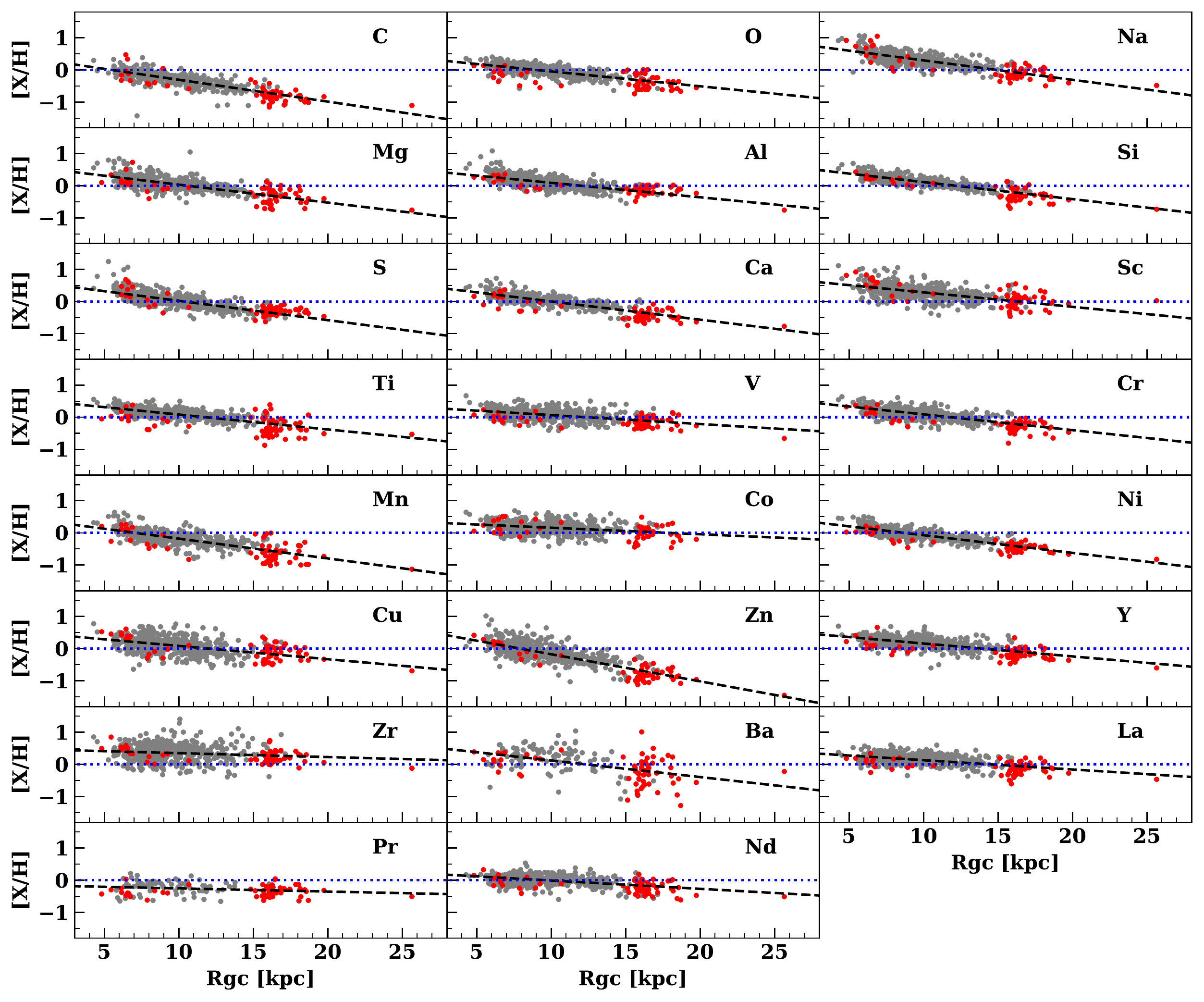}
    \caption{Galactic gradient for all the estimated abundances for our sample of stars (in red) and the literature stars (in gray, see section ~\ref{literature}). The black dashed line represents the linear fit computed using both literature and our stars, while the dotted blue horizontal line highlights the solar abundance. The coefficients of the fits are listed in Table ~\ref{tab:slopes}. }
	\label{fig:grad_elem}
\end{figure*}

The relation between DCEPs metal abundance and Galactocentric radius can be investigated for elements other than iron. We noted hints of the same break at $R_{GC}$=9.25 kpc that was suggested for the iron in several elements of his group, especially in V, Cr, Mn and Ni. However, the two-lines fits for these elements were not convincing, so that we only report single line fits for these and all elements other than iron. The result of this procedure is shown in Fig.~\ref{fig:grad_elem} and Table~\ref{tab:slopes}.

\input{fits_results}

\section{Discussion}

\subsection{Iron abundance disc gradient}
Several works focused their attention on the radial gradient of the MW disc, using as tracers either DCEPs or other objects such as OCs, groups of stars that formed at the same time from the same material and therefore have similar distances from the sun as well as the same chemical composition and kinematics. They have been used to study Galactic chemical trends since \citet{janes1979evidence} where it has been shown how they could be treated as tracers of the Galactic iron gradient. Furthermore, OCs span a wide range of ages, from Myr to Gyr, making them optimal tools for Galactic evolution studies as well. Since DCEPs are young stars (spanning approximately the age range between 20 and 500 Myr) we compared, when possible, results from OCs studies that split their sample into different age intervals.

Some estimates for the metal gradient of the Galactic disc, based on both DCEPs and OCs studies are listed in Table~\ref{tab:radialGradient}, and compared graphically with our work in Fig.~\ref{fig:literature}, showing the results only relative to the data and not to the binned points, since they are comparable within the errors.

We point out that most of the previous gradient estimations come from objects with distances up to $\approx$10-15 kpc, so for the following discussion we took in consideration, except when specified, the gradients from the whole sample and the "inner" slope ($R_{GC}<9.25 kpc$) from the two-fit case. In more detail:

\begin{itemize}
    \item \citet{Genovali2014} estimated two slopes, first considering UVES, NARVAL and FEROS spectra only and then including also the literature data. The slope obtained in the second case is steeper than the first and in good agreement with our results for both single and two-fit lines (when considering our "inner" slope). Although the same agreement was found with \citet{Luck2011} and \citet{da2022new}, other recent studies \citep{Luck2018, lemasle2018milky, Ripepi2022} estimated flatter slopes, but comparable within 2$\sigma$.  It is interesting the method used by \citet{lemasle2018milky}, where the metallicity of Galactic F/1O\footnote{F/1O indicates DCEPs pulsating simultaneously in the fundamental and first overtone mode.} DCEPs was estimated from the $P_1/P_0$ ratio. The low number of stars used should increase in future Gaia releases.
    
    \item \citet{cunha2016chemical}, studying APOGEE OCs, found a generally flatter slope than those provided by our and other literature DCEPs studies. It is worth mentioning the presence in \citet{cunha2016chemical}'s work of an object distant 25 kpc. Following the possible break at about 10-12 kpc hypothesized for OCs as well (see e.g. \citet{frinchaboy2013open, yong2012elemental, magrini2010open}, the sample was divided into an inner region, where the slope of the gradient is as steep as ours, and a flatter outer slope (not shown in Fig.~\ref{fig:literature}), in contrast with the results obtained using the DCEPs as tracers. Moreover, fitting only young clusters (age$<$500 Myr) resulted in a very low slope, which is in complete disagreement with all DCEPs results. But, since this result is based on only 7 OCs, it has a very limited statistical value.
 \begin{figure*}
	\includegraphics[width=15cm]{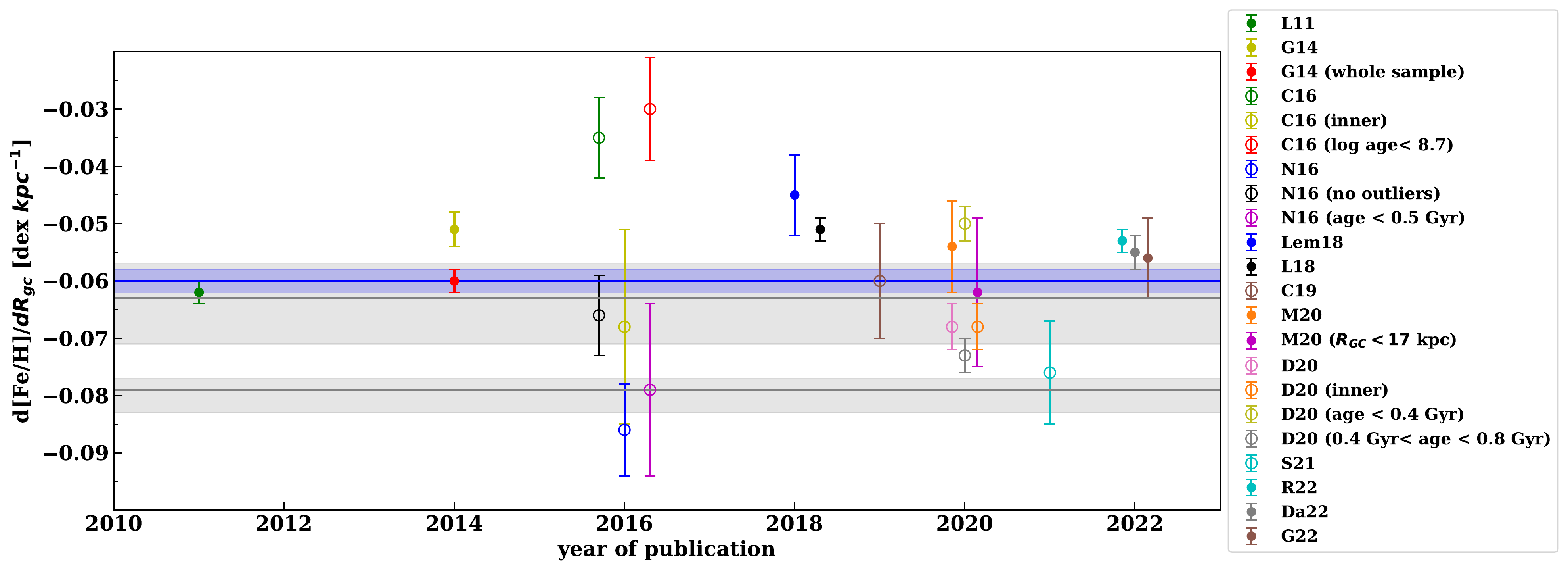}
    \caption{Radial gradient comparison between our work and previous ones listed in Table~\ref{tab:radialGradient}. Open circles symbols represent results from OCs works, filled circles symbols from DCEPs ones. Colors for different works (or same work but different sample configuration) are described in detail in the legend on the right, with reference acronyms the same as Table~\ref{tab:radialGradient}. Results are plotted in order of year publication on the x-axis. In order to avoid overlapping, results from the same work have been slightly shifted on the x-axis. Blue and gray horizontal lines represent the results in our work in the case of one and two-line fitting on the data, while shaded regions correspond to the 1$\sigma$ dispersion around the respective results.}
	\label{fig:literature}
\end{figure*}
   
    \item \citet{netopil2016metallicity} studied the iron gradient using about 100  OCs in the MW disc. They estimated the gradient in different ways, subdividing their sample according to the $R_{GC}$ or age (see Table~\ref{tab:radialGradient}). Using all the OCs within 12 kpc they found a slope significantly steeper than ours both using the whole sample and for $R_{GC}<9.25$ kpc. The same is true for the comparison with the other literature DCEPs results. The agreement with all DCEPs results improves when \citet{netopil2016metallicity} exclude outliers from the calculation. Similarly to  \citet{cunha2016chemical}, and at odds with DCEPs results, also \citet{netopil2016metallicity} found that the iron gradient flattens for $R_{GC}>12$ kpc. However, this result is based on only 12 objects and it should be verified with a more consistent sample.     
    Finally, \citet{netopil2016metallicity} divided their sample into three different age bins. Considering the sample with age<0.5 Myr, that is the age range spanned by DCEPs, they find a steep slope, similar to what we find for $R_{GC}>9.25$ kpc, but in complete disagreement with \citet{cunha2016chemical} in the same OCs age range. This occurrence seems to suggest that the results based on young OCs are still significantly sample dependent. 
    
    \item \citet{casamiquela2019occaso} investigated the Galactic gradient using new high-resolution spectroscopy for 18 OCs which complemented with a compilation of data by \citet{carrera2019open}, finding a present-day gradient for the youngest clusters (having $R_{GC}<13$ kpc) in good agreement with our results for the whole and the inner DCEPs samples. 
    
    \item Studying OCs from SDSS/APOGEE DR16, \citet{donor2020open} fitted the data using a two-fit line, obtaining a break at $R_{GC}$ 13.9 kpc, after which the slope is comparable with a null one, thus confirming the results by \citet{cunha2016chemical} and \citet{netopil2016metallicity}, but at variance with ours. However, again this result is based on only a few OCs, which in addition are much older than the typical DCEPs and thus could have migrated from other disc regions, thus not being representative of the present-day iron gradient. As for previous studies involving OCs, the sample was divided in 4 bins of age. The results found for ages lower than 0.4 Gyr and between 0.4 and 0.8 Gyrs are sensitively different from each other, but both in agreement with our inner slope within 2$\sigma$.
    \item  On the far side of the MW, beyond the bulge,  \citet{Minniti2020} characterized 30 DCEPs using near-infrared VVV photometry, and estimated two different slopes using either the whole sample or DCEPs with $R_{GC}\leq 17$ kpc. Both values are in agreement with our inner slope.
    \item  \citet{spina2021galah}, used OCs data from the APOGEE and GALAH surveys to estimate the iron gradient up to $R_{GC}\sim$13.9 kpc, obtaining a slope steeper than ours, but still in agreement within 1$\sigma$ with our "inner" slope and 2$\sigma$ with the whole sample slope.
    \item It is worth mentioning two last papers: \citet{Gaia2022gaia} based on the {\it Gaia} DR3 results and \citet{randich2022gaia}, which review the outcomes of the {\it Gaia} ESO survey (GES). 
    In the former, the radial metallicity gradient was estimated for 503 open clusters. Splitting the sample in 4 groups depending on ages, they found a flattening of the gradient slope going toward larger ages (see their sect. 8 for more details). In Table~\ref{tab:radialGradient} and Fig.~\ref{fig:literature} we report the radial gradient result obtained by \citet{Gaia2022gaia} with the whole OC sample, which is in good agreement with our outcome.  
    
    \citet{randich2022gaia}  presented an overview of the GES survey results for the Galactic OCs, comprising a study of the radial gradient. They confirm the flattening of the slope at higher distances, but again the gradient greatly depends on the age of the adopted clusters (see their fig. 26).
\end{itemize}

\subsection{Disc gradient for elements other than iron}

\begin{figure*}
	\includegraphics[width=16cm,bb=18 290 592 618]{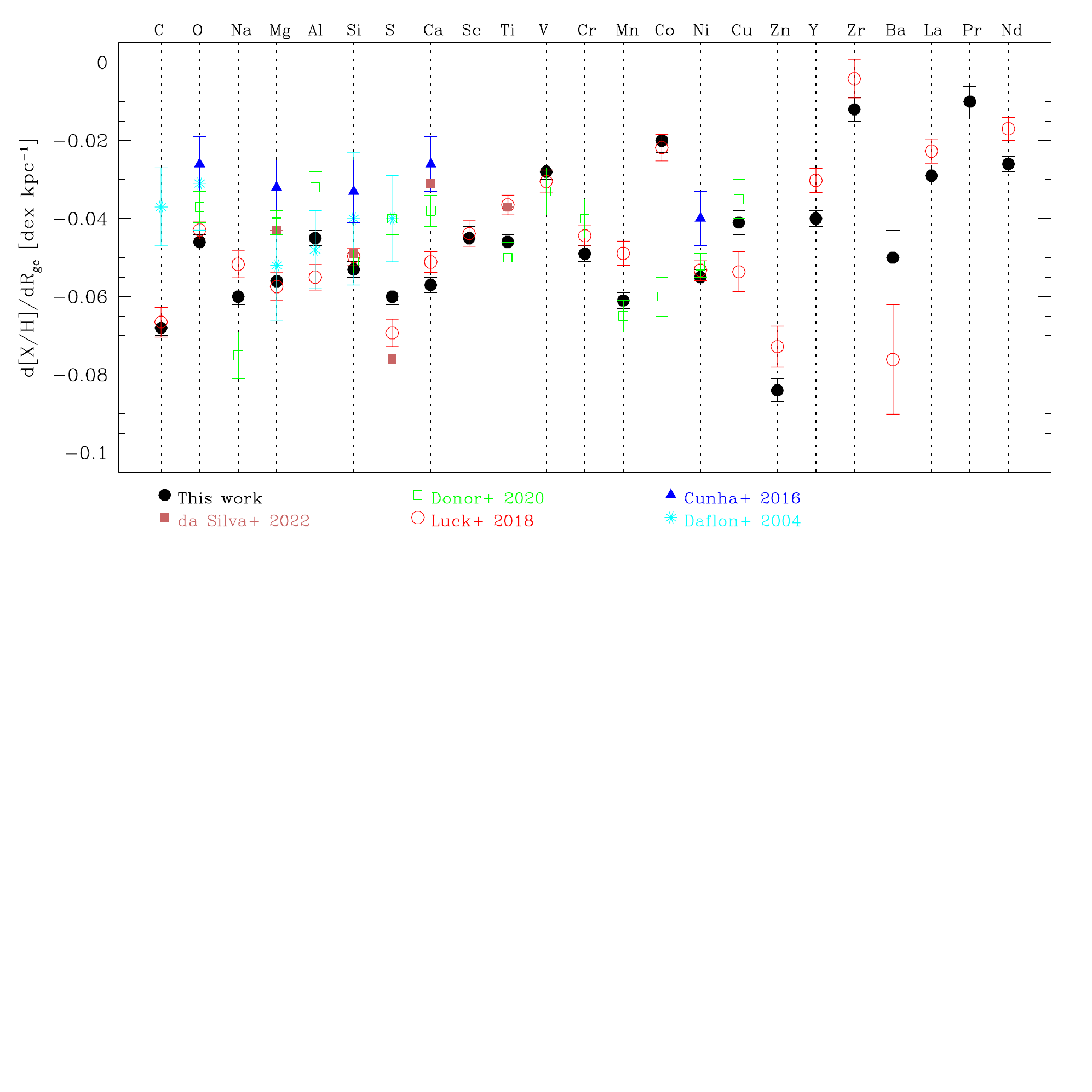}
    \caption{Radial gradients for each chemical species inferred in this study compared with literature.}
	\label{fig:grad_others}
\end{figure*}

In this section we summarize and discuss gradients inferred from the other elements shown in Fig.~\ref{fig:grad_elem} and compare our results with other studies. 
In Fig.~\ref{fig:grad_others} we compare gradients derived in this study with results obtained by different authors using different tracers, for instance OCs \citep{donor2020open,cunha2016chemical}, B-type stars \citep{daflon2004}, and DCEPs \citep{da2022new,Luck2018}. To be comparable with our results, since DCEPs are young objects, we selected from \citet{donor2020open} only clusters younger than 0.4 Gyr.

For light elements such as carbon our results are consistent with \citet{Luck2018} while \citet{daflon2004} analysing a sample of B-type stars found a lower value. This perhaps could be due to the evolution in DCEPs, that alters the abundance of carbon in non homogeneous ways after the first dredge-up. 
For oxygen, our gradient is consistent with all the determinations found in literature, even when B-type stars are used as tracers as in \citet{daflon2004}. This is not surprising since oxygen abundance in DCEPs does not alter during the first dredge-up. As for carbon, sodium abundance is altered by evolution in DCEPs via the Na-Ne cycle \citep{sasselov1986normal}, so our gradient is consistent with \citet[][]{Luck2018}'s value while \citet[][]{donor2020open} found a steeper gradient.

Gradients for $\alpha$-elements derived in this paper are consistent at least within 2$\sigma$ with other studies, an exception is sulfur for which results from open cluster and B-type stars are not consistent with those from DCEPs. For the other light elements, namely aluminum and scandium, results from various sources are in good agreement.   

With the exception of cobalt, our gradients for iron-peak elements are consistent with the literature. For cobalt, \citet[][]{donor2020open} found a steeper slope. 

Also for heavy elements (Z > 30) our slopes are in agreement with the literature, at least at 2$\sigma$ level.

\section{Summary}

In the context of the C-MetaLL project, we presented high-resolution UVES@VLT spectra for a sample of 65 DCEPs mostly located in the outskirts of the MW disc, at Galactocentric distances larger than 15 kpc. 
We analysed the observed spectra in detail to derive the main parameters for each target, such as effective temperature, surface gravity, microturbulence and radial velocity. It is worth noticing that we have carried out the effective temperature estimate using two different procedures: by adopting the LDR method and by minimizing the correlation between the abundance and the excitation potential. The two different approaches provided consistent results.  

On this basis, we have derived the chemical abundances of 24 species for which we have detected and measured spectral lines. Studying their distribution we showed how they are sensibly under abundant with respect to the sun, except for Sc, Co and Zr, having nonetheless (as for the other elements) a mean abundance lower than the literature one.

Our sample includes objects in the range $-1.1<$[Fe/H]$<+0.3$ dex, but the majority of the stars have iron abundances lower than about $-$0.3 dex, constituting the most metal-poor DCEPs ever studied with high-resolution spectroscopy and extending the metallicity range of the MW DCEPs even below the metallicity of the Small Magellanic Cloud (SMC) DCEPs \citep[see][]{Romaniello2008}. We analyzed all the chemical abundances as a function of iron. All of them either decrease or remain constant as iron increases, with the only exception of Zn that shows a positive trend.

We complemented our sample with recent literature results to obtain a data-set including 637 confirmed DCEPs with individual metallicities from high-resolution spectroscopy. For all these objects, we adopted \gaia\ EDR3 photometry and the iron abundance to calculate accurate distances based on the PWZ relation in the \gaia\ bands by using the same method as in \citet{Ripepi2022}. The resulting spatial distribution of our outer sample clearly depicts the shape of a spiral arm extending for about 60 deg in azimuth at approximately constant Galactocentric distance between 16-18 kpc. Based on the comparison with  \citet{minniti2021using} disc model, we identify this spiral arm as the Outer. 

The distribution of DCEPs in height below/above the Galactic disc plane reveals a larger scatter  for $\phi>0$ and $R_{GC}>10$ kpc while the disc warp starts to be visible at  at $R_{GC}>10$ kpc. Our data allows us to conclude that the disc continues to bend with the same slope also between 15-18 kpc and possibly up to 25 kpc, even if we have only one star at this distance and no one between 20 and 25 kpc. On the other hand, for $\phi<0$ the warp is barely visible, but we do not have enough data beyond $R_{GC}>13$ kpc, so that no firm conclusion can be driven. Also the metallicity distribution appears different  moving from negative to positive $\phi$ values as in the latter the iron abundance seems to be more metal poor at fixed $R_{GC}$ values for $R_{GC}$ larger than $\sim$12 kpc.       

We studied the iron abundance Galactic radial gradient, which can now be studied up to 20 kpc from the Galactic centre. The analysis of the data plausibly revealed the presence of an already suggested break in the distribution around 9.25 kpc which led us to divide the DCEP data-set into inner and outer samples. A linear fit to the two samples allows us to estimate two different slopes: $-0.063 \pm 0.007$ and $-0.079 \pm 0.003$ dex kpc$^{-1}$ for the inner and outer samples, respectively. The two slopes are different at more than 1 $\sigma$. A fit to the entire sample provides instead a slope of $-0.060 \pm 0.002$ dex kpc$^{-1}$, which is in very good agreement with previous literature determinations of the gradient using DCEPs, all conducted without any separation in the sample. Since the rms are comparable, we conclude that the hypothesis of the presence of the break cannot be ruled out. Future works with a more homogeneous and extended sample will allow us to further test the plausibility of the presence of such a break.

We also carried out a detailed comparison between our results and the radial metallicity gradients estimated using the OCs. As a result we found good agreement between the inner slope and most of the previous works. On the other hand, our outer slope reflects a steeper trend, opposite to what is found for OCs for which, in general, the relation becomes almost flat. However, if we restrict the comparison to OCs young enough to have not undergone a significant migration, we find a good agreement with the work by \citet[][for OCs with Age$<$0.5 Gyr]{netopil2016metallicity} and \citet[][for OCs with 0.4$<$Age$<$0.8 Gyr]{donor2020open}. A good agreement for the outer slope is also found with the work by \citet{spina2021galah} which does not show any slope difference with the age of the investigated OCs.\\ 
Finally, we studied Galactic radial gradients for the elements other than iron. Our gradients are consistent with the literature.

\section*{Acknowledgements}

We thank the Referee for her/his helpful and insight comments, which helped to improve the manuscript.
This work has made use of data from the European Space Agency
(ESA) mission Gaia (https://www.cosmos.esa.int/gaia), processed
by the Gaia Data Processing and Analysis Consortium (DPAC, https:
//www.cosmos.esa.int/web/gaia/dpac/consortium). Funding for the
DPAC has been provided by national institutions, in particular the
institutions participating in the Gaia Multilateral Agreement. In
particular, the Italian participation in DPAC has been supported by
Istituto Nazionale di Astrofisica (INAF) and the Agenzia Spaziale
Italiana (ASI) through grants I/037/08/0, I/058/10/0, 2014-025-R.0,
and 2014-025-R.1.2015 to INAF (PI: M.G. Lattanzi).
This research has made use of the
SIMBAD database, operated at CDS, Strasbourg, France.

\section*{Data Availability}

The only proprietary data used in this paper are represented by the
UVES spectra. They are freely retrievable from the ESO archive also in the fully reduced and calibrated form. Any additional data request can be directed to the corresponding author (E. Trentin).



\bibliographystyle{mnras}
\bibliography{example} 



\appendix

\section{Abundances for target stars}

\input{T22_abund_first}

\input{T22_abund_second}

\input{T22_abund_third}

In this appendix we present the complete list of the chemical abundances for the 24 species detected in our stars. 


\bsp	
\label{lastpage}
\end{document}

%% file: T22_ra_dec.tex
\begin{table*}
\caption{Properties of the observed stars. From left to right, the columns report: Identification of the star; source ID from Gaia; right ascension and declination at J2000; period in days; pulsation mode  (fundamental F, first overtone 1O, mixed$-$mode (both first and second overtone 1O/2O); G magnitude; distance (in kpc); both exposure time (in seconds) and period of observation (P105 or P106); signal to noise ration at the wavelengths: 5050 and 6300 Å.}
\label{tab:radec}
\begin{adjustbox}{width=\textwidth}
\begin{tabular}{|l|c|r|r|r|c|c|c|r|c|r|}
\hline
  \multicolumn{1}{|c|}{Star} &
  \multicolumn{1}{c|}{GaiaID} &
  \multicolumn{1}{c|}{Ra } &
  \multicolumn{1}{c|}{Dec} &
  \multicolumn{1}{c|}{P} &
  \multicolumn{1}{c|}{Mode} &
  \multicolumn{1}{c|}{$G_{mag}$ } &
  \multicolumn{1}{c|}{distance } &
  \multicolumn{1}{c|}{$T_{exp}$} &
  \multicolumn{1}{c|}{S/N} \\
  & &(J2000)~~~&(J2000)~~~&(days)~~~~~& &(mag)&(kpc)~~&(s)~~~~~~~\\
\hline
ASAS J060450+1021.9 & 3329807983320140160 & 91.20833 & 10.36348 & 3.075512 & 1O & 12.403 & 8.44 $\pm$ 0.58 & 900 P106 & 69/90\\ 
ASAS J062939-1840.5 & 2940212053953709312 & 97.41343 & $-$18.67407 & 16.942502 & F & 12.803 & 19.82 $\pm$ 1.23 & 1200 P106 & 48/83\\ 
ASAS J064001-0754.8 & 3099655288815382912 & 100.00495 & $-$7.91417 & 1.604003 & 1O & 13.194 & 8.37 $\pm$ 0.50 & 1800 P106 & 63/83\\ 
ASAS J065758-1521.4 & 2947876298535964416 & 104.49359 & $-$15.35740 & 2.415850 & 1O & 13.219 & 10.16 $\pm$ 0.60 & 1500 P106 & 58/71\\ 
ASAS J074401-3008.4 & 5598852026287511424 & 116.00046 & $-$30.14147 & 3.390517 & F & 13.248 & 9.80 $\pm$ 0.62 & 1800 P106 & 89/112\\ 
ASAS J074925-3814.4 & 5538569613358406144 & 117.35519 & $-$38.23936 & 10.503835 & F & 12.675 & 11.93 $\pm$ 0.69 & 1200 P106 & 58/85\\ 
ASAS J084127-4353.6 & 5522522863932402432 & 130.36186 & $-$43.89295 & 25.364784 & F & 12.027 & 5.82 $\pm$ 0.38 & 1500 P105 & 42/87\\ 
ASAS J164120-4739.6 & 5942521668514052608 & 250.33354 & $-$47.66079 & 13.018153 & F & 11.653 & 4.02 $\pm$ 0.23 & 1200 P105 & 10/62\\ 
ASAS SN-J061713.86+022837.1 & 3124796657276655488 & 94.30773 & 2.47701 & 2.019309 & F & 14.447 & 9.41 $\pm$ 0.64 & 3000 P106 & 49/70\\ 
ASAS SN-J063841.36-034927.7 & 3104095494729372032 & 99.67234 & $-$3.82435 & 3.860779 & F & 14.003 & 10.45 $\pm$ 0.65 & 2400 P106 & 42/64\\ 
ASAS SN-J072739.70-252241.1 & 5613685331497869312 & 111.91542 & $-$25.37809 & 2.760085 & 1O & 13.879 & 10.30 $\pm$ 0.67 & 1800 P106 & 40/57\\ 
ASAS SN-J074354.86-323013.7 & 5594991812757424768 & 115.97856 & $-$32.50378 & 3.149330 & F & 14.855 & 11.00 $\pm$ 0.76 & 3000 P106 & 20/43\\ 
ASAS SN-J091822.17-542444.5 & 5310669788148987520 & 139.59224 & $-$54.41227 & 13.120770 & F & 13.671 & 16.84 $\pm$ 1.05 & 1800 P106 & 45/81\\ 
ATLAS J102.7978-10.2541 & 3050117174686674560 & 102.79785 & $-$10.25419 & 4.908117 & 1O & 13.758 & 10.43 $\pm$ 0.68 & 1800 P106 & 42/72\\ 
ATLAS J106.7120-14.0234 & 3044465577537521792 & 106.71207 & $-$14.02341 & 1.394987 & 1O & 14.707 & 9.30 $\pm$ 0.61 & 3000 P106 & 33/54\\ 
ATLAS J113.8534-31.0749 & 5599120204043561600 & 113.85342 & $-$31.07500 & 1.748419 & 1O & 13.583 & 10.50 $\pm$ 0.65 & 1800 P106 & 54/64\\ 
BQ Vel & 5516452460238707968 & 125.52707 & $-$47.30396 & 3.372479 & F & 14.251 & 13.85 $\pm$ 0.92 & 2400 P106 & 43/63\\ 
GDS J133950.2-634049 & 5864243376958232576 & 204.95960 & $-$63.68047 & 3.791978 & 1O & 12.354 & 3.36 $\pm$ 0.21 & 3000 P105 & 40/70\\ 
OGLE GD-CEP-0029 & 3344577418076306944 & 93.72745 & 13.87880 & 5.992352 & F & 14.975 & 6.71 $\pm$ 0.43 & 3000 P106 & 11/31\\ 
OGLE GD-CEP-0089 & 2936063665309240576 & 107.55532 & $-$15.19940 & 1.975315 & 1O & 13.548 & 9.16 $\pm$ 0.61 & 1800 P106 & 49/67\\ 
OGLE GD-CEP-0120 & 5614916921960857088 & 114.04436 & $-$25.36404 & 7.959546 & F & 14.498 & 14.31 $\pm$ 0.87 & 3000 P106 & 20/46\\ 
OGLE GD-CEP-0123 & 5615958881027868544 & 115.24714 & $-$22.49207 & 1.399290 & F & 15.521 & 12.83 $\pm$ 0.78 & 4800 P106 & 29/49\\ 
OGLE GD-CEP-0127 & 5586922973654572416 & 116.29912 & $-$36.97476 & 13.766689 & F & 12.217 & 10.92 $\pm$ 0.65 & 1800 P106 & 103/154\\ 
OGLE GD-CEP-0134 & 5594793935021492992 & 117.62511 & $-$32.77881 & 1.978211 & 1O & 15.703 & 10.75 $\pm$ 0.71 & 4800 P106 & 17/34\\ 
OGLE GD-CEP-0156 & 5546003793739552896 & 119.74496 & $-$34.07070 & 2.488076 & 1O & 14.639 & 11.85 $\pm$ 0.70 & 3000 P106 & 31/53\\ 
OGLE GD-CEP-0159 & 5534744416711687424 & 119.86020 & $-$40.02831 & 2.368497 & F & 14.827 & 12.39 $\pm$ 0.76 & 3000 P106 & 32/54\\ 
OGLE GD-CEP-0162 & 5534048803808721664 & 120.14347 & $-$41.68488 & 3.516367 & F & 14.569 & 11.91 $\pm$ 0.69 & 3000 P106 & 39/61\\ 
OGLE GD-CEP-0168 & 5543972789600867328 & 121.20575 & $-$38.05453 & 2.116514 & 1O & 15.065 & 12.01 $\pm$ 0.71 & 3000 P106 & 26/42\\ 
OGLE GD-CEP-0176 & 5533237604747205120 & 122.05544 & $-$43.25774 & 1.502544 & 1O & 13.934 & 12.01 $\pm$ 0.78 & 2400 P106 & 54/73\\ 
OGLE GD-CEP-0179 & 5540666145820286848 & 122.38303 & $-$39.20109 & 2.236739 & 1O & 15.040 & 12.48 $\pm$ 0.90 & 3000 P106 & 24/41\\ 
OGLE GD-CEP-0181 & 5521137101325930624 & 122.45845 & $-$43.97273 & 2.342181 & F & 14.327 & 12.71 $\pm$ 0.79 & 3000 P106 & 53/67\\ 
OGLE GD-CEP-0185 & 5520935444020919296 & 122.93821 & $-$44.17211 & 2.540510 & F & 15.004 & 13.05 $\pm$ 0.88 & 3600 P106 & 29/57\\ 
OGLE GD-CEP-0186 & 5533407170051636608 & 122.95172 & $-$42.05165 & 4.764552 & F & 14.786 & 13.17 $\pm$ 0.82 & 3000 P106 & 22/45\\ 
OGLE GD-CEP-0196 & 5520993099661921536 & 123.96589 & $-$44.01581 & 2.808806 & F & 15.302 & 12.51 $\pm$ 0.79 & 3600 P106 & 31/48\\ 
OGLE GD-CEP-0206 & 5519997491885313280 & 125.34584 & $-$44.51655 & 1.681608 & 1O & 15.388 & 12.14 $\pm$ 0.84 & 3600 P106 & 16/31\\ 
OGLE GD-CEP-0213 & 5516789701069413504 & 125.61290 & $-$46.21602 & 1.446897 & 1O & 14.776 & 12.99 $\pm$ 0.84 & 3000 P106 & 36/50\\ 
OGLE GD-CEP-0214 & 5543722994310379136 & 125.64924 & $-$34.40032 & 2.039692 & 1O & 14.330 & 14.30 $\pm$ 0.93 & 3000 P106 & 54/74\\ 
OGLE GD-CEP-0224 & 5515778734493663232 & 127.31511 & $-$46.74925 & 2.716676 & F & 14.837 & 12.79 $\pm$ 0.83 & 3000 P106 & 45/64\\ 
OGLE GD-CEP-0228 & 5515442352654094976 & 127.55723 & $-$48.20902 & 2.596410 & F & 15.176 & 15.45 $\pm$ 1.00 & 3600 P106 & 33/48\\ 
OGLE GD-CEP-0247 & 5329302696287713152 & 129.39239 & $-$47.57711 & 1.649005 & F & 15.682 & 14.36 $\pm$ 0.91 & 4800 P106 & 23/38\\ 
OGLE GD-CEP-0252 & 5322878146768508800 & 129.73195 & $-$49.77017 & 3.184702 & F & 14.617 & 13.92 $\pm$ 0.85 & 3000 P106 & 48/71\\ 
OGLE GD-CEP-0271 & 5318708901756763520 & 131.02836 & $-$51.91946 & 2.984789 & F & 15.076 & 16.25 $\pm$ 1.03 & 3600 P106 & 31/46\\ 
OGLE GD-CEP-0316 & 5310969267622303872 & 136.56212 & $-$54.93719 & 2.927582 & F & 14.586 & 16.84 $\pm$ 1.01 & 3000 P106 & 40/58\\ 
OGLE GD-CEP-0342 & 5310717135870439424 & 139.24329 & $-$53.96897 & 2.920501 & 1O & 14.904 & 14.48 $\pm$ 0.85 & 3000 P106 & 29/44\\ 
OGLE GD-CEP-0348 & 5310642025484526208 & 139.98489 & $-$54.44281 & 2.404135 & F & 15.225 & 14.67 $\pm$ 0.88 & 3600 P106 & 36/49\\ 
OGLE GD-CEP-0353 & 5306782361701536128 & 140.15438 & $-$56.85402 & 4.862639 & F & 13.944 & 14.89 $\pm$ 0.91 & 1800 P106 & 41/54\\ 
OGLE GD-CEP-0516 & 5255256669866274816 & 156.87433 & $-$59.35961 & 0.394959 & 1O/2O & 12.462 & 2.72 $\pm$ 0.18 & 3000 P105 & 81/105\\ 
OGLE GD-CEP-0568 & 5350409780509559040 & 160.82905 & $-$59.19156 & 45.449567 & F & 12.227 & 5.20 $\pm$ 0.34 & 1500 P105 & 14/53\\ 
OGLE GD-CEP-0575 & 5350281236399037696 & 161.68646 & $-$60.26828 & 6.611049 & F & 12.048 & 2.80 $\pm$ 0.18 & 1500 P105 & 22/54\\ 
OGLE GD-CEP-0889 & 5859114571927308672 & 198.46073 & $-$64.44122 & 45.179934 & F & 13.232 & 10.65 $\pm$ 0.70 & 3000 P105 & 18/46\\ 
OGLE GD-CEP-0974 & 5854021702729341952 & 213.76026 & $-$62.73593 & 12.540476 & F & 12.675 & 3.23 $\pm$ 0.20 & 2400 P105 & 16/63\\ 
OGLE GD-CEP-0996 & 5878506555352293888 & 219.96841 & $-$60.56111 & 6.750565 & F & 12.727 & 2.17 $\pm$ 0.15 & 453 P105 & 10/21\\ 
OGLE GD-CEP-1012 & 5877982466278766208 & 223.59745 & $-$60.45781 & 15.967508 & F & 12.268 & 3.53 $\pm$ 0.23 & 1800 P105 & 36/52\\ 
OGLE GD-CEP-1111 & 5932882731154933888 & 240.67672 & $-$53.54162 & 4.594834 & F & 13.219 & 2.25 $\pm$ 0.16 & 3000 P105 & 24/57\\ 
OGLE GD-CEP-1210 & 4256467552765653376 & 279.99422 & $-$5.83193 & 35.823278 & F & 12.380 & 3.40 $\pm$ 0.24 & 2100 P105 & 19/37\\ 
OGLE GD-CEP-1285 & 3329873163744496000 & 92.63155 & 9.90423 & 1.523069 & 1O & 13.362 & 7.09 $\pm$ 0.51 & 1800 P106 & 59/80\\ 
OGLE GD-CEP-1311 & 2949613084534085760 & 103.66101 & $-$13.45176 & 3.868596 & F & 13.447 & 9.17 $\pm$ 0.53 & 1800 P106 & 42/62\\ 
OGLE GD-CEP-1337 & 5613275454180075264 & 112.04712 & $-$26.49072 & 1.307503 & 1O & 15.423 & 13.22 $\pm$ 0.84 & 4500 P106 & 33/53\\ 
V1253 Cen & 6153387928308811264 & 189.51591 & $-$38.52351 & 4.320929 & F & 12.069 & 10.95 $\pm$ 0.62 & 1500 P105 & 55/57\\ 
V1819 Ori & 3343252261748847104 & 88.61981 & 12.53213 & 3.149580 & F & 12.527 & 7.61 $\pm$ 0.46 & 1200 P106 & 91/114\\ 
V418 CMa & 2936165984303583360 & 105.87696 & $-$15.54335 & 3.522430 & F & 13.495 & 9.48 $\pm$ 0.58 & 1800 P106 & 68/90\\ 
V459 Sct & 4153177128348076032 & 276.46232 & $-$12.33285 & 5.762921 & F & 11.467 & 2.32 $\pm$ 0.15 & 1500 P105 & 43/93\\ 
V480 Aql & 4285878256152184832 & 282.64531 & 7.12600 & 18.998974 & F & 10.999 & 2.67 $\pm$ 0.18 & 2100 P105 & 25/92\\ 
V881 Cen & 5865214662485305216 & 201.84005 & $-$63.01964 & 15.217374 & F & 12.013 & 5.33 $\pm$ 0.31 & 1500 P105 & 18/52\\ 
VX CMa & 2928096226097916160 & 109.16267 & $-$22.18043 & 2.043935 & F & 14.789 & 9.61 $\pm$ 0.58 & 3000 P106 & 27/44\\ 
\hline\end{tabular}
\end{adjustbox}
\end{table*}

%% file: T22_atm_err.tex
\begin{table*}
\caption{Atmospheric parameters estimated for each star. The various columns provide: identification, Julian date at the middle of the observation, effective temperature, gravity, microturbulent velocity, broadening parameter and radial velocities. }
\label{tab:atm_err}
\begin{adjustbox}{width=0.77\textwidth}
\begin{tabular}{|l|c|c|r|c|c|r|r}
\hline
  \multicolumn{1}{|c|}{Star} &
  \multicolumn{1}{|c|}{HJD} &
  \multicolumn{1}{c|}{Teff} &
  \multicolumn{1}{c|}{log $g$} &
  \multicolumn{1}{c|}{$\xi$ } &
  \multicolumn{1}{c|}{$v_{br}$ } &
  \multicolumn{1}{c|}{$v_{rad}$ }\\
  &(days) & (K)&(dex)~~~~~~&(km s$^{-1}$)&(km s$^{-1}$)&(km s$^{-1}$)\\
\hline
 ASAS J060450+1021.9         & 59202.5964 & 6409 $\pm$ 220 & 2.2  $\pm$ 0.1 & 3.0 $\pm$ 0.4 & 15 $\pm$ 1 &  36.1 $\pm$ 0.3\\
 ASAS J062939-1840.5         & 59196.5666 & 5100 $\pm$ 83~~ & 1.0  $\pm$ 0.1 & 4.2 $\pm$ 0.3 & 22 $\pm$ 1 & 145.4 $\pm$ 0.3\\
 ASAS J064001-0754.8         & 59202.7323 & 6137 $\pm$ 169 & 1.8  $\pm$ 0.1 & 2.3 $\pm$ 0.4 & 12 $\pm$ 1 &  98.3 $\pm$ 0.2\\
 ASAS J065758-1521.4         & 59201.6212 & 6266 $\pm$ 140 & 2.0  $\pm$ 0.1 & 3.0 $\pm$ 0.4 & 12 $\pm$ 1 & 113.2 $\pm$ 0.1\\
 ASAS J074401-3008.4         & 59197.8179 & 6444 $\pm$ 306 & 1.5  $\pm$ 0.1 & 2.9 $\pm$ 0.4 & ~~8  $\pm$ 1 & 101.1 $\pm$ 0.1\\
 ASAS J074925-3814.4         & 59196.6319 & 5838 $\pm$ 144 & 1.4  $\pm$ 0.1 & 4.4 $\pm$ 0.5 & 21 $\pm$ 2 & 127.7 $\pm$ 0.3\\
 ASAS J084127-4353.6         & 59270.6998 & 6167 $\pm$ 159 & 0.5  $\pm$ 0.1 & 3.5 $\pm$ 0.4 & 15 $\pm$ 1 &  48.7 $\pm$ 0.2\\
 ASAS J164120-4739.6         & 59424.6099 & 4890 $\pm$ 120 & 0.8  $\pm$ 0.1 & 2.9 $\pm$ 0.5 & 11 $\pm$ 1 & $-$29.9 $\pm$ 0.1\\
 ASAS SN-J061713.86+022837.1 & 59198.5960 & 6629 $\pm$ 328 & 1.6  $\pm$ 0.1 & 3.2 $\pm$ 0.4 & 13 $\pm$ 1 & 30.7 $\pm$ 0.2\\
 ASAS SN-J063841.36-034927.7 & 59199.5975 & 6259 $\pm$ 291 & 1.9  $\pm$ 0.1 & 3.1 $\pm$ 0.4 & 11 $\pm$ 1 & 96.7 $\pm$ 0.1\\
 ASAS SN-J072739.70-252241.1 & 59202.6672 & 6382 $\pm$ 202 & 2.0  $\pm$ 0.1 & 2.8 $\pm$ 0.4 & 10 $\pm$ 1 & 113.8 $\pm$ 0.1\\
 ASAS SN-J074354.86-323013.7 & 59198.6348 & 5981 $\pm$ 220 & 2.0  $\pm$ 0.1 & 3.7 $\pm$ 0.4 & 15 $\pm$ 1 & 171.0 $\pm$ 0.3\\
 ASAS SN-J091822.17-542444.5 & 59196.7584 & 5243 $\pm$ 142 & 0.2  $\pm$ 0.1 & 2.8 $\pm$ 0.2 & 10 $\pm$ 1 & 135.3 $\pm$ 0.1\\
 ATLAS J102.7978-10.2541     & 59196.5892 & 6042 $\pm$ 177 & 1.7  $\pm$ 0.1 & 3.4 $\pm$ 0.4 & 13 $\pm$ 1 & 79.3 $\pm$ 0.1\\
 ATLAS J106.7120-14.0234     & 59202.6341 & 6304 $\pm$ 159 & 1.5  $\pm$ 0.1 & 3.2 $\pm$ 0.2 & 13 $\pm$ 1 & 72.6 $\pm$ 0.1\\
 ATLAS J113.8534-31.0749     & 59201.6430 & 6244 $\pm$ 168 & 2.1  $\pm$ 0.1 & 3.3 $\pm$ 0.4 & 21 $\pm$ 1 & 116.9 $\pm$ 0.4\\
 BQ Vel                      & 59198.8353  & 5847 $\pm$ 220 & 1.6 $\pm$ 0.1 & 3.1 $\pm$ 0.4 & 12 $\pm$ 1 & 139.9 $\pm$ 0.2\\
 GDS J133950.2-634049        & 59294.7621 & 6146 $\pm$ 107 & 1.1  $\pm$ 0.1 & 3.0 $\pm$ 0.4 & 12 $\pm$ 2 & $-$68.4 $\pm$ 0.1\\
 OGLE GD-CEP-0029            & 59197.6209 & 5817 $\pm$ 274 & 0.9  $\pm$ 0.1 & 2.7 $\pm$ 0.4 & 11 $\pm$ 1 & 8.3 $\pm$ 0.2\\
 OGLE GD-CEP-0089            & 59201.5986 & 6240 $\pm$ 175 & 1.8  $\pm$ 0.1 & 3.1 $\pm$ 0.4 & 11 $\pm$ 1 & 100.4 $\pm$ 0.1\\
 OGLE GD-CEP-0120            & 59197.6572 & 5435 $\pm$ 237 & 0.9  $\pm$ 0.1 & 3.2 $\pm$ 0.4 & 15 $\pm$ 2 & 150.9 $\pm$ 0.1\\
 OGLE GD-CEP-0123            & 59198.7220 & 5990 $\pm$ 174 & 1.8  $\pm$ 0.1 & 3.3 $\pm$ 0.4 & 12 $\pm$ 1 & 148.7 $\pm$ 0.2\\
 OGLE GD-CEP-0127            & 59196.6970 & 5202 $\pm$ 176 & 0.7  $\pm$ 0.1 & 3.2 $\pm$ 0.4 & 15 $\pm$ 1 & 130.1 $\pm$ 0.1\\
 OGLE GD-CEP-0134            & 59201.7136 & 5732 $\pm$ 285 & 1.6  $\pm$ 0.1 & 3.2 $\pm$ 0.4 & 17 $\pm$ 1 & 127.5 $\pm$ 0.2\\
 OGLE GD-CEP-0156            & 59202.6981 & 6230 $\pm$ 268 & 1.6  $\pm$ 0.1 & 2.3 $\pm$ 0.4 & ~~6 $\pm$ 1 & 131.2 $\pm$ 0.1\\
 OGLE GD-CEP-0159            & 59197.7810 & 5910 $\pm$ 161 & 2.2  $\pm$ 0.1 & 3.8 $\pm$ 0.4 & 13 $\pm$ 1 & 145.8 $\pm$ 0.2\\
 OGLE GD-CEP-0162            & 59197.6942 & 6133 $\pm$ 145 & 1.6  $\pm$ 0.1 & 3.1 $\pm$ 0.4 & 11 $\pm$ 1 &  90.1 $\pm$ 0.1\\
 OGLE GD-CEP-0168            & 59202.7614 & 6069 $\pm$ 197 & 1.6  $\pm$ 0.1 & 2.9 $\pm$ 0.4 & 13 $\pm$ 1 & 132.6 $\pm$ 0.1\\
 OGLE GD-CEP-0176            & 59201.8437 & 6201 $\pm$ 207 & 1.4  $\pm$ 0.1 & 2.2 $\pm$ 0.4 & ~~7 $\pm$ 2 & 154.8 $\pm$ 0.1\\
 OGLE GD-CEP-0179            & 59201.8097 & 6188 $\pm$ 192 & 1.7  $\pm$ 0.1 & 2.9 $\pm$ 0.4 & 10 $\pm$ 1 & 125.6 $\pm$ 0.1\\
 OGLE GD-CEP-0181            & 59198.6715 & 5982 $\pm$ 261 & 1.3  $\pm$ 0.1 & 2.4 $\pm$ 0.4 & 13 $\pm$ 1 & 124.8 $\pm$ 0.2\\
 OGLE GD-CEP-0185            & 59196.7954 & 5737 $\pm$ 190 & 1.5  $\pm$ 0.1 & 3.3 $\pm$ 0.4 & 11 $\pm$ 1 & 154.1 $\pm$ 0.1\\
 OGLE GD-CEP-0186            & 59196.6597 & 5723 $\pm$ 246 & 1.7  $\pm$ 0.1 & 2.9 $\pm$ 0.4 & 12 $\pm$ 1 & 128.1 $\pm$ 0.1\\
 OGLE GD-CEP-0196            & 59197.7388 & 6011 $\pm$ 180 & 1.3  $\pm$ 0.1 & 3.3 $\pm$ 0.4 & 11 $\pm$ 1 & 107.5 $\pm$ 0.1\\
 OGLE GD-CEP-0206            & 59201.7645 & 6192 $\pm$ 240 & 1.9  $\pm$ 0.1 & 2.7 $\pm$ 0.4 & ~~9 $\pm$ 1 & 119.7 $\pm$ 0.1\\
 OGLE GD-CEP-0213            & 59202.7983 & 6342 $\pm$ 162 & 1.4  $\pm$ 0.1 & 3.0 $\pm$ 0.2 & 10 $\pm$ 1 & 146.8 $\pm$ 0.2\\
 OGLE GD-CEP-0214            & 59202.8390 & 6720 $\pm$ 186 & 1.6  $\pm$ 0.1 & 2.9 $\pm$ 0.3 & ~~8 $\pm$ 1 & 150.1 $\pm$ 0.2\\
 OGLE GD-CEP-0224            & 59196.7282 & 6815 $\pm$ 211 & 1.8  $\pm$ 0.1 & 3.4 $\pm$ 0.2 & 14 $\pm$ 1 & 122.5 $\pm$ 0.3\\
 OGLE GD-CEP-0228            & 59198.7974 & 5935 $\pm$ 121 & 1.9  $\pm$ 0.1 & 3.1 $\pm$ 0.4 & 10 $\pm$ 1 & 142.2 $\pm$ 0.2\\
 OGLE GD-CEP-0247            & 59199.6770 & 5899 $\pm$ 209 & 1.7  $\pm$ 0.1 & 2.5 $\pm$ 0.3 & ~~7 $\pm$ 1 & 145.8 $\pm$ 0.1\\
 OGLE GD-CEP-0252            & 59196.8361 & 6418 $\pm$ 203 & 2.1  $\pm$ 0.1 & 3.8 $\pm$ 0.5 & 12 $\pm$ 1 &  99.5 $\pm$ 0.2\\
 OGLE GD-CEP-0271            & 59199.7473 & 5914 $\pm$ 180 & 1.5  $\pm$ 0.1 & 3.2 $\pm$ 0.5 & 14 $\pm$ 1 & 158.0 $\pm$ 0.2\\
 OGLE GD-CEP-0316            & 59199.7894 & 5881 $\pm$ 221 & 2.0  $\pm$ 0.1 & 3.7 $\pm$ 0.4 & 15 $\pm$ 1 & 159.7 $\pm$ 0.2\\
 OGLE GD-CEP-0342            & 59203.8370 & 6096 $\pm$ 191 & 1.0  $\pm$ 0.1 & 2.5 $\pm$ 0.4 & 13 $\pm$ 1 & 120.4 $\pm$ 0.1\\
 OGLE GD-CEP-0348            & 59199.8311 & 6152 $\pm$ 160 & 1.3  $\pm$ 0.1 & 2.3 $\pm$ 0.4 & 12 $\pm$ 1 & 112.2 $\pm$ 0.1\\
 OGLE GD-CEP-0353            & 59197.8401 & 5725 $\pm$ 239 & 1.4  $\pm$ 0.1 & 3.3 $\pm$ 0.4 & 11 $\pm$ 1 & 118.2 $\pm$ 0.1\\
 OGLE GD-CEP-0516            & 59217.6507 & 6367 $\pm$ 151 & 1.5  $\pm$ 0.1 & 2.4 $\pm$ 0.2 & 13 $\pm$ 1 & $-$9.0 $\pm$ 0.2\\
 OGLE GD-CEP-0568            & 59271.6099 & 5086 $\pm$ 118 & 0.2  $\pm$ 0.1 & 3.8 $\pm$ 0.5 & 15 $\pm$ 1 & $-$37.4 $\pm$ 0.2\\
 OGLE GD-CEP-0575            & 59271.6462 & 5352 $\pm$ 161 & 0.9  $\pm$ 0.1 & 4.4 $\pm$ 0.5 & 18 $\pm$ 1 & 12.7 $\pm$ 0.1\\
 OGLE GD-CEP-0889            & 59294.7240 & 5670 $\pm$ 116 & 0.4  $\pm$ 0.1 & 5.3 $\pm$ 0.2 & 29 $\pm$ 1 & 48.1 $\pm$ 0.8\\
 OGLE GD-CEP-0974            & 59458.5146 & 4984 $\pm$ 117 & 0.5  $\pm$ 0.1 & 3.0 $\pm$ 0.4 & 11 $\pm$ 1 & $-$44.3 $\pm$ 0.3\\
 OGLE GD-CEP-0996            & 59294.7855 & 5506 $\pm$ 175 & 1.0  $\pm$ 0.1 & 3.0 $\pm$ 0.5 & 11 $\pm$ 1 & $-$40.1 $\pm$ 2.8\\
 OGLE GD-CEP-1012            & 59471.5399 & 5038 $\pm$ 119 & 0.5  $\pm$ 0.1 & 3.0 $\pm$ 0.4 & 13 $\pm$ 1 & $-$55.8 $\pm$ 0.3\\
 OGLE GD-CEP-1111            & 59471.5722 & 6239 $\pm$ 386 & 1.4  $\pm$ 0.1 & 3.5 $\pm$ 0.3 & 11 $\pm$ 1 & $-$92.4 $\pm$ 0.3\\
 OGLE GD-CEP-1210            & 59464.5956 & 5047 $\pm$ 156 & 0.9  $\pm$ 0.1 & 4.4 $\pm$ 0.3 & 13 $\pm$ 1 & 70.7 $\pm$ 0.6\\
 OGLE GD-CEP-1285            & 59201.6706 & 6038 $\pm$ 105 & 0.9  $\pm$ 0.1 & 3.4 $\pm$ 0.2 & 26 $\pm$ 3 & 46.2 $\pm$ 0.4\\
 OGLE GD-CEP-1311            & 59196.6139 & 5856 $\pm$ 154 & 1.5  $\pm$ 0.1 & 3.8 $\pm$ 0.4 & 18 $\pm$ 1 & 121.9 $\pm$ 0.2\\
 OGLE GD-CEP-1337            & 59203.7932 & 5977 $\pm$ 151 & 0.6  $\pm$ 0.1 & 1.9 $\pm$ 0.2 & ~~9 $\pm$ 1 & 145.2 $\pm$ 0.1\\
 V1253 Cen                   & 59294.6819 & 5670 $\pm$ 193 & 1.5  $\pm$ 0.1 & 3.8 $\pm$ 0.4 & 16 $\pm$ 1 & 119.1$\pm$ 0.3\\
 V1819 Ori                   & 59199.7206 & 6342 $\pm$ 95~~ & 1.2 $\pm$ 0.1 & 3.7 $\pm$ 0.1 & 19 $\pm$ 1 & 27.2 $\pm$ 0.3\\
 V418 CMa                    & 59198.7652 & 6299 $\pm$ 261 & 1.8  $\pm$ 0.1 & 3.8 $\pm$ 0.5 & 13 $\pm$ 1 & 105.5 $\pm$ 0.2\\
 V459 Sct                    & 59370.9059 & 5972 $\pm$ 233 & 1.1  $\pm$ 0.1 & 3.4 $\pm$ 0.2 & 11 $\pm$ 1 & $-$4.1 $\pm$ 0.1\\
 V480 Aql                    & 59464.6236 & 5019 $\pm$ 122 & 1.0  $\pm$ 0.1 & 3.6 $\pm$ 0.8 & 12 $\pm$ 1 & 29.1 $\pm$ 0.1\\
 V881 Cen                    & 59294.6322 & 5064 $\pm$ 120 & 0.3  $\pm$ 0.1 & 3.2 $\pm$ 0.4 & 11 $\pm$ 1 & $-$34.5 $\pm$ 0.1\\
 VX CMa                      & 59199.6303 & 5820 $\pm$ 106 & 1.2  $\pm$ 0.1 & 3.7 $\pm$ 0.2 & 14 $\pm$ 1 & 128.4 $\pm$ 0.2\\
\hline\end{tabular}
\end{adjustbox}
\end{table*}

%% file: fits_results.tex
\begin{table}
\centering
\caption{Coefficients of the linear fit of the form $[X/H]=\alpha_{rgc}\times R_{gc}+\beta_{rgc}$ with relative dispersion relative to Fig~\ref{fig:grad_elem}.}
\label{tab:slopes}
\footnotesize\setlength{\tabcolsep}{3pt}
\begin{tabular}{lrrrc}
\hline
  \multicolumn{1}{|c|}{Element} &
  \multicolumn{1}{c|}{$\alpha_{rgc}$} &
  \multicolumn{1}{c|}{$\beta_{rgc}$} &
  \multicolumn{1}{c|}{rms} \\
    \multicolumn{1}{|c|}{} &
  \multicolumn{1}{c|}{(dex kpc$^{-1}$)} &
  \multicolumn{1}{c|}{(dex)} &
  \multicolumn{1}{c|}{(dex)} & 
  \multicolumn{1}{c|}{} \\
\hline
C & $-$0.068$\pm$0.002 & 0.371$\pm$0.023 & 0.14  \\ 
O & $-$0.046$\pm$0.002 & 0.414$\pm$0.022 & 0.12   \\
Na & $-$0.060$\pm$0.002 & 0.901$\pm$0.022 & 0.15    \\
Mg & $-$0.056$\pm$0.002 & 0.588$\pm$0.024 & 0.17    \\
Al & $-$0.045$\pm$0.002 & 0.538$\pm$0.022 & 0.14    \\
Si & $-$0.053$\pm$0.002 & 0.640$\pm$0.022 & 0.10    \\
S & $-$0.060$\pm$0.002 & 0.618$\pm$0.023 & 0.15     \\
Ca & $-$0.057$\pm$0.002 & 0.564$\pm$0.022 & 0.13    \\
Sc & $-$0.045$\pm$0.003 & 0.731$\pm$0.028 & 0.20    \\
Ti & $-$0.046$\pm$0.002 & 0.543$\pm$0.022 & 0.14    \\
V & $-$0.028$\pm$0.002 & 0.341$\pm$0.023 & 0.16    \\
Cr & $-$0.049$\pm$0.002 & 0.577$\pm$0.022 & 0.14   \\
Mn & $-$0.061$\pm$0.002 & 0.436$\pm$0.025 & 0.17     \\
Co & $-$0.020$\pm$0.003 & 0.363$\pm$0.027 & 0.18        \\
Ni & $-$0.055$\pm$0.002 & 0.479$\pm$0.022 & 0.11  \\
Cu & $-$0.041$\pm$0.003 & 0.497$\pm$0.034 & 0.23       \\
Zn & $-$0.084$\pm$0.003 & 0.669$\pm$0.031 & 0.21       \\
Y & $-$0.040$\pm$0.002 & 0.562$\pm$0.021 & 0.14       \\
Zr & $-$0.012$\pm$0.003 & 0.474$\pm$0.036 & 0.25     \\
Ba & $-$0.051$\pm$0.007 & 0.636$\pm$0.087 & 0.38      \\
La & $-$0.029$\pm$0.002 & 0.421$\pm$0.021 & 0.15     \\
Pr & $-$0.010$\pm$0.004 & $-$0.157$\pm$0.046 & 0.19      \\ 
Nd & $-$0.026$\pm$0.002 & 0.247$\pm$0.021 & 0.14      \\
\hline\end{tabular}
\end{table}

%% file: T22_abund_first.tex
\begin{table*}
\caption{Abundances expressed in solar terms for the chemical elements detected in our targets. In the first part of the table elements from C to Ca are shown.}
\label{tab:abundances}
\begin{adjustbox}{width=\textwidth}
\begin{tabular}{|l|r|r|r|r|r|r|r|r|}
\hline
  \multicolumn{1}{|c|}{Star} &
  \multicolumn{1}{c|}{[C/H]} &
  \multicolumn{1}{c|}{[O/H]} &
  \multicolumn{1}{c|}{[Na/H]} &
  \multicolumn{1}{c|}{[Mg/H]} &
  \multicolumn{1}{c|}{[Al/H]} &
  \multicolumn{1}{c|}{[Si/H]} &
  \multicolumn{1}{c|}{[S/H]} &
  \multicolumn{1}{c|}{[Ca/H]}\\
\hline
ASAS J060450+1021.9 & $-$0.83 $\pm$ 0.05 & $-$0.61 $\pm$ 0.16 & $-$0.07 $\pm$ 0.04 & $-$0.40 $\pm$ 0.16 & $-$0.17 $\pm$ 0.15 & $-$0.44 $\pm$ 0.15 & $-$0.41 $\pm$ 0.03 & $-$0.46 $\pm$ 0.07\\ 
ASAS J062939-1840.5 & $-$1.10 $\pm$ 0.10 & ---~~~~~~~ & $-$0.48 $\pm$ 0.19 & $-$0.76 $\pm$ 0.11 & $-$0.76 $\pm$ 0.15 & $-$0.73 $\pm$ 0.20 & ---~~~~~~~ & $-$0.77 $\pm$ 0.14\\ 
ASAS J064001-0754.8 & $-$0.99 $\pm$ 0.05 & $-$0.49 $\pm$ 0.16 & $-$0.22 $\pm$ 0.04 & $-$0.52 $\pm$ 0.16 & $-$0.48 $\pm$ 0.15 & $-$0.40 $\pm$ 0.20 & $-$0.36 $\pm$ 0.03 & $-$0.51 $\pm$ 0.04\\ 
ASAS J065758-1521.4 & $-$0.70 $\pm$ 0.05 & $-$0.42 $\pm$ 0.16 & $-$0.06 $\pm$ 0.04 & $-$0.33 $\pm$ 0.16 & $-$0.20 $\pm$ 0.15 & $-$0.32 $\pm$ 0.09 & $-$0.25 $\pm$ 0.03 & $-$0.46 $\pm$ 0.11\\ 
ASAS J074401-3008.4 & $-$0.79 $\pm$ 0.05 & $-$0.45 $\pm$ 0.16 & $-$0.20 $\pm$ 0.04 & $-$0.65 $\pm$ 0.16 & $-$0.23 $\pm$ 0.15 & $-$0.32 $\pm$ 0.11 & $-$0.38 $\pm$ 0.03 & $-$0.53 $\pm$ 0.04\\ 
ASAS J074925-3814.4 & ---~~~~~~~ & $-$0.61 $\pm$ 0.16 & $-$0.11 $\pm$ 0.07 & $-$0.15 $\pm$ 0.16 & $-$0.17 $\pm$ 0.15 & $-$0.22 $\pm$ 0.05 & $-$0.28 $\pm$ 0.03 & $-$0.50 $\pm$ 0.05\\ 
ASAS J084127-4353.6 & $-$0.59 $\pm$ 0.05 & $-$0.49 $\pm$ 0.16 & $-$0.00 $\pm$ 0.04 & $-$0.05 $\pm$ 0.16 & 0.02 $\pm$ 0.15 & 0.08 $\pm$ 0.06 & $-$0.18 $\pm$ 0.03 & $-$0.15 $\pm$ 0.04\\ 
ASAS J164120-4739.6 & ---~~~~~~~ & 0.14 $\pm$ 0.16 & 0.92 $\pm$ 0.05 & 0.10 $\pm$ 0.16 & 0.27 $\pm$ 0.15 & ---~~~~~~~ & ---~~~~~~~ & 0.16 $\pm$ 0.05\\ 
ASAS SN-J061713.86+022837.1 & $-$0.97 $\pm$ 0.10 & $-$0.24 $\pm$ 0.53 & $-$0.29 $\pm$ 0.10 & $-$0.27 $\pm$ 0.35 & 0.02 $\pm$ 0.03 & $-$0.54 $\pm$ 0.05 & $-$0.29 $\pm$ 0.03 & $-$0.59 $\pm$ 0.19\\ 
ASAS SN-J063841.36-034927.7 & $-$0.63 $\pm$ 0.05 & $-$0.36 $\pm$ 0.16 & 0.00 $\pm$ 0.04 & $-$0.24 $\pm$ 0.16 & ---~~~~~~~ & $-$0.26 $\pm$ 0.15 & $-$0.25 $\pm$ 0.03 & $-$0.20 $\pm$ 0.04\\ 
ASAS SN-J072739.70-252241.1 & $-$0.72 $\pm$ 0.05 & $-$0.36 $\pm$ 0.16 & $-$0.10 $\pm$ 0.04 & $-$0.46 $\pm$ 0.16 & $-$0.10 $\pm$ 0.15 & $-$0.25 $\pm$ 0.12 & $-$0.29 $\pm$ 0.15 & $-$0.34 $\pm$ 0.04\\ 
ASAS SN-J074354.86-323013.7 & $-$0.42 $\pm$ 0.12 & ---~~~~~~~ & 0.01 $\pm$ 0.04 & $-$0.27 $\pm$ 0.16 & ---~~~~~~~ & $-$0.17 $\pm$ 0.12 & $-$0.17 $\pm$ 0.03 & $-$0.59 $\pm$ 0.15\\ 
ASAS SN-J091822.17-542444.5 & ---~~~~~~~ & $-$0.64 $\pm$ 0.16 & 0.08 $\pm$ 0.04 & $-$0.02 $\pm$ 0.16 & $-$0.28 $\pm$ 0.15 & $-$0.25 $\pm$ 0.09 & $-$0.43 $\pm$ 0.03 & $-$0.23 $\pm$ 0.04\\ 
ATLAS J102.7978-10.2541 & $-$0.78 $\pm$ 0.05 & $-$0.61 $\pm$ 0.16 & $-$0.03 $\pm$ 0.04 & $-$0.11 $\pm$ 0.16 & $-$0.23 $\pm$ 0.15 & $-$0.29 $\pm$ 0.15 & $-$0.31 $\pm$ 0.03 & $-$0.29 $\pm$ 0.05\\ 
ATLAS J106.7120-14.0234 & $-$0.77 $\pm$ 0.20 & $-$0.08 $\pm$ 0.15 & $-$0.11 $\pm$ 0.07 & 0.05 $\pm$ 0.10 & ---~~~~~~~ & $-$0.36 $\pm$ 0.18 & $-$0.32 $\pm$ 0.22 & $-$0.47 $\pm$ 0.11\\ 
ATLAS J113.8534-31.0749 & $-$0.82 $\pm$ 0.05 & $-$0.36 $\pm$ 0.16 & $-$0.22 $\pm$ 0.04 & $-$0.52 $\pm$ 0.16 & $-$0.10 $\pm$ 0.15 & $-$0.38 $\pm$ 0.17 & $-$0.47 $\pm$ 0.03 & $-$0.45 $\pm$ 0.04\\ 
BQ Vel & $-$0.79 $\pm$ 0.10 & $-$0.24 $\pm$ 0.16 & $-$0.12 $\pm$ 0.04 & $-$0.11 $\pm$ 0.16 & $-$0.23 $\pm$ 0.15 & $-$0.07 $\pm$ 0.09 & $-$0.35 $\pm$ 0.15 & $-$0.42 $\pm$ 0.07\\ 
GDS J133950.2-634049 & $-$0.33 $\pm$ 0.05 & $-$0.17 $\pm$ 0.16 & 0.42 $\pm$ 0.05 & 0.10 $\pm$ 0.16 & 0.18 $\pm$ 0.15 & 0.26 $\pm$ 0.13 & 0.14 $\pm$ 0.03 & 0.13 $\pm$ 0.04\\ 
OGLE GD-CEP-0029 & $-$0.30 $\pm$ 0.21 & $-$0.05 $\pm$ 0.16 & $-$0.14 $\pm$ 0.04 & $-$0.21 $\pm$ 0.16 & $-$0.10 $\pm$ 0.15 & $-$0.02 $\pm$ 0.05 & $-$0.27 $\pm$ 0.03 & $-$0.54 $\pm$ 0.07\\ 
OGLE GD-CEP-0089 & $-$0.81 $\pm$ 0.05 & $-$0.36 $\pm$ 0.16 & $-$0.18 $\pm$ 0.04 & $-$0.52 $\pm$ 0.16 & $-$0.23 $\pm$ 0.15 & $-$0.38 $\pm$ 0.17 & $-$0.31 $\pm$ 0.03 & $-$0.39 $\pm$ 0.04\\ 
OGLE GD-CEP-0120 & $-$0.83 $\pm$ 0.16 & $-$0.55 $\pm$ 0.16 & $-$0.40 $\pm$ 0.04 & $-$0.40 $\pm$ 0.16 & $-$0.23 $\pm$ 0.15 & $-$0.44 $\pm$ 0.10 & $-$0.47 $\pm$ 0.03 & $-$0.64 $\pm$ 0.12\\ 
OGLE GD-CEP-0123 & $-$0.91 $\pm$ 0.09 & $-$0.36 $\pm$ 0.16 & $-$0.20 $\pm$ 0.04 & $-$0.52 $\pm$ 0.16 & $-$0.10 $\pm$ 0.15 & $-$0.34 $\pm$ 0.09 & $-$0.28 $\pm$ 0.03 & $-$0.49 $\pm$ 0.04\\ 
OGLE GD-CEP-0127 & $-$0.76 $\pm$ 0.06 & $-$0.74 $\pm$ 0.16 & 0.15 $\pm$ 0.04 & $-$0.08 $\pm$ 0.16 & $-$0.12 $\pm$ 0.15 & 0.13 $\pm$ 0.20 & $-$0.29 $\pm$ 0.03 & $-$0.21 $\pm$ 0.04\\ 
OGLE GD-CEP-0134 & $-$0.66 $\pm$ 0.06 & $-$0.11 $\pm$ 0.16 & $-$0.25 $\pm$ 0.04 & $-$0.52 $\pm$ 0.16 & $-$0.11 $\pm$ 0.15 & $-$0.32 $\pm$ 0.15 & $-$0.25 $\pm$ 0.03 & $-$0.59 $\pm$ 0.04\\ 
OGLE GD-CEP-0156 & $-$0.89 $\pm$ 0.05 & $-$0.55 $\pm$ 0.16 & $-$0.22 $\pm$ 0.04 & $-$0.48 $\pm$ 0.16 & $-$0.23 $\pm$ 0.15 & $-$0.38 $\pm$ 0.09 & $-$0.44 $\pm$ 0.03 & $-$0.57 $\pm$ 0.04\\ 
OGLE GD-CEP-0159 & $-$0.86 $\pm$ 0.07 & 0.01 $\pm$ 0.16 & $-$0.14 $\pm$ 0.04 & $-$0.24 $\pm$ 0.16 & $-$0.04 $\pm$ 0.15 & $-$0.37 $\pm$ 0.04 & $-$0.33 $\pm$ 0.03 & $-$0.34 $\pm$ 0.15\\ 
OGLE GD-CEP-0162 & $-$0.85 $\pm$ 0.05 & $-$0.61 $\pm$ 0.16 & $-$0.19 $\pm$ 0.04 & $-$0.52 $\pm$ 0.16 & $-$0.10 $\pm$ 0.15 & $-$0.47 $\pm$ 0.18 & $-$0.42 $\pm$ 0.03 & $-$0.43 $\pm$ 0.04\\ 
OGLE GD-CEP-0168 & $-$0.68 $\pm$ 0.06 & $-$0.11 $\pm$ 0.16 & $-$0.08 $\pm$ 0.04 & $-$0.21 $\pm$ 0.16 & $-$0.10 $\pm$ 0.15 & $-$0.07 $\pm$ 0.16 & $-$0.22 $\pm$ 0.03 & $-$0.48 $\pm$ 0.17\\ 
OGLE GD-CEP-0176 & $-$0.85 $\pm$ 0.05 & $-$0.61 $\pm$ 0.16 & $-$0.25 $\pm$ 0.04 & $-$0.71 $\pm$ 0.16 & $-$0.35 $\pm$ 0.15 & $-$0.47 $\pm$ 0.10 & $-$0.42 $\pm$ 0.15 & $-$0.55 $\pm$ 0.04\\ 
OGLE GD-CEP-0179 & $-$0.66 $\pm$ 0.06 & $-$0.36 $\pm$ 0.16 & 0.04 $\pm$ 0.04 & $-$0.33 $\pm$ 0.16 & 0.02 $\pm$ 0.15 & $-$0.27 $\pm$ 0.14 & $-$0.19 $\pm$ 0.03 & $-$0.21 $\pm$ 0.04\\ 
OGLE GD-CEP-0181 & $-$1.00 $\pm$ 0.05 & $-$0.49 $\pm$ 0.16 & $-$0.27 $\pm$ 0.04 & $-$0.74 $\pm$ 0.16 & $-$0.29 $\pm$ 0.15 & $-$0.27 $\pm$ 0.14 & $-$0.41 $\pm$ 0.03 & $-$0.68 $\pm$ 0.14\\ 
OGLE GD-CEP-0185 & $-$0.88 $\pm$ 0.05 & ---~~~~~~~ & $-$0.13 $\pm$ 0.04 & $-$0.33 $\pm$ 0.16 & $-$0.10 $\pm$ 0.15 & $-$0.32 $\pm$ 0.11 & $-$0.33 $\pm$ 0.03 & $-$0.40 $\pm$ 0.04\\ 
OGLE GD-CEP-0186 & $-$0.74 $\pm$ 0.07 & $-$0.24 $\pm$ 0.16 & 0.21 $\pm$ 0.05 & 0.01 $\pm$ 0.16 & $-$0.04 $\pm$ 0.15 & $-$0.07 $\pm$ 0.19 & $-$0.13 $\pm$ 0.03 & $-$0.08 $\pm$ 0.04\\ 
OGLE GD-CEP-0196 & $-$0.85 $\pm$ 0.07 & $-$0.30 $\pm$ 0.16 & $-$0.25 $\pm$ 0.04 & $-$0.55 $\pm$ 0.16 & $-$0.23 $\pm$ 0.15 & $-$0.27 $\pm$ 0.15 & $-$0.25 $\pm$ 0.04 & $-$0.54 $\pm$ 0.04\\ 
OGLE GD-CEP-0206 & $-$0.55 $\pm$ 0.05 & $-$0.11 $\pm$ 0.16 & $-$0.08 $\pm$ 0.04 & $-$0.27 $\pm$ 0.16 & $-$0.10 $\pm$ 0.15 & 0.13 $\pm$ 0.13 & $-$0.14 $\pm$ 0.03 & $-$0.30 $\pm$ 0.04\\ 
OGLE GD-CEP-0213 & $-$0.89 $\pm$ 0.21 & 0.01 $\pm$ 0.15 & $-$0.24 $\pm$ 0.06 & $-$0.67 $\pm$ 0.04 & $-$0.10 $\pm$ 0.10 & $-$0.47 $\pm$ 0.17 & $-$0.42 $\pm$ 0.03 & $-$0.61 $\pm$ 0.07\\ 
OGLE GD-CEP-0214 & $-$1.00 $\pm$ 0.05 & $-$0.58 $\pm$ 0.16 & $-$0.30 $\pm$ 0.04 & $-$0.71 $\pm$ 0.16 & $-$0.17 $\pm$ 0.15 & $-$0.57 $\pm$ 0.13 & $-$0.34 $\pm$ 0.03 & $-$0.57 $\pm$ 0.04\\ 
OGLE GD-CEP-0224 & $-$0.72 $\pm$ 0.12 & ---~~~~~~~ & $-$0.19 $\pm$ 0.06 & 0.12 $\pm$ 0.16 & ---~~~~~~~ & $-$0.07 $\pm$ 0.15 & $-$0.23 $\pm$ 0.07 & $-$0.38 $\pm$ 0.14\\ 
OGLE GD-CEP-0228 & $-$0.81 $\pm$ 0.05 & $-$0.49 $\pm$ 0.16 & $-$0.10 $\pm$ 0.04 & $-$0.36 $\pm$ 0.16 & $-$0.04 $\pm$ 0.15 & $-$0.38 $\pm$ 0.11 & $-$0.28 $\pm$ 0.03 & $-$0.47 $\pm$ 0.08\\ 
OGLE GD-CEP-0247 & $-$1.09 $\pm$ 0.05 & $-$0.24 $\pm$ 0.16 & 0.12 $\pm$ 0.04 & $-$0.33 $\pm$ 0.16 & 0.02 $\pm$ 0.15 & 0.03 $\pm$ 0.18 & $-$0.10 $\pm$ 0.03 & $-$0.25 $\pm$ 0.05\\ 
OGLE GD-CEP-0252 & $-$0.71 $\pm$ 0.05 & $-$0.49 $\pm$ 0.16 & $-$0.05 $\pm$ 0.04 & $-$0.27 $\pm$ 0.16 & 0.02 $\pm$ 0.15 & $-$0.29 $\pm$ 0.06 & $-$0.28 $\pm$ 0.03 & $-$0.39 $\pm$ 0.08\\ 
OGLE GD-CEP-0271 & $-$0.91 $\pm$ 0.16 & $-$0.36 $\pm$ 0.16 & $-$0.50 $\pm$ 0.11 & $-$0.52 $\pm$ 0.16 & 0.02 $\pm$ 0.15 & $-$0.27 $\pm$ 0.20 & $-$0.20 $\pm$ 0.06 & $-$0.51 $\pm$ 0.04\\ 
OGLE GD-CEP-0316 & $-$0.80 $\pm$ 0.07 & $-$0.61 $\pm$ 0.16 & 0.04 $\pm$ 0.04 & $-$0.15 $\pm$ 0.16 & ---~~~~~~~ & $-$0.28 $\pm$ 0.12 & $-$0.21 $\pm$ 0.03 & $-$0.29 $\pm$ 0.04\\ 
OGLE GD-CEP-0342 & $-$1.15 $\pm$ 0.05 & $-$0.36 $\pm$ 0.16 & 0.18 $\pm$ 0.04 & $-$0.27 $\pm$ 0.16 & $-$0.04 $\pm$ 0.15 & $-$0.17 $\pm$ 0.11 & $-$0.27 $\pm$ 0.03 & $-$0.35 $\pm$ 0.04\\ 
OGLE GD-CEP-0348 & $-$0.92 $\pm$ 0.05 & $-$0.36 $\pm$ 0.16 & $-$0.24 $\pm$ 0.04 & $-$0.71 $\pm$ 0.16 & 0.02 $\pm$ 0.15 & $-$0.32 $\pm$ 0.20 & $-$0.13 $\pm$ 0.06 & $-$0.44 $\pm$ 0.05\\ 
OGLE GD-CEP-0353 & $-$1.09 $\pm$ 0.14 & $-$0.36 $\pm$ 0.16 & 0.10 $\pm$ 0.04 & $-$0.05 $\pm$ 0.16 & $-$0.10 $\pm$ 0.15 & $-$0.07 $\pm$ 0.04 & $-$0.25 $\pm$ 0.03 & $-$0.29 $\pm$ 0.05\\ 
OGLE GD-CEP-0516 & $-$0.45 $\pm$ 0.05 & $-$0.14 $\pm$ 0.16 & $-$0.04 $\pm$ 0.04 & $-$0.40 $\pm$ 0.16 & 0.02 $\pm$ 0.15 & 0.08 $\pm$ 0.11 & $-$0.17 $\pm$ 0.03 & $-$0.30 $\pm$ 0.11\\ 
OGLE GD-CEP-0568 & $-$0.40 $\pm$ 0.46 & $-$0.05 $\pm$ 0.16 & 0.29 $\pm$ 0.04 & 0.04 $\pm$ 0.16 & $-$0.17 $\pm$ 0.15 & 0.23 $\pm$ 0.15 & $-$0.11 $\pm$ 0.03 & 0.01 $\pm$ 0.05\\ 
OGLE GD-CEP-0575 & $-$0.41 $\pm$ 0.16 & $-$0.49 $\pm$ 0.16 & 0.07 $\pm$ 0.04 & $-$0.18 $\pm$ 0.16 & $-$0.04 $\pm$ 0.15 & 0.18 $\pm$ 0.13 & 0.04 $\pm$ 0.03 & $-$0.31 $\pm$ 0.04\\ 
OGLE GD-CEP-0889 & 0.05 $\pm$ 0.20 & $-$0.39 $\pm$ 0.06 & 0.41 $\pm$ 0.14 & $-$0.11 $\pm$ 0.11 & $-$0.07 $\pm$ 0.10 & 0.03 $\pm$ 0.06 & $-$0.36 $\pm$ 0.03 & $-$0.30 $\pm$ 0.17\\ 
OGLE GD-CEP-0974 & 0.34 $\pm$ 0.16 & $-$0.11 $\pm$ 0.16 & 0.74 $\pm$ 0.05 & 0.10 $\pm$ 0.16 & 0.24 $\pm$ 0.15 & 0.23 $\pm$ 0.30 & 0.37 $\pm$ 0.03 & 0.32 $\pm$ 0.12\\ 
OGLE GD-CEP-0996 & ---~~~~~~~ & 0.14 $\pm$ 0.16 & 1.05 $\pm$ 0.08 & 0.73 $\pm$ 0.16 & 0.36 $\pm$ 0.15 & ---~~~~~~~ & 0.45 $\pm$ 0.03 & 0.36 $\pm$ 0.04\\ 
OGLE GD-CEP-1012 & $-$0.16 $\pm$ 0.16 & $-$0.05 $\pm$ 0.16 & 0.45 $\pm$ 0.04 & 0.10 $\pm$ 0.16 & 0.11 $\pm$ 0.15 & 0.33 $\pm$ 0.20 & 0.47 $\pm$ 0.04 & 0.23 $\pm$ 0.05\\ 
OGLE GD-CEP-1111 & $-$0.03 $\pm$ 0.19 & $-$0.36 $\pm$ 0.16 & 0.24 $\pm$ 0.08 & 0.51 $\pm$ 1.02 & 0.21 $\pm$ 0.10 & 0.24 $\pm$ 0.09 & 0.18 $\pm$ 0.13 & $-$0.23 $\pm$ 0.08\\ 
OGLE GD-CEP-1210 & ---~~~~~~~ & 0.14 $\pm$ 0.16 & 0.73 $\pm$ 0.04 & 0.34 $\pm$ 0.16 & 0.24 $\pm$ 0.15 & 0.43 $\pm$ 0.30 & ---~~~~~~~ & $-$0.10 $\pm$ 0.04\\ 
OGLE GD-CEP-1285 & $-$0.39 $\pm$ 0.14 & $-$0.00 $\pm$ 0.21 & $-$0.36 $\pm$ 0.05 & $-$0.32 $\pm$ 0.10 & $-$0.10 $\pm$ 0.10 & $-$0.35 $\pm$ 0.09 & $-$0.59 $\pm$ 0.18 & $-$0.74 $\pm$ 0.12\\ 
OGLE GD-CEP-1311 & $-$0.75 $\pm$ 0.16 & $-$0.61 $\pm$ 0.16 & $-$0.14 $\pm$ 0.04 & $-$0.11 $\pm$ 0.16 & $-$0.10 $\pm$ 0.15 & $-$0.27 $\pm$ 0.05 & $-$0.24 $\pm$ 0.03 & $-$0.41 $\pm$ 0.07\\ 
OGLE GD-CEP-1337 & $-$1.01 $\pm$ 0.05 & $-$0.66 $\pm$ 0.29 & $-$0.30 $\pm$ 0.06 & $-$0.40 $\pm$ 0.44 & $-$0.10 $\pm$ 0.18 & $-$0.57 $\pm$ 0.10 & $-$0.37 $\pm$ 0.03 & $-$0.68 $\pm$ 0.15\\ 
V1253 Cen & $-$0.50 $\pm$ 0.05 & $-$0.55 $\pm$ 0.16 & 0.18 $\pm$ 0.04 & $-$0.08 $\pm$ 0.16 & $-$0.10 $\pm$ 0.15 & $-$0.02 $\pm$ 0.17 & 0.25 $\pm$ 0.03 & $-$0.03 $\pm$ 0.04\\ 
V1819 Ori & $-$0.60 $\pm$ 0.27 & $-$0.17 $\pm$ 0.10 & $-$0.25 $\pm$ 0.11 & $-$0.51 $\pm$ 0.14 & $-$0.02 $\pm$ 0.19 & $-$0.62 $\pm$ 0.14 & $-$0.48 $\pm$ 0.09 & $-$0.60 $\pm$ 0.13\\ 
V418 CMa & $-$0.78 $\pm$ 0.16 & $-$0.30 $\pm$ 0.16 & $-$0.10 $\pm$ 0.04 & $-$0.58 $\pm$ 0.16 & $-$0.02 $\pm$ 0.15 & $-$0.27 $\pm$ 0.11 & $-$0.37 $\pm$ 0.03 & $-$0.66 $\pm$ 0.12\\ 
V459 Sct & $-$0.32 $\pm$ 0.05 & $-$0.24 $\pm$ 0.16 & 0.68 $\pm$ 0.05 & 0.17 $\pm$ 0.16 & 0.33 $\pm$ 0.15 & 0.22 $\pm$ 0.08 & 0.21 $\pm$ 0.03 & 0.19 $\pm$ 0.04\\ 
V480 Aql & 0.47 $\pm$ 0.06 & 0.08 $\pm$ 0.16 & 0.91 $\pm$ 0.04 & 0.10 $\pm$ 0.16 & 0.33 $\pm$ 0.15 & 0.21 $\pm$ 0.05 & 0.68 $\pm$ 0.04 & 0.14 $\pm$ 0.04\\ 
V881 Cen & $-$0.12 $\pm$ 0.07 & $-$0.05 $\pm$ 0.16 & 0.76 $\pm$ 0.04 & 0.04 $\pm$ 0.16 & 0.24 $\pm$ 0.15 & 0.23 $\pm$ 0.20 & 0.59 $\pm$ 0.03 & 0.24 $\pm$ 0.11\\ 
VX CMa & ---~~~~~~~ & $-$0.49 $\pm$ 0.05 & $-$0.40 $\pm$ 0.16 & $-$0.07 $\pm$ 0.77 & $-$0.10 $\pm$ 0.10 & $-$0.70 $\pm$ 0.20 & $-$0.63 $\pm$ 0.22 & $-$0.61 $\pm$ 0.10\\ 

\hline\end{tabular}
\end{adjustbox}
\end{table*}

%% file: T22_abund_second.tex
\begin{table*}
\ContinuedFloat  
\caption{Second part of the table, with chemical elements from Sc to Ni.}
\begin{adjustbox}{width=\textwidth}
\begin{tabular}{|l|r|r|r|r|r|r|r|r|}
\hline
  \multicolumn{1}{|c|}{Star} &
  \multicolumn{1}{c|}{[Sc/H]} &
  \multicolumn{1}{c|}{[Ti/H]} &
  \multicolumn{1}{c|}{[V/H]} &
  \multicolumn{1}{c|}{[Cr/H]} &
  \multicolumn{1}{c|}{[Mn/H]} &
  \multicolumn{1}{c|}{[Fe/H]} &
  \multicolumn{1}{c|}{[Co/H]} &
  \multicolumn{1}{c|}{[Ni/H]}\\
\hline
ASAS J060450+1021.9 & 0.05 $\pm$ 0.04 & $-$0.41 $\pm$ 0.07 & $-$0.03 $\pm$ 0.08 & $-$0.05 $\pm$ 0.05 & $-$0.80 $\pm$ 0.04 & $-$0.50 $\pm$ 0.18 & $-$0.11 $\pm$ 0.14 & $-$0.47 $\pm$ 0.16\\ 
ASAS J062939-1840.5 & 0.02 $\pm$ 0.11 & $-$0.53 $\pm$ 0.16 & $-$0.66 $\pm$ 0.18 & ---~~~~~~~ & $-$1.13 $\pm$ 0.07 & $-$1.10 $\pm$ 0.19 & ---~~~~~~~ & $-$0.82 $\pm$ 0.14\\ 
ASAS J064001-0754.8 & $-$0.15 $\pm$ 0.04 & $-$0.58 $\pm$ 0.17 & $-$0.10 $\pm$ 0.08 & $-$0.34 $\pm$ 0.05 & $-$0.95 $\pm$ 0.16 & $-$0.76 $\pm$ 0.12 & $-$0.35 $\pm$ 0.16 & $-$0.57 $\pm$ 0.16\\ 
ASAS J065758-1521.4 & 0.11 $\pm$ 0.04 & $-$0.37 $\pm$ 0.11 & $-$0.02 $\pm$ 0.08 & $-$0.16 $\pm$ 0.04 & $-$0.58 $\pm$ 0.55 & $-$0.48 $\pm$ 0.11 & 0.01 $\pm$ 0.16 & $-$0.47 $\pm$ 0.16\\ 
ASAS J074401-3008.4 & $-$0.20 $\pm$ 0.04 & $-$0.64 $\pm$ 0.15 & $-$0.17 $\pm$ 0.08 & $-$0.22 $\pm$ 0.05 & $-$0.76 $\pm$ 0.16 & $-$0.70 $\pm$ 0.15 & $-$0.28 $\pm$ 0.17 & $-$0.66 $\pm$ 0.16\\ 
ASAS J074925-814.4 & 0.19 $\pm$ 0.04 & $-$0.38 $\pm$ 0.12 & $-$0.27 $\pm$ 0.08 & $-$0.34 $\pm$ 0.18 & $-$0.68 $\pm$ 0.58 & $-$0.58 $\pm$ 0.10 & 0.07 $\pm$ 0.10 & $-$0.47 $\pm$ 0.16\\ 
ASAS J084127-4353.6 & 0.23 $\pm$ 0.04 & $-$0.28 $\pm$ 0.21 & $-$0.34 $\pm$ 0.09 & $-$0.15 $\pm$ 0.18 & $-$0.83 $\pm$ 0.08 & $-$0.28 $\pm$ 0.16 & 0.33 $\pm$ 0.17 & $-$0.28 $\pm$ 0.16\\ 
ASAS J164120-4739.6 & 0.81 $\pm$ 0.05 & $-$0.05 $\pm$ 0.14 & 0.07 $\pm$ 0.09 & 0.32 $\pm$ 0.18 & 0.21 $\pm$ 0.16 & 0.18 $\pm$ 0.19 & 0.06 $\pm$ 0.16 & 0.03 $\pm$ 0.16\\ 
ASAS SN-J061713.86+022837.1 & $-$0.33 $\pm$ 0.14 & $-$0.69 $\pm$ 0.08 & $-$0.30 $\pm$ 0.18 & $-$0.60 $\pm$ 0.07 & $-$0.32 $\pm$ 0.16 & $-$0.72 $\pm$ 0.16 & ---~~~~~~~ & $-$0.38 $\pm$ 0.14\\ 
ASAS SN-J063841.36-034927.7 & 0.33 $\pm$ 0.04 & $-$0.20 $\pm$ 0.19 & $-$0.09 $\pm$ 0.08 & $-$0.09 $\pm$ 0.06 & $-$0.78 $\pm$ 0.05 & $-$0.35 $\pm$ 0.14 & 0.26 $\pm$ 0.11 & $-$0.44 $\pm$ 0.16\\ 
ASAS SN-J072739.70-252241.1 & 0.05 $\pm$ 0.04 & $-$0.34 $\pm$ 0.13 & 0.02 $\pm$ 0.09 & $-$0.16 $\pm$ 0.09 & $-$0.89 $\pm$ 0.05 & $-$0.48 $\pm$ 0.16 & ---~~~~~~~ & $-$0.53 $\pm$ 0.16\\ 
ASAS SN-J074354.86-323013.7 & 0.08 $\pm$ 0.04 & $-$0.15 $\pm$ 0.17 & 0.13 $\pm$ 0.09 & $-$0.20 $\pm$ 0.19 & $-$0.74 $\pm$ 0.04 & $-$0.43 $\pm$ 0.18 & 0.49 $\pm$ 0.14 & $-$0.60 $\pm$ 0.16\\ 
ASAS SN-J091822.17-542444.5 & 0.10 $\pm$ 0.04 & $-$0.65 $\pm$ 0.09 & $-$0.38 $\pm$ 0.08 & $-$0.19 $\pm$ 0.05 & $-$0.57 $\pm$ 0.42 & $-$0.40 $\pm$ 0.14 & $-$0.46 $\pm$ 0.15 & $-$0.47 $\pm$ 0.16\\ 
ATLAS J102.7978-10.2541 & 0.14 $\pm$ 0.04 & $-$0.34 $\pm$ 0.13 & $-$0.09 $\pm$ 0.09 & $-$0.19 $\pm$ 0.05 & $-$0.92 $\pm$ 0.05 & $-$0.41 $\pm$ 0.16 & 0.22 $\pm$ 0.20 & $-$0.38 $\pm$ 0.16\\ 
ATLAS J106.7120-14.0234 & $-$0.18 $\pm$ 0.11 & 0.39 $\pm$ 0.09 & 0.04 $\pm$ 0.12 & $-$0.52 $\pm$ 0.18 & ---~~~~~~~ & $-$0.58 $\pm$ 0.12 & ---~~~~~~~ & $-$0.35 $\pm$ 0.16\\ 
ATLAS J113.8534-31.0749 & $-$0.13 $\pm$ 0.04 & $-$0.57 $\pm$ 0.12 & $-$0.11 $\pm$ 0.09 & $-$0.41 $\pm$ 0.05 & $-$0.70 $\pm$ 0.16 & $-$0.66 $\pm$ 0.14 & $-$0.38 $\pm$ 0.09 & $-$0.41 $\pm$ 0.16\\ 
BQ Vel & 0.14 $\pm$ 0.04 & $-$0.40 $\pm$ 0.13 & $-$0.35 $\pm$ 0.09 & $-$0.34 $\pm$ 0.05 & $-$0.61 $\pm$ 0.52 & $-$0.50 $\pm$ 0.16 & $-$0.05 $\pm$ 0.16 & $-$0.47 $\pm$ 0.16\\ 
GDS J133950.2-634049 & 0.47 $\pm$ 0.04 & 0.03 $\pm$ 0.19 & $-$0.16 $\pm$ 0.08 & 0.11 $\pm$ 0.10 & $-$0.23 $\pm$ 0.04 & 0.05 $\pm$ 0.14 & 0.51 $\pm$ 0.49 & 0.03 $\pm$ 0.16\\ 
OGLE GD-CEP-0029 & 0.12 $\pm$ 0.09 & $-$0.01 $\pm$ 0.15 & $-$0.21 $\pm$ 0.08 & $-$0.18 $\pm$ 0.14 & $-$0.62 $\pm$ 0.04 & $-$0.44 $\pm$ 0.16 & 0.02 $\pm$ 0.18 & $-$0.22 $\pm$ 0.16\\ 
OGLE GD-CEP-0089 & 0.03 $\pm$ 0.04 & $-$0.35 $\pm$ 0.16 & $-$0.28 $\pm$ 0.08 & $-$0.26 $\pm$ 0.13 & $-$0.56 $\pm$ 0.47 & $-$0.54 $\pm$ 0.16 & $-$0.04 $\pm$ 0.13 & $-$0.47 $\pm$ 0.16\\ 
OGLE GD-CEP-0120 & $-$0.08 $\pm$ 0.04 & $-$0.51 $\pm$ 0.16 & $-$0.27 $\pm$ 0.12 & $-$0.47 $\pm$ 0.09 & $-$0.73 $\pm$ 0.42 & $-$0.69 $\pm$ 0.13 & $-$0.20 $\pm$ 0.17 & $-$0.66 $\pm$ 0.16\\ 
OGLE GD-CEP-0123 & $-$0.06 $\pm$ 0.04 & $-$0.40 $\pm$ 0.13 & 0.07 $\pm$ 0.08 & $-$0.31 $\pm$ 0.09 & $-$0.98 $\pm$ 0.05 & $-$0.63 $\pm$ 0.13 & $-$0.10 $\pm$ 0.19 & $-$0.60 $\pm$ 0.16\\ 
OGLE GD-CEP-0127 & 0.19 $\pm$ 0.04 & $-$0.43 $\pm$ 0.17 & $-$0.37 $\pm$ 0.08 & $-$0.18 $\pm$ 0.04 & $-$0.40 $\pm$ 0.39 & $-$0.34 $\pm$ 0.12 & $-$0.44 $\pm$ 0.11 & $-$0.47 $\pm$ 0.16\\ 
OGLE GD-CEP-0134 & $-$0.39 $\pm$ 0.04 & $-$0.01 $\pm$ 0.18 & 0.01 $\pm$ 0.10 & $-$0.23 $\pm$ 0.05 & $-$0.95 $\pm$ 0.10 & $-$0.62 $\pm$ 0.16 & ---~~~~~~~ & $-$0.41 $\pm$ 0.16\\ 
OGLE GD-CEP-0156 & $-$0.34 $\pm$ 0.04 & $-$0.55 $\pm$ 0.21 & $-$0.29 $\pm$ 0.09 & $-$0.28 $\pm$ 0.07 & $-$0.96 $\pm$ 0.04 & $-$0.65 $\pm$ 0.11 & 0.12 $\pm$ 0.17 & $-$0.60 $\pm$ 0.16\\ 
OGLE GD-CEP-0159 & 0.12 $\pm$ 0.04 & $-$0.32 $\pm$ 0.14 & 0.03 $\pm$ 0.10 & $-$0.28 $\pm$ 0.04 & $-$0.72 $\pm$ 0.10 & $-$0.51 $\pm$ 0.13 & 0.16 $\pm$ 0.17 & $-$0.47 $\pm$ 0.16\\ 
OGLE GD-CEP-0162 & $-$0.18 $\pm$ 0.04 & $-$0.48 $\pm$ 0.19 & $-$0.21 $\pm$ 0.08 & $-$0.38 $\pm$ 0.14 & $-$0.90 $\pm$ 0.06 & $-$0.62 $\pm$ 0.10 & 0.20 $\pm$ 0.12 & $-$0.50 $\pm$ 0.16\\ 
OGLE GD-CEP-0168 & 0.04 $\pm$ 0.04 & $-$0.09 $\pm$ 0.17 & 0.00 $\pm$ 0.08 & $-$0.21 $\pm$ 0.09 & $-$0.79 $\pm$ 0.04 & $-$0.42 $\pm$ 0.11 & 0.15 $\pm$ 0.18 & $-$0.34 $\pm$ 0.16\\ 
OGLE GD-CEP-0176 & $-$0.32 $\pm$ 0.04 & $-$0.87 $\pm$ 0.11 & $-$0.26 $\pm$ 0.08 & $-$0.34 $\pm$ 0.09 & $-$0.54 $\pm$ 0.55 & $-$0.74 $\pm$ 0.16 & $-$0.15 $\pm$ 0.14 & $-$0.72 $\pm$ 0.16\\ 
OGLE GD-CEP-0179 & 0.06 $\pm$ 0.04 & $-$0.17 $\pm$ 0.16 & $-$0.04 $\pm$ 0.08 & $-$0.15 $\pm$ 0.09 & $-$0.64 $\pm$ 0.04 & $-$0.32 $\pm$ 0.20 & ---~~~~~~~ & $-$0.28 $\pm$ 0.16\\ 
OGLE GD-CEP-0181 & $-$0.23 $\pm$ 0.04 & $-$0.61 $\pm$ 0.18 & $-$0.33 $\pm$ 0.08 & $-$0.28 $\pm$ 0.05 & $-$0.73 $\pm$ 0.69 & $-$0.74 $\pm$ 0.14 & 0.01 $\pm$ 0.15 & $-$0.60 $\pm$ 0.16\\ 
OGLE GD-CEP-0185 & $-$0.04 $\pm$ 0.04 & $-$0.44 $\pm$ 0.17 & $-$0.16 $\pm$ 0.08 & $-$0.25 $\pm$ 0.05 & $-$0.53 $\pm$ 0.41 & $-$0.54 $\pm$ 0.17 & $-$0.11 $\pm$ 0.17 & $-$0.50 $\pm$ 0.16\\ 
OGLE GD-CEP-0186 & 0.42 $\pm$ 0.04 & $-$0.05 $\pm$ 0.13 & 0.06 $\pm$ 0.08 & $-$0.03 $\pm$ 0.35 & $-$0.59 $\pm$ 0.04 & $-$0.16 $\pm$ 0.14 & 0.04 $\pm$ 0.17 & $-$0.22 $\pm$ 0.16\\ 
OGLE GD-CEP-0196 & $-$0.19 $\pm$ 0.04 & $-$0.62 $\pm$ 0.18 & $-$0.36 $\pm$ 0.09 & $-$0.45 $\pm$ 0.05 & $-$0.82 $\pm$ 0.04 & $-$0.73 $\pm$ 0.15 & 0.15 $\pm$ 0.13 & $-$0.60 $\pm$ 0.16\\ 
OGLE GD-CEP-0206 & 0.15 $\pm$ 0.04 & $-$0.02 $\pm$ 0.17 & $-$0.16 $\pm$ 0.09 & $-$0.18 $\pm$ 0.19 & $-$0.61 $\pm$ 0.04 & $-$0.37 $\pm$ 0.18 & ---~~~~~~~ & $-$0.34 $\pm$ 0.16\\ 
OGLE GD-CEP-0213 & 0.55 $\pm$ 0.07 & 0.26 $\pm$ 0.17 & $-$0.27 $\pm$ 0.21 & $-$0.33 $\pm$ 0.14 & $-$0.00 $\pm$ 0.19 & $-$0.78 $\pm$ 0.16 & 0.07 $\pm$ 0.13 & $-$0.45 $\pm$ 0.15\\ 
OGLE GD-CEP-0214 & $-$0.35 $\pm$ 0.04 & $-$0.66 $\pm$ 0.20 & $-$0.25 $\pm$ 0.15 & $-$0.34 $\pm$ 0.07 & $-$0.29 $\pm$ 0.34 & $-$0.69 $\pm$ 0.16 & $-$0.07 $\pm$ 0.17 & $-$0.60 $\pm$ 0.16\\ 
OGLE GD-CEP-0224 & $-$0.05 $\pm$ 0.07 & 0.14 $\pm$ 0.16 & $-$0.13 $\pm$ 0.15 & $-$0.22 $\pm$ 0.18 & $-$0.02 $\pm$ 0.10 & $-$0.51 $\pm$ 0.13 & ---~~~~~~~ & $-$0.27 $\pm$ 0.08\\ 
OGLE GD-CEP-0228 & 0.12 $\pm$ 0.04 & $-$0.33 $\pm$ 0.15 & $-$0.08 $\pm$ 0.08 & $-$0.17 $\pm$ 0.05 & $-$0.39 $\pm$ 0.35 & $-$0.47 $\pm$ 0.16 & $-$0.04 $\pm$ 0.17 & $-$0.47 $\pm$ 0.16\\ 
OGLE GD-CEP-0247 & 0.03 $\pm$ 0.04 & $-$0.13 $\pm$ 0.18 & $-$0.07 $\pm$ 0.08 & $-$0.18 $\pm$ 0.11 & $-$0.57 $\pm$ 0.04 & $-$0.36 $\pm$ 0.15 & 0.24 $\pm$ 0.17 & $-$0.41 $\pm$ 0.16\\ 
OGLE GD-CEP-0252 & 0.13 $\pm$ 0.04 & $-$0.18 $\pm$ 0.17 & $-$0.12 $\pm$ 0.08 & $-$0.11 $\pm$ 0.08 & $-$0.82 $\pm$ 0.08 & $-$0.39 $\pm$ 0.13 & ---~~~~~~~ & $-$0.38 $\pm$ 0.16\\ 
OGLE GD-CEP-0271 & $-$0.27 $\pm$ 0.04 & $-$0.39 $\pm$ 0.16 & 0.11 $\pm$ 0.08 & $-$0.52 $\pm$ 0.04 & $-$1.00 $\pm$ 0.05 & $-$0.66 $\pm$ 0.15 & $-$0.29 $\pm$ 0.20 & $-$0.47 $\pm$ 0.16\\ 
OGLE GD-CEP-0316 & 0.29 $\pm$ 0.04 & $-$0.17 $\pm$ 0.14 & 0.14 $\pm$ 0.08 & $-$0.18 $\pm$ 0.04 & $-$0.40 $\pm$ 0.42 & $-$0.33 $\pm$ 0.12 & 0.30 $\pm$ 0.15 & $-$0.41 $\pm$ 0.16\\ 
OGLE GD-CEP-0342 & $-$0.11 $\pm$ 0.06 & $-$0.45 $\pm$ 0.12 & $-$0.34 $\pm$ 0.08 & $-$0.29 $\pm$ 0.05 & $-$0.93 $\pm$ 0.08 & $-$0.59 $\pm$ 0.11 & 0.18 $\pm$ 0.18 & $-$0.53 $\pm$ 0.16\\ 
OGLE GD-CEP-0348 & $-$0.22 $\pm$ 0.04 & $-$0.54 $\pm$ 0.19 & $-$0.34 $\pm$ 0.08 & $-$0.42 $\pm$ 0.18 & $-$1.02 $\pm$ 0.04 & $-$0.63 $\pm$ 0.12 & 0.12 $\pm$ 0.17 & $-$0.53 $\pm$ 0.16\\ 
OGLE GD-CEP-0353 & 0.25 $\pm$ 0.04 & $-$0.22 $\pm$ 0.20 & $-$0.20 $\pm$ 0.08 & $-$0.14 $\pm$ 0.05 & $-$0.41 $\pm$ 0.33 & $-$0.41 $\pm$ 0.13 & 0.07 $\pm$ 0.17 & $-$0.41 $\pm$ 0.16\\ 
OGLE GD-CEP-0516 & 0.00 $\pm$ 0.04 & $-$0.39 $\pm$ 0.21 & $-$0.02 $\pm$ 0.11 & $-$0.08 $\pm$ 0.05 & $-$0.48 $\pm$ 0.10 & $-$0.54 $\pm$ 0.16 & 0.35 $\pm$ 0.15 & $-$0.32 $\pm$ 0.16\\ 
OGLE GD-CEP-0568 & 0.53 $\pm$ 0.04 & $-$0.27 $\pm$ 0.19 & $-$0.14 $\pm$ 0.08 & $-$0.12 $\pm$ 0.27 & $-$0.41 $\pm$ 0.06 & $-$0.20 $\pm$ 0.16 & 0.02 $\pm$ 0.13 & $-$0.25 $\pm$ 0.16\\ 
OGLE GD-CEP-0575 & 0.17 $\pm$ 0.04 & $-$0.38 $\pm$ 0.19 & $-$0.26 $\pm$ 0.09 & $-$0.18 $\pm$ 1.06 & $-$0.35 $\pm$ 0.22 & $-$0.37 $\pm$ 0.15 & $-$0.12 $\pm$ 0.16 & $-$0.22 $\pm$ 0.16\\ 
OGLE GD-CEP-0889 & $-$0.00 $\pm$ 0.16 & $-$0.05 $\pm$ 0.14 & 0.18 $\pm$ 0.11 & $-$0.30 $\pm$ 0.27 & $-$0.05 $\pm$ 0.05 & $-$0.64 $\pm$ 0.18 & 0.43 $\pm$ 0.10 & $-$0.45 $\pm$ 0.19\\ 
OGLE GD-CEP-0974 & 0.74 $\pm$ 0.04 & 0.02 $\pm$ 0.17 & $-$0.05 $\pm$ 0.09 & 0.23 $\pm$ 0.35 & 0.11 $\pm$ 0.04 & 0.15 $\pm$ 0.14 & 0.11 $\pm$ 0.17 & $-$0.03 $\pm$ 0.16\\ 
OGLE GD-CEP-0996 & 0.59 $\pm$ 0.04 & 0.38 $\pm$ 0.15 & $-$0.01 $\pm$ 0.08 & 0.39 $\pm$ 0.10 & 0.18 $\pm$ 0.06 & 0.28 $\pm$ 0.18 & 0.51 $\pm$ 0.17 & 0.17 $\pm$ 0.16\\ 
OGLE GD-CEP-1012 & 0.83 $\pm$ 0.05 & $-$0.05 $\pm$ 0.19 & $-$0.13 $\pm$ 0.11 & 0.10 $\pm$ 0.05 & 0.26 $\pm$ 0.16 & 0.12 $\pm$ 0.16 & 0.01 $\pm$ 0.13 & 0.03 $\pm$ 0.16\\ 
OGLE GD-CEP-1111 & 0.43 $\pm$ 0.09 & 0.33 $\pm$ 0.15 & 0.02 $\pm$ 0.19 & 0.15 $\pm$ 0.16 & 0.27 $\pm$ 0.15 & $-$0.09 $\pm$ 0.12 & 0.43 $\pm$ 0.18 & 0.15 $\pm$ 0.11\\ 
OGLE GD-CEP-1210 & 0.92 $\pm$ 0.04 & 0.03 $\pm$ 0.14 & 0.23 $\pm$ 0.08 & 0.36 $\pm$ 0.18 & $-$0.26 $\pm$ 0.10 & $-$0.00 $\pm$ 0.20 & 0.24 $\pm$ 0.17 & 0.03 $\pm$ 0.16\\ 
OGLE GD-CEP-1285 & 0.37 $\pm$ 0.11 & 0.25 $\pm$ 0.17 & $-$0.23 $\pm$ 0.11 & ---~~~~~~~ & $-$0.03 $\pm$ 0.19 & $-$1.07 $\pm$ 0.17 & ---~~~~~~~ & $-$0.59 $\pm$ 0.19\\ 
OGLE GD-CEP-1311 & 0.11 $\pm$ 0.04 & $-$0.44 $\pm$ 0.12 & $-$0.09 $\pm$ 0.08 & $-$0.32 $\pm$ 0.05 & $-$0.56 $\pm$ 0.45 & $-$0.47 $\pm$ 0.10 & $-$0.03 $\pm$ 0.17 & $-$0.41 $\pm$ 0.16\\ 
OGLE GD-CEP-1337 & 0.00 $\pm$ 0.06 & 0.07 $\pm$ 0.14 & $-$0.43 $\pm$ 0.14 & $-$0.65 $\pm$ 0.07 & $-$0.98 $\pm$ 0.20 & $-$0.93 $\pm$ 0.17 & $-$0.23 $\pm$ 0.16 & $-$0.62 $\pm$ 0.14\\ 
V1253 Cen & 0.31 $\pm$ 0.04 & $-$0.13 $\pm$ 0.20 & $-$0.08 $\pm$ 0.09 & $-$0.06 $\pm$ 0.05 & $-$0.50 $\pm$ 0.04 & $-$0.26 $\pm$ 0.14 & 0.15 $\pm$ 0.20 & $-$0.22 $\pm$ 0.16\\ 
V1819 Ori & 0.50 $\pm$ 0.05 & $-$0.07 $\pm$ 0.19 & $-$0.17 $\pm$ 0.21 & $-$0.81 $\pm$ 0.27 & $-$0.13 $\pm$ 0.19 & $-$0.85 $\pm$ 0.15 & $-$0.30 $\pm$ 0.17 & $-$0.49 $\pm$ 0.11\\ 
V418 CMa & $-$0.07 $\pm$ 0.04 & $-$0.42 $\pm$ 0.15 & $-$0.23 $\pm$ 0.10 & $-$0.41 $\pm$ 0.19 & $-$0.74 $\pm$ 0.69 & $-$0.64 $\pm$ 0.13 & $-$0.16 $\pm$ 0.17 & $-$0.56 $\pm$ 0.16\\ 
V459 Sct & 0.51 $\pm$ 0.04 & 0.18 $\pm$ 0.17 & 0.02 $\pm$ 0.08 & 0.30 $\pm$ 0.04 & 0.11 $\pm$ 0.11 & 0.22 $\pm$ 0.16 & 0.38 $\pm$ 0.18 & 0.22 $\pm$ 0.16\\ 
V480 Aql & 0.70 $\pm$ 0.04 & 0.03 $\pm$ 0.19 & 0.15 $\pm$ 0.09 & 0.27 $\pm$ 0.35 & 0.11 $\pm$ 0.08 & 0.17 $\pm$ 0.16 & 0.15 $\pm$ 0.17 & $-$0.03 $\pm$ 0.16\\ 
V881 Cen & 0.64 $\pm$ 0.04 & $-$0.11 $\pm$ 0.18 & $-$0.09 $\pm$ 0.09 & 0.19 $\pm$ 0.18 & 0.04 $\pm$ 0.10 & 0.07 $\pm$ 0.11 & $-$0.02 $\pm$ 0.17 & 0.09 $\pm$ 0.16\\ 
VX CMa & $-$0.46 $\pm$ 0.06 & 0.18 $\pm$ 0.13 & $-$0.34 $\pm$ 0.14 & $-$0.50 $\pm$ 0.13 & $-$0.04 $\pm$ 0.04 & $-$0.92 $\pm$ 0.17 & $-$0.14 $\pm$ 0.21 & $-$0.44 $\pm$ 0.14\\ 
\hline\end{tabular}
\end{adjustbox}
\end{table*}

%% file: T22_abund_third.tex
\begin{table*}
\ContinuedFloat  
\caption{Third part of the table. Abundances for chemical elements from Cu to Nd.}
\begin{adjustbox}{width=\textwidth}
\begin{tabular}{|l|r|r|r|r|r|r|r|r|}
\hline
  \multicolumn{1}{|c|}{Star} &
  \multicolumn{1}{c|}{[Cu/H]} &
  \multicolumn{1}{c|}{[Zn]/H]} &
  \multicolumn{1}{c|}{[Y/H]} &
  \multicolumn{1}{c|}{[Zr/H]} &
  \multicolumn{1}{c|}{[Ba/H]} &
  \multicolumn{1}{c|}{[La/H]} &
  \multicolumn{1}{c|}{[Pr/H]} &
  \multicolumn{1}{c|}{[Nd/H]}\\
\hline
ASAS J060450+1021.9 & 0.01 $\pm$ 0.05 & $-$0.80 $\pm$ 0.16 & $-$0.04 $\pm$ 0.04 & 0.43 $\pm$ 0.16 & $-$0.18 $\pm$ 0.14 & $-$0.09 $\pm$ 0.04 & 0.04 $\pm$ 0.16 & $-$0.08 $\pm$ 0.05\\ 
ASAS J062939-1840.5 & $-$0.69 $\pm$ 0.06 & $-$1.45 $\pm$ 0.11 & $-$0.60 $\pm$ 0.04 & $-$0.12 $\pm$ 0.10 & $-$0.22 $\pm$ 0.14 & $-$0.47 $\pm$ 0.04 & $-$0.51 $\pm$ 0.17 & $-$0.51 $\pm$ 0.05\\ 
ASAS J064001-0754.8 & $-$0.42 $\pm$ 0.07 & $-$0.99 $\pm$ 0.16 & $-$0.27 $\pm$ 0.04 & 0.33 $\pm$ 0.16 & $-$0.21 $\pm$ 0.18 & $-$0.32 $\pm$ 0.05 & $-$0.33 $\pm$ 0.16 & $-$0.44 $\pm$ 0.05\\ 
ASAS J065758-1521.4 & $-$0.12 $\pm$ 0.08 & $-$0.93 $\pm$ 0.16 & $-$0.04 $\pm$ 0.04 & 0.42 $\pm$ 0.16 & 0.24 $\pm$ 0.14 & $-$0.01 $\pm$ 0.04 & $-$0.39 $\pm$ 0.16 & $-$0.15 $\pm$ 0.05\\ 
ASAS J074401-3008.4 & $-$0.23 $\pm$ 0.04 & $-$0.91 $\pm$ 0.16 & $-$0.23 $\pm$ 0.04 & 0.14 $\pm$ 0.16 & $-$0.44 $\pm$ 0.13 & $-$0.35 $\pm$ 0.04 & $-$0.51 $\pm$ 0.16 & $-$0.38 $\pm$ 0.04\\ 
ASAS J074925-3814.4 & $-$0.47 $\pm$ 0.34 & $-$0.46 $\pm$ 0.16 & $-$0.16 $\pm$ 0.06 & 0.15 $\pm$ 0.16 & 0.08 $\pm$ 0.09 & $-$0.09 $\pm$ 0.04 & $-$0.22 $\pm$ 0.16 & $-$0.43 $\pm$ 0.54\\ 
ASAS J084127-4353.6 & 0.10 $\pm$ 0.04 & $-$0.21 $\pm$ 0.16 & 0.09 $\pm$ 0.04 & 0.11 $\pm$ 0.16 & 0.45 $\pm$ 0.10 & $-$0.05 $\pm$ 0.04 & $-$0.14 $\pm$ 0.16 & $-$0.11 $\pm$ 0.06\\ 
ASAS J164120-4739.6 & 0.53 $\pm$ 0.20 & 0.42 $\pm$ 0.16 & 0.22 $\pm$ 0.06 & 0.50 $\pm$ 0.16 & 0.39 $\pm$ 0.10 & 0.19 $\pm$ 0.13 & $-$0.43 $\pm$ 0.16 & 0.15 $\pm$ 0.05\\ 
ASAS SN-J061713.86+022837.1 & 0.15 $\pm$ 0.16 & $-$0.90 $\pm$ 0.11 & $-$0.20 $\pm$ 0.04 & 0.21 $\pm$ 0.04 & $-$0.88 $\pm$ 0.10 & 0.05 $\pm$ 0.04 & ---~~~~~~~ & $-$0.38 $\pm$ 0.11\\ 
ASAS SN-J063841.36-034927.7 & 0.04 $\pm$ 0.06 & $-$0.63 $\pm$ 0.16 & 0.08 $\pm$ 0.04 & 0.41 $\pm$ 0.16 & 0.28 $\pm$ 0.14 & 0.20 $\pm$ 0.04 & $-$0.13 $\pm$ 0.16 & $-$0.02 $\pm$ 0.12\\ 
ASAS SN-J072739.70-252241.1 & $-$0.01 $\pm$ 0.09 & $-$0.83 $\pm$ 0.16 & 0.00 $\pm$ 0.05 & 0.41 $\pm$ 0.16 & 0.05 $\pm$ 0.09 & $-$0.04 $\pm$ 0.04 & $-$0.20 $\pm$ 0.16 & 0.00 $\pm$ 0.05\\ 
ASAS SN-J074354.86-323013.7 & 0.08 $\pm$ 0.25 & $-$0.71 $\pm$ 0.16 & ---~~~~~~~ & 0.70 $\pm$ 0.16 & 1.01 $\pm$ 0.09 & ---~~~~~~~ & ---~~~~~~~ & ---~~~~~~~\\ 
ASAS SN-J091822.17-542444.5 & $-$0.21 $\pm$ 0.11 & $-$0.87 $\pm$ 0.16 & $-$0.27 $\pm$ 0.06 & $-$0.11 $\pm$ 0.16 & $-$0.36 $\pm$ 0.15 & $-$0.19 $\pm$ 0.06 & $-$0.64 $\pm$ 0.16 & $-$0.27 $\pm$ 0.04\\ 
ATLAS J102.7978-10.2541 & $-$0.04 $\pm$ 0.06 & $-$0.73 $\pm$ 0.16 & $-$0.15 $\pm$ 0.04 & 0.20 $\pm$ 0.16 & 0.14 $\pm$ 0.15 & $-$0.05 $\pm$ 0.04 & $-$0.28 $\pm$ 0.16 & $-$0.12 $\pm$ 0.04\\ 
ATLAS J106.7120-14.0234 & $-$0.10 $\pm$ 0.10 & $-$0.52 $\pm$ 0.11 & 0.34 $\pm$ 0.11 & 0.74 $\pm$ 0.12 & 0.33 $\pm$ 0.11 & $-$0.18 $\pm$ 0.05 & $-$0.53 $\pm$ 0.40 & 0.03 $\pm$ 0.12\\ 
ATLAS J113.8534-31.0749 & $-$0.35 $\pm$ 0.16 & $-$0.99 $\pm$ 0.16 & $-$0.25 $\pm$ 0.05 & 0.33 $\pm$ 0.16 & $-$0.27 $\pm$ 0.14 & $-$0.18 $\pm$ 0.04 & ---~~~~~~~ & $-$0.50 $\pm$ 0.07\\ 
BQ Vel & $-$0.38 $\pm$ 0.09 & $-$0.85 $\pm$ 0.16 & $-$0.22 $\pm$ 0.04 & 0.13 $\pm$ 0.16 & 0.04 $\pm$ 0.11 & $-$0.14 $\pm$ 0.04 & $-$0.24 $\pm$ 0.16 & $-$0.52 $\pm$ 0.48\\ 
GDS J133950.2-634049 & 0.38 $\pm$ 0.05 & 0.10 $\pm$ 0.16 & 0.10 $\pm$ 0.04 & 0.34 $\pm$ 0.16 & 0.33 $\pm$ 0.19 & $-$0.07 $\pm$ 0.04 & $-$0.51 $\pm$ 0.16 & $-$0.16 $\pm$ 0.07\\ 
OGLE GD-CEP-0029 & 0.11 $\pm$ 0.04 & $-$0.74 $\pm$ 0.16 & $-$0.12 $\pm$ 0.10 & 0.14 $\pm$ 0.16 & 0.23 $\pm$ 0.13 & $-$0.07 $\pm$ 0.06 & $-$0.28 $\pm$ 0.16 & $-$0.18 $\pm$ 0.08\\ 
OGLE GD-CEP-0089 & $-$0.04 $\pm$ 0.05 & $-$0.85 $\pm$ 0.16 & $-$0.07 $\pm$ 0.04 & 0.17 $\pm$ 0.16 & $-$0.29 $\pm$ 0.11 & $-$0.08 $\pm$ 0.04 & $-$0.26 $\pm$ 0.16 & $-$0.08 $\pm$ 0.04\\ 
OGLE GD-CEP-0120 & $-$0.33 $\pm$ 0.06 & $-$0.96 $\pm$ 0.16 & $-$0.36 $\pm$ 0.04 & 0.06 $\pm$ 0.16 & $-$0.56 $\pm$ 0.10 & $-$0.27 $\pm$ 0.04 & $-$0.32 $\pm$ 0.16 & $-$0.47 $\pm$ 0.05\\ 
OGLE GD-CEP-0123 & $-$0.17 $\pm$ 0.04 & $-$0.85 $\pm$ 0.16 & $-$0.21 $\pm$ 0.04 & 0.30 $\pm$ 0.16 & $-$0.45 $\pm$ 0.10 & $-$0.11 $\pm$ 0.05 & $-$0.35 $\pm$ 0.16 & $-$0.23 $\pm$ 0.04\\ 
OGLE GD-CEP-0127 & $-$0.10 $\pm$ 0.32 & $-$0.33 $\pm$ 0.16 & $-$0.20 $\pm$ 0.08 & 0.00 $\pm$ 0.16 & $-$0.05 $\pm$ 0.14 & 0.05 $\pm$ 0.06 & $-$0.49 $\pm$ 0.16 & $-$0.00 $\pm$ 0.04\\ 
OGLE GD-CEP-0134 & 0.30 $\pm$ 0.04 & $-$0.90 $\pm$ 0.16 & $-$0.20 $\pm$ 0.09 & 0.27 $\pm$ 0.16 & $-$0.92 $\pm$ 0.13 & $-$0.21 $\pm$ 0.10 & ---~~~~~~~ & $-$0.13 $\pm$ 0.06\\ 
OGLE GD-CEP-0156 & $-$0.17 $\pm$ 0.06 & $-$1.05 $\pm$ 0.16 & $-$0.19 $\pm$ 0.06 & 0.15 $\pm$ 0.16 & $-$0.61 $\pm$ 0.14 & $-$0.32 $\pm$ 0.07 & $-$0.39 $\pm$ 0.16 & $-$0.42 $\pm$ 0.05\\ 
OGLE GD-CEP-0159 & $-$0.13 $\pm$ 0.08 & $-$0.83 $\pm$ 0.16 & $-$0.08 $\pm$ 0.04 & 0.22 $\pm$ 0.16 & $-$0.02 $\pm$ 0.10 & $-$0.01 $\pm$ 0.04 & $-$0.39 $\pm$ 0.16 & $-$0.28 $\pm$ 0.32\\ 
OGLE GD-CEP-0162 & $-$0.09 $\pm$ 0.04 & $-$0.85 $\pm$ 0.16 & $-$0.26 $\pm$ 0.04 & 0.15 $\pm$ 0.16 & $-$0.46 $\pm$ 0.14 & $-$0.61 $\pm$ 0.07 & $-$0.51 $\pm$ 0.16 & $-$0.18 $\pm$ 0.05\\ 
OGLE GD-CEP-0168 & 0.05 $\pm$ 0.05 & $-$0.83 $\pm$ 0.16 & $-$0.10 $\pm$ 0.04 & 0.35 $\pm$ 0.16 & $-$0.19 $\pm$ 0.17 & $-$0.07 $\pm$ 0.05 & $-$0.51 $\pm$ 0.16 & $-$0.49 $\pm$ 0.49\\ 
OGLE GD-CEP-0176 & $-$0.34 $\pm$ 0.04 & $-$1.01 $\pm$ 0.16 & $-$0.27 $\pm$ 0.04 & 0.05 $\pm$ 0.16 & $-$0.83 $\pm$ 0.15 & $-$0.49 $\pm$ 0.04 & $-$0.45 $\pm$ 0.16 & $-$0.44 $\pm$ 0.05\\ 
OGLE GD-CEP-0179 & 0.21 $\pm$ 0.05 & $-$0.77 $\pm$ 0.16 & $-$0.04 $\pm$ 0.04 & 0.40 $\pm$ 0.16 & $-$0.11 $\pm$ 0.09 & 0.10 $\pm$ 0.05 & $-$0.38 $\pm$ 0.16 & $-$0.14 $\pm$ 0.09\\ 
OGLE GD-CEP-0181 & $-$0.50 $\pm$ 0.06 & $-$1.05 $\pm$ 0.16 & $-$0.32 $\pm$ 0.04 & 0.05 $\pm$ 0.16 & $-$0.63 $\pm$ 0.14 & $-$0.30 $\pm$ 0.04 & $-$0.15 $\pm$ 0.16 & $-$0.48 $\pm$ 0.04\\ 
OGLE GD-CEP-0185 & $-$0.29 $\pm$ 0.07 & $-$0.82 $\pm$ 0.16 & $-$0.23 $\pm$ 0.04 & 0.13 $\pm$ 0.16 & $-$0.32 $\pm$ 0.10 & $-$0.20 $\pm$ 0.04 & $-$0.36 $\pm$ 0.16 & $-$0.30 $\pm$ 0.11\\ 
OGLE GD-CEP-0186 & 0.00 $\pm$ 0.10 & $-$0.46 $\pm$ 0.16 & 0.05 $\pm$ 0.04 & 0.43 $\pm$ 0.16 & 0.50 $\pm$ 0.11 & 0.20 $\pm$ 0.04 & $-$0.26 $\pm$ 0.16 & 0.03 $\pm$ 0.06\\ 
OGLE GD-CEP-0196 & $-$0.26 $\pm$ 0.04 & $-$0.90 $\pm$ 0.16 & $-$0.36 $\pm$ 0.04 & 0.00 $\pm$ 0.16 & $-$0.75 $\pm$ 0.10 & $-$0.34 $\pm$ 0.04 & $-$0.22 $\pm$ 0.16 & $-$0.33 $\pm$ 0.04\\ 
OGLE GD-CEP-0206 & 0.36 $\pm$ 0.04 & $-$0.71 $\pm$ 0.16 & 0.05 $\pm$ 0.04 & 0.38 $\pm$ 0.16 & $-$0.11 $\pm$ 0.11 & 0.05 $\pm$ 0.04 & $-$0.13 $\pm$ 0.16 & 0.04 $\pm$ 0.05\\ 
OGLE GD-CEP-0213 & $-$0.22 $\pm$ 0.06 & $-$1.01 $\pm$ 0.11 & $-$0.32 $\pm$ 0.05 & 0.11 $\pm$ 1.06 & $-$0.53 $\pm$ 0.16 & $-$0.21 $\pm$ 0.15 & $-$0.42 $\pm$ 0.16 & $-$0.03 $\pm$ 0.14\\ 
OGLE GD-CEP-0214 & 0.03 $\pm$ 0.04 & $-$0.87 $\pm$ 0.16 & $-$0.35 $\pm$ 0.04 & 0.09 $\pm$ 0.16 & $-$0.95 $\pm$ 0.10 & $-$0.40 $\pm$ 0.12 & $-$0.30 $\pm$ 0.16 & $-$0.57 $\pm$ 0.06\\ 
OGLE GD-CEP-0224 & 0.08 $\pm$ 0.12 & $-$0.76 $\pm$ 0.11 & $-$0.15 $\pm$ 0.04 & 0.21 $\pm$ 0.14 & $-$0.44 $\pm$ 0.16 & 0.10 $\pm$ 0.11 & $-$0.20 $\pm$ 0.16 & 0.19 $\pm$ 0.11\\ 
OGLE GD-CEP-0228 & $-$0.12 $\pm$ 0.06 & $-$0.71 $\pm$ 0.16 & $-$0.04 $\pm$ 0.04 & 0.32 $\pm$ 0.16 & $-$0.10 $\pm$ 0.11 & 0.04 $\pm$ 0.04 & $-$0.13 $\pm$ 0.16 & $-$0.01 $\pm$ 0.04\\ 
OGLE GD-CEP-0247 & 0.10 $\pm$ 0.05 & $-$0.71 $\pm$ 0.16 & $-$0.10 $\pm$ 0.04 & 0.18 $\pm$ 0.16 & $-$0.27 $\pm$ 0.11 & $-$0.11 $\pm$ 0.04 & $-$0.26 $\pm$ 0.16 & $-$0.24 $\pm$ 0.04\\ 
OGLE GD-CEP-0252 & 0.21 $\pm$ 0.05 & $-$0.58 $\pm$ 0.16 & 0.04 $\pm$ 0.04 & 0.38 $\pm$ 0.16 & 0.20 $\pm$ 0.09 & 0.09 $\pm$ 0.04 & 0.02 $\pm$ 0.16 & $-$0.30 $\pm$ 0.66\\ 
OGLE GD-CEP-0271 & $-$0.36 $\pm$ 0.06 & $-$0.96 $\pm$ 0.16 & $-$0.30 $\pm$ 0.05 & 0.27 $\pm$ 0.16 & $-$0.58 $\pm$ 0.13 & $-$0.23 $\pm$ 0.07 & $-$0.51 $\pm$ 0.16 & $-$0.34 $\pm$ 0.06\\ 
OGLE GD-CEP-0316 & $-$0.06 $\pm$ 0.15 & $-$0.58 $\pm$ 0.16 & 0.01 $\pm$ 0.04 & 0.29 $\pm$ 0.16 & 0.23 $\pm$ 0.12 & 0.09 $\pm$ 0.04 & $-$0.26 $\pm$ 0.16 & 0.02 $\pm$ 0.05\\ 
OGLE GD-CEP-0342 & $-$0.23 $\pm$ 0.06 & $-$0.93 $\pm$ 0.16 & $-$0.30 $\pm$ 0.05 & 0.04 $\pm$ 0.16 & $-$0.15 $\pm$ 0.16 & $-$0.40 $\pm$ 0.04 & $-$0.49 $\pm$ 0.16 & $-$0.31 $\pm$ 0.04\\ 
OGLE GD-CEP-0348 & $-$0.20 $\pm$ 0.07 & $-$0.93 $\pm$ 0.16 & $-$0.42 $\pm$ 0.04 & $-$0.01 $\pm$ 0.16 & $-$0.69 $\pm$ 0.35 & $-$0.34 $\pm$ 0.05 & $-$0.24 $\pm$ 0.16 & $-$0.34 $\pm$ 0.04\\ 
OGLE GD-CEP-0353 & $-$0.20 $\pm$ 0.08 & $-$0.65 $\pm$ 0.16 & $-$0.15 $\pm$ 0.04 & 0.25 $\pm$ 0.16 & $-$0.08 $\pm$ 0.09 & $-$0.03 $\pm$ 0.04 & $-$0.38 $\pm$ 0.16 & $-$0.17 $\pm$ 0.08\\ 
OGLE GD-CEP-0516 & $-$0.18 $\pm$ 0.05 & $-$0.33 $\pm$ 0.16 & $-$0.07 $\pm$ 0.04 & 0.24 $\pm$ 0.16 & $-$0.36 $\pm$ 0.12 & 0.17 $\pm$ 0.05 & ---~~~~~~~ & $-$0.41 $\pm$ 0.04\\ 
OGLE GD-CEP-0568 & $-$0.09 $\pm$ 0.28 & $-$0.09 $\pm$ 0.16 & 0.06 $\pm$ 0.04 & 0.01 $\pm$ 0.16 & 0.30 $\pm$ 0.14 & 0.20 $\pm$ 0.07 & $-$0.33 $\pm$ 0.16 & 0.09 $\pm$ 0.04\\ 
OGLE GD-CEP-0575 & $-$0.30 $\pm$ 0.05 & $-$0.16 $\pm$ 0.16 & $-$0.19 $\pm$ 0.04 & 0.07 $\pm$ 0.16 & $-$0.31 $\pm$ 0.09 & $-$0.15 $\pm$ 0.04 & $-$0.62 $\pm$ 0.16 & $-$0.25 $\pm$ 0.04\\ 
OGLE GD-CEP-0889 & $-$0.29 $\pm$ 0.35 & $-$0.24 $\pm$ 0.11 & $-$0.11 $\pm$ 0.12 & 0.29 $\pm$ 0.09 & 0.18 $\pm$ 0.20 & $-$0.09 $\pm$ 0.14 & ---~~~~~~~ & $-$0.24 $\pm$ 0.16\\ 
OGLE GD-CEP-0974 & 0.40 $\pm$ 0.07 & 0.17 $\pm$ 0.16 & 0.08 $\pm$ 0.05 & 0.51 $\pm$ 0.16 & $-$0.02 $\pm$ 0.12 & 0.08 $\pm$ 0.08 & $-$0.40 $\pm$ 0.16 & $-$0.02 $\pm$ 0.05\\ 
OGLE GD-CEP-0996 & ---~~~~~~~ & ---~~~~~~~ & 0.66 $\pm$ 0.16 & 0.33 $\pm$ 0.16 & 0.37 $\pm$ 0.45 & ---~~~~~~~ & ---~~~~~~~ & 0.05 $\pm$ 0.16\\ 
OGLE GD-CEP-1012 & 0.43 $\pm$ 0.19 & 0.13 $\pm$ 0.16 & 0.00 $\pm$ 0.09 & 0.46 $\pm$ 0.16 & 0.08 $\pm$ 0.19 & 0.15 $\pm$ 0.11 & $-$0.38 $\pm$ 0.16 & 0.06 $\pm$ 0.07\\ 
OGLE GD-CEP-1111 & 0.61 $\pm$ 0.08 & ---~~~~~~~ & 0.09 $\pm$ 0.09 & 0.59 $\pm$ 0.27 & $-$0.24 $\pm$ 0.41 & $-$0.25 $\pm$ 0.07 & 0.03 $\pm$ 0.16 & 0.17 $\pm$ 0.17\\ 
OGLE GD-CEP-1210 & 0.46 $\pm$ 0.12 & 0.29 $\pm$ 0.16 & 0.41 $\pm$ 0.19 & 0.85 $\pm$ 0.16 & 0.15 $\pm$ 0.17 & 0.19 $\pm$ 0.09 & $-$0.30 $\pm$ 0.16 & 0.33 $\pm$ 0.05\\ 
OGLE GD-CEP-1285 & $-$0.48 $\pm$ 0.12 & $-$1.01 $\pm$ 0.11 & $-$0.44 $\pm$ 0.11 & 0.17 $\pm$ 0.11 & $-$1.11 $\pm$ 0.16 & 0.20 $\pm$ 0.04 & ---~~~~~~~ & $-$0.25 $\pm$ 0.17\\ 
OGLE GD-CEP-1311 & $-$0.44 $\pm$ 0.36 & $-$0.77 $\pm$ 0.16 & $-$0.25 $\pm$ 0.04 & 0.15 $\pm$ 0.16 & 0.08 $\pm$ 0.11 & $-$0.13 $\pm$ 0.04 & $-$0.51 $\pm$ 0.16 & $-$0.29 $\pm$ 0.04\\ 
OGLE GD-CEP-1337 & $-$0.36 $\pm$ 0.12 & $-$1.08 $\pm$ 0.11 & $-$0.34 $\pm$ 0.11 & ---~~~~~~~ & $-$1.28 $\pm$ 0.11 & $-$0.13 $\pm$ 0.15 & $-$0.63 $\pm$ 0.16 & $-$0.61 $\pm$ 0.16\\ 
V1253 Cen & $-$0.01 $\pm$ 0.07 & $-$0.51 $\pm$ 0.16 & 0.00 $\pm$ 0.06 & 0.33 $\pm$ 0.16 & 0.17 $\pm$ 0.11 & $-$0.04 $\pm$ 0.04 & $-$0.40 $\pm$ 0.16 & $-$0.11 $\pm$ 0.04\\ 
V1819 Ori & $-$0.48 $\pm$ 0.05 & $-$1.02 $\pm$ 0.11 & $-$0.22 $\pm$ 0.31 & ---~~~~~~~ & $-$0.61 $\pm$ 0.19 & $-$0.29 $\pm$ 0.33 & $-$0.63 $\pm$ 0.16 & $-$0.22 $\pm$ 0.15\\ 
V418 CMa & $-$0.21 $\pm$ 0.06 & $-$0.85 $\pm$ 0.16 & $-$0.33 $\pm$ 0.04 & 0.10 $\pm$ 0.16 & $-$0.33 $\pm$ 0.11 & $-$0.23 $\pm$ 0.04 & $-$0.24 $\pm$ 0.16 & $-$0.27 $\pm$ 0.04\\ 
V459 Sct & 0.49 $\pm$ 0.08 & 0.23 $\pm$ 0.16 & 0.17 $\pm$ 0.04 & 0.54 $\pm$ 0.16 & 0.15 $\pm$ 0.13 & 0.06 $\pm$ 0.04 & $-$0.26 $\pm$ 0.16 & $-$0.17 $\pm$ 0.10\\ 
V480 Aql & 0.30 $\pm$ 0.25 & 0.20 $\pm$ 0.16 & 0.32 $\pm$ 0.10 & 0.53 $\pm$ 0.16 & 0.36 $\pm$ 0.09 & 0.34 $\pm$ 0.05 & $-$0.48 $\pm$ 0.16 & 0.15 $\pm$ 0.04\\ 
V881 Cen & 0.30 $\pm$ 0.09 & 0.17 $\pm$ 0.16 & 0.10 $\pm$ 0.05 & 0.39 $\pm$ 0.16 & 0.14 $\pm$ 0.12 & 0.14 $\pm$ 0.08 & $-$0.41 $\pm$ 0.16 & $-$0.08 $\pm$ 0.04\\ 
VX CMa & $-$0.45 $\pm$ 0.18 & $-$1.12 $\pm$ 0.11 & $-$0.47 $\pm$ 0.11 & 0.08 $\pm$ 0.40 & $-$0.96 $\pm$ 0.17 & $-$0.50 $\pm$ 0.06 & $-$0.48 $\pm$ 0.16 & $-$0.33 $\pm$ 0.13\\ 
\hline\end{tabular}
\end{adjustbox}
\end{table*}

%% file: T22.bbl
\begin{thebibliography}{}
\makeatletter
\relax
\def\mn@urlcharsother{\let\do\@makeother \do\$\do\&\do\#\do\^\do\_\do\%\do\~}
\def\mn@doi{\begingroup\mn@urlcharsother \@ifnextchar [ {\mn@doi@}
  {\mn@doi@[]}}
\def\mn@doi@[#1]#2{\def\@tempa{#1}\ifx\@tempa\@empty \href
  {http://dx.doi.org/#2} {doi:#2}\else \href {http://dx.doi.org/#2} {#1}\fi
  \endgroup}
\def\mn@eprint#1#2{\mn@eprint@#1:#2::\@nil}
\def\mn@eprint@arXiv#1{\href {http://arxiv.org/abs/#1} {{\tt arXiv:#1}}}
\def\mn@eprint@dblp#1{\href {http://dblp.uni-trier.de/rec/bibtex/#1.xml}
  {dblp:#1}}
\def\mn@eprint@#1:#2:#3:#4\@nil{\def\@tempa {#1}\def\@tempb {#2}\def\@tempc
  {#3}\ifx \@tempc \@empty \let \@tempc \@tempb \let \@tempb \@tempa \fi \ifx
  \@tempb \@empty \def\@tempb {arXiv}\fi \@ifundefined
  {mn@eprint@\@tempb}{\@tempb:\@tempc}{\expandafter \expandafter \csname
  mn@eprint@\@tempb\endcsname \expandafter{\@tempc}}}

\bibitem[\protect\citeauthoryear{{Andrievsky}, {L{\'e}pine}, {Korotin}, {Luck},
  {Kovtyukh}  \& {Maciel}}{{Andrievsky} et~al.}{2013}]{andy2013}
{Andrievsky} S.~M.,  {L{\'e}pine} J.~R.~D.,  {Korotin} S.~A.,  {Luck} R.~E.,
  {Kovtyukh} V.~V.,   {Maciel} W.~J.,  2013, \mn@doi [\mnras]
  {10.1093/mnras/sts270}, \href
  {https://ui.adsabs.harvard.edu/abs/2013MNRAS.428.3252A} {428, 3252}

\bibitem[\protect\citeauthoryear{Argast, Samland, Thielemann  \& Qian}{Argast
  et~al.}{2004}]{argast2004neutron}
Argast D.,  Samland M.,  Thielemann F.-K.,   Qian Y.-Z.,  2004, Astronomy \&
  Astrophysics, 416, 997

\bibitem[\protect\citeauthoryear{{Bono}, {Marconi}  \& {Stellingwerf}}{{Bono}
  et~al.}{1999}]{Caputo1999}
{Bono} G.,  {Marconi} M.,   {Stellingwerf} R.~F.,  1999, \mn@doi [\apjs]
  {10.1086/313207}, \href
  {https://ui.adsabs.harvard.edu/abs/1999ApJS..122..167B} {122, 167}

\bibitem[\protect\citeauthoryear{{Breuval} et~al.,}{{Breuval}
  et~al.}{2021}]{Breuval2021}
{Breuval} L.,  et~al., 2021, \mn@doi [\apj] {10.3847/1538-4357/abf0ae}, \href
  {https://ui.adsabs.harvard.edu/abs/2021ApJ...913...38B} {913, 38}

\bibitem[\protect\citeauthoryear{{Cappellari} et~al.,}{{Cappellari}
  et~al.}{2013}]{Cappellari2013}
{Cappellari} M.,  et~al., 2013, \mn@doi [\mnras] {10.1093/mnras/stt562}, \href
  {https://ui.adsabs.harvard.edu/abs/2013MNRAS.432.1709C} {432, 1709}

\bibitem[\protect\citeauthoryear{{Caputo}, {Marconi}, {Musella}  \&
  {Santolamazza}}{{Caputo} et~al.}{2000}]{Caputo2000}
{Caputo} F.,  {Marconi} M.,  {Musella} I.,   {Santolamazza} P.,  2000, \aap,
  \href {https://ui.adsabs.harvard.edu/abs/2000A&A...359.1059C} {359, 1059}

\bibitem[\protect\citeauthoryear{Carrera et~al.,}{Carrera
  et~al.}{2019}]{carrera2019open}
Carrera R.,  et~al., 2019, Astronomy \& Astrophysics, 623, A80

\bibitem[\protect\citeauthoryear{Casamiquela et~al.,}{Casamiquela
  et~al.}{2019}]{casamiquela2019occaso}
Casamiquela L.,  et~al., 2019, Monthly Notices of the Royal Astronomical
  Society, 490, 1821

\bibitem[\protect\citeauthoryear{Castelli \& Hubrig}{Castelli \&
  Hubrig}{2004}]{castelli2004spectroscopic}
Castelli F.,  Hubrig S.,  2004, Astronomy \& Astrophysics, 425, 263

\bibitem[\protect\citeauthoryear{{Catanzaro} et~al.,}{{Catanzaro}
  et~al.}{2020}]{Catanzaro2020}
{Catanzaro} G.,  et~al., 2020, \mn@doi [\aap] {10.1051/0004-6361/202038486},
  \href {https://ui.adsabs.harvard.edu/abs/2020A&A...639L...4C} {639, L4}

\bibitem[\protect\citeauthoryear{{Clayton}}{{Clayton}}{2003}]{clay03}
{Clayton} D.,  2003, {Handbook of Isotopes in the Cosmos}

\bibitem[\protect\citeauthoryear{{Clementini} et~al.,}{{Clementini}
  et~al.}{2016}]{Clementini2016}
{Clementini} G.,  et~al., 2016, \mn@doi [\aap] {10.1051/0004-6361/201629583},
  \href {https://ui.adsabs.harvard.edu/abs/2016A&A...595A.133C} {595, A133}

\bibitem[\protect\citeauthoryear{Cowan, Sneden, Lawler, Aprahamian, Wiescher,
  Langanke, Mart{\'\i}nez-Pinedo  \& Thielemann}{Cowan
  et~al.}{2021}]{cowan2021origin}
Cowan J.~J.,  Sneden C.,  Lawler J.~E.,  Aprahamian A.,  Wiescher M.,  Langanke
  K.,  Mart{\'\i}nez-Pinedo G.,   Thielemann F.-K.,  2021, Reviews of Modern
  Physics, 93, 015002

\bibitem[\protect\citeauthoryear{Cunha et~al.,}{Cunha
  et~al.}{2016}]{cunha2016chemical}
Cunha K.,  et~al., 2016, Astronomische Nachrichten, 337, 922

\bibitem[\protect\citeauthoryear{{Daflon} \& {Cunha}}{{Daflon} \&
  {Cunha}}{2004}]{daflon2004}
{Daflon} S.,  {Cunha} K.,  2004, \mn@doi [\apj] {10.1086/425607}, \href
  {https://ui.adsabs.harvard.edu/abs/2004ApJ...617.1115D} {617, 1115}

\bibitem[\protect\citeauthoryear{{Dainotti}, {De Simone}, {Schiavone},
  {Montani}, {Rinaldi}  \& {Lambiase}}{{Dainotti} et~al.}{2021}]{Dainotti2021}
{Dainotti} M.~G.,  {De Simone} B.,  {Schiavone} T.,  {Montani} G.,  {Rinaldi}
  E.,   {Lambiase} G.,  2021, \mn@doi [\apj] {10.3847/1538-4357/abeb73}, \href
  {https://ui.adsabs.harvard.edu/abs/2021ApJ...912..150D} {912, 150}

\bibitem[\protect\citeauthoryear{Donor et~al.,}{Donor
  et~al.}{2020}]{donor2020open}
Donor J.,  et~al., 2020, The Astronomical Journal, 159, 199

\bibitem[\protect\citeauthoryear{{Freedman}}{{Freedman}}{2021}]{Freedman2021}
{Freedman} W.~L.,  2021, \mn@doi [\apj] {10.3847/1538-4357/ac0e95}, \href
  {https://ui.adsabs.harvard.edu/abs/2021ApJ...919...16F} {919, 16}

\bibitem[\protect\citeauthoryear{{Freedman}, {Madore}, {Scowcroft}, {Burns},
  {Monson}, {Persson}, {Seibert}  \& {Rigby}}{{Freedman}
  et~al.}{2012}]{Freedman2012}
{Freedman} W.~L.,  {Madore} B.~F.,  {Scowcroft} V.,  {Burns} C.,  {Monson} A.,
  {Persson} S.~E.,  {Seibert} M.,   {Rigby} J.,  2012, \mn@doi [\apj]
  {10.1088/0004-637X/758/1/24}, \href
  {https://ui.adsabs.harvard.edu/abs/2012ApJ...758...24F} {758, 24}

\bibitem[\protect\citeauthoryear{Frinchaboy et~al.,}{Frinchaboy
  et~al.}{2013}]{frinchaboy2013open}
Frinchaboy P.~M.,  et~al., 2013, The Astrophysical Journal Letters, 777, L1

\bibitem[\protect\citeauthoryear{{Gaia Collaboration} et~al.,}{{Gaia
  Collaboration} et~al.}{2016}]{Gaia2016}
{Gaia Collaboration} et~al., 2016, \mn@doi [\aap]
  {10.1051/0004-6361/201629272}, \href
  {https://ui.adsabs.harvard.edu/abs/2016A&A...595A...1G} {595, A1}

\bibitem[\protect\citeauthoryear{{Gaia Collaboration} et~al.,}{{Gaia
  Collaboration} et~al.}{2017}]{Gaia2017}
{Gaia Collaboration} et~al., 2017, \mn@doi [\aap]
  {10.1051/0004-6361/201629925}, \href
  {https://ui.adsabs.harvard.edu/abs/2017A&A...605A..79G} {605, A79}

\bibitem[\protect\citeauthoryear{{Gaia Collaboration} et~al.,}{{Gaia
  Collaboration} et~al.}{2018}]{Gaia2018}
{Gaia Collaboration} et~al., 2018, \mn@doi [\aap]
  {10.1051/0004-6361/201833051}, \href
  {https://ui.adsabs.harvard.edu/abs/2018A&A...616A...1G} {616, A1}

\bibitem[\protect\citeauthoryear{{Gaia Collaboration} et~al.,}{{Gaia
  Collaboration} et~al.}{2021}]{Gaia2021}
{Gaia Collaboration} et~al., 2021, \mn@doi [\aap]
  {10.1051/0004-6361/202039657}, \href
  {https://ui.adsabs.harvard.edu/abs/2021A&A...649A...1G} {649, A1}

\bibitem[\protect\citeauthoryear{{Gaia Collaboration} et~al.,}{{Gaia
  Collaboration} et~al.}{2022}]{Gaia2022gaia}
{Gaia Collaboration} et~al., 2022, arXiv e-prints, \href
  {https://ui.adsabs.harvard.edu/abs/2022arXiv220605534G} {p. arXiv:2206.05534}

\bibitem[\protect\citeauthoryear{{Genovali} et~al.,}{{Genovali}
  et~al.}{2014}]{Genovali2014}
{Genovali} K.,  et~al., 2014, \mn@doi [\aap] {10.1051/0004-6361/201323198},
  \href {https://ui.adsabs.harvard.edu/abs/2014A&A...566A..37G} {566, A37}

\bibitem[\protect\citeauthoryear{{Gordon}, {Clayton}, {Misselt}, {Landolt}  \&
  {Wolff}}{{Gordon} et~al.}{2003}]{Gordon2003}
{Gordon} K.~D.,  {Clayton} G.~C.,  {Misselt} K.~A.,  {Landolt} A.~U.,   {Wolff}
  M.~J.,  2003, \mn@doi [\apj] {10.1086/376774}, \href
  {https://ui.adsabs.harvard.edu/abs/2003ApJ...594..279G} {594, 279}

\bibitem[\protect\citeauthoryear{{Gravity Collaboration} et~al.,}{{Gravity
  Collaboration} et~al.}{2022}]{Gravity2022}
{Gravity Collaboration} et~al., 2022, \mn@doi [\aap]
  {10.1051/0004-6361/202142465}, \href
  {https://ui.adsabs.harvard.edu/abs/2022A&A...657L..12G} {657, L12}

\bibitem[\protect\citeauthoryear{{Groenewegen}}{{Groenewegen}}{2018}]{Groenewegen2018}
{Groenewegen} M.~A.~T.,  2018, \mn@doi [\aap] {10.1051/0004-6361/201833478},
  \href {https://ui.adsabs.harvard.edu/abs/2018A&A...619A...8G} {619, A8}

\bibitem[\protect\citeauthoryear{{Hayden}, {Recio-Blanco}, {de Laverny},
  {Mikolaitis}  \& {Worley}}{{Hayden} et~al.}{2017}]{hayden2017}
{Hayden} M.~R.,  {Recio-Blanco} A.,  {de Laverny} P.,  {Mikolaitis} S.,
  {Worley} C.~C.,  2017, \mn@doi [\aap] {10.1051/0004-6361/201731494}, \href
  {https://ui.adsabs.harvard.edu/abs/2017A&A...608L...1H} {608, L1}

\bibitem[\protect\citeauthoryear{Iben~Jr}{Iben~Jr}{1967}]{iben1967stellar}
Iben~Jr I.,  1967, Annual Review of Astronomy and Astrophysics, 5, 571

\bibitem[\protect\citeauthoryear{Janes}{Janes}{1979}]{janes1979evidence}
Janes K.,  1979, The Astrophysical Journal Supplement Series, 39, 135

\bibitem[\protect\citeauthoryear{{Jordi} et~al.,}{{Jordi}
  et~al.}{2010}]{Jordi2010}
{Jordi} C.,  et~al., 2010, \mn@doi [\aap] {10.1051/0004-6361/201015441}, \href
  {https://ui.adsabs.harvard.edu/abs/2010A&A...523A..48J} {523, A48}

\bibitem[\protect\citeauthoryear{Karakas \& Lugaro}{Karakas \&
  Lugaro}{2016}]{karakas2016stellar}
Karakas A.~I.,  Lugaro M.,  2016, The Astrophysical Journal, 825, 26

\bibitem[\protect\citeauthoryear{Korobkin, Rosswog, Arcones  \&
  Winteler}{Korobkin et~al.}{2012}]{korobkin2012astrophysical}
Korobkin O.,  Rosswog S.,  Arcones A.,   Winteler C.,  2012, Monthly Notices of
  the Royal Astronomical Society, 426, 1940

\bibitem[\protect\citeauthoryear{Kovtyukh \& Gorlova}{Kovtyukh \&
  Gorlova}{2000}]{kovtyukh2000precise}
Kovtyukh V.,  Gorlova N.,  2000, Astronomy and Astrophysics, 358, 587

\bibitem[\protect\citeauthoryear{{Kovtyukh} et~al.,}{{Kovtyukh}
  et~al.}{2022}]{Kovtyukh2022}
{Kovtyukh} V.,  et~al., 2022, \mn@doi [\mnras] {10.1093/mnras/stab3530}, \href
  {https://ui.adsabs.harvard.edu/abs/2022MNRAS.510.1894K} {510, 1894}

\bibitem[\protect\citeauthoryear{Kurucz}{Kurucz}{1993}]{kurucz1993new}
Kurucz R.~L.,  1993, in International Astronomical Union Colloquium. pp 87--97

\bibitem[\protect\citeauthoryear{Kurucz}{Kurucz}{1995}]{kurucz1995kurucz}
Kurucz R.,  1995, Atomic Line List

\bibitem[\protect\citeauthoryear{Kurucz \& Avrett}{Kurucz \&
  Avrett}{1981}]{kurucz1981solar}
Kurucz R.~L.,  Avrett E.~H.,  1981, SAO Special Report, 391

\bibitem[\protect\citeauthoryear{{Leavitt} \& {Pickering}}{{Leavitt} \&
  {Pickering}}{1912}]{Leavitt1912}
{Leavitt} H.~S.,  {Pickering} E.~C.,  1912, Harvard College Observatory
  Circular, \href {https://ui.adsabs.harvard.edu/abs/1912HarCi.173....1L} {173,
  1}

\bibitem[\protect\citeauthoryear{Lemasle et~al.,}{Lemasle
  et~al.}{2018}]{lemasle2018milky}
Lemasle B.,  et~al., 2018, Astronomy \& Astrophysics, 618, A160

\bibitem[\protect\citeauthoryear{{Luck}}{{Luck}}{2018}]{Luck2018}
{Luck} R.~E.,  2018, \mn@doi [\aj] {10.3847/1538-3881/aadcac}, \href
  {https://ui.adsabs.harvard.edu/abs/2018AJ....156..171L} {156, 171}

\bibitem[\protect\citeauthoryear{{Luck} \& {Lambert}}{{Luck} \&
  {Lambert}}{2011}]{Luck2011}
{Luck} R.~E.,  {Lambert} D.~L.,  2011, \mn@doi [\aj]
  {10.1088/0004-6256/142/4/136}, \href
  {https://ui.adsabs.harvard.edu/abs/2011AJ....142..136L} {142, 136}

\bibitem[\protect\citeauthoryear{{Madore}}{{Madore}}{1982}]{Madore1982}
{Madore} B.~F.,  1982, \mn@doi [\apj] {10.1086/159659}, \href
  {https://ui.adsabs.harvard.edu/abs/1982ApJ...253..575M} {253, 575}

\bibitem[\protect\citeauthoryear{Magrini, Randich, Zoccali, Jilkova, Carraro,
  Galli, Maiorca  \& Busso}{Magrini et~al.}{2010}]{magrini2010open}
Magrini L.,  Randich S.,  Zoccali M.,  Jilkova L.,  Carraro G.,  Galli D.,
  Maiorca E.,   Busso M.,  2010, Astronomy \& Astrophysics, 523, A11

\bibitem[\protect\citeauthoryear{{Matteucci}}{{Matteucci}}{2021}]{matteucci2021}
{Matteucci} F.,  2021, \mn@doi [\aapr] {10.1007/s00159-021-00133-8}, \href
  {https://ui.adsabs.harvard.edu/abs/2021A&ARv..29....5M} {29, 5}

\bibitem[\protect\citeauthoryear{{Minniti} et~al.,}{{Minniti}
  et~al.}{2020}]{Minniti2020}
{Minniti} J.~H.,  et~al., 2020, \mn@doi [\aap] {10.1051/0004-6361/202037575},
  \href {https://ui.adsabs.harvard.edu/abs/2020A&A...640A..92M} {640, A92}

\bibitem[\protect\citeauthoryear{Minniti et~al.,}{Minniti
  et~al.}{2021}]{minniti2021using}
Minniti J.,  et~al., 2021, arXiv preprint arXiv:2107.03464

\bibitem[\protect\citeauthoryear{Mucciarelli \& Bonifacio}{Mucciarelli \&
  Bonifacio}{2020}]{mucciarelli2020facing}
Mucciarelli A.,  Bonifacio P.,  2020, Astronomy \& Astrophysics, 640, A87

\bibitem[\protect\citeauthoryear{Netopil, Paunzen, Heiter  \& Soubiran}{Netopil
  et~al.}{2016}]{netopil2016metallicity}
Netopil M.,  Paunzen E.,  Heiter U.,   Soubiran C.,  2016, Astronomy \&
  Astrophysics, 585, A150

\bibitem[\protect\citeauthoryear{{Planck Collaboration} et~al.,}{{Planck
  Collaboration} et~al.}{2020}]{Planck2020}
{Planck Collaboration} et~al., 2020, \mn@doi [\aap]
  {10.1051/0004-6361/201833910}, \href
  {https://ui.adsabs.harvard.edu/abs/2020A&A...641A...6P} {641, A6}

\bibitem[\protect\citeauthoryear{{Poggio} et~al.,}{{Poggio}
  et~al.}{2021}]{Poggio2021}
{Poggio} E.,  et~al., 2021, \mn@doi [\aap] {10.1051/0004-6361/202140687}, \href
  {https://ui.adsabs.harvard.edu/abs/2021A&A...651A.104P} {651, A104}

\bibitem[\protect\citeauthoryear{Ralchenko \& Reader}{Ralchenko \&
  Reader}{2019}]{ralchenko2019nist}
Ralchenko A.~K.,  Reader J.,  2019, National Institute of Standards and
  Technology, Gaithersburg, MD. DOI: https://doi. org/10.18434/T4W30F

\bibitem[\protect\citeauthoryear{Randich et~al.,}{Randich
  et~al.}{2022}]{randich2022gaia}
Randich S.,  et~al., 2022, arXiv preprint arXiv:2206.02901

\bibitem[\protect\citeauthoryear{{Riess} et~al.,}{{Riess}
  et~al.}{2016}]{Riess2016}
{Riess} A.~G.,  et~al., 2016, \mn@doi [\apj] {10.3847/0004-637X/826/1/56},
  \href {https://ui.adsabs.harvard.edu/abs/2016ApJ...826...56R} {826, 56}

\bibitem[\protect\citeauthoryear{{Riess}, {Casertano}, {Yuan}, {Macri}  \&
  {Scolnic}}{{Riess} et~al.}{2019}]{Riess2019}
{Riess} A.~G.,  {Casertano} S.,  {Yuan} W.,  {Macri} L.~M.,   {Scolnic} D.,
  2019, \mn@doi [\apj] {10.3847/1538-4357/ab1422}, \href
  {https://ui.adsabs.harvard.edu/abs/2019ApJ...876...85R} {876, 85}

\bibitem[\protect\citeauthoryear{{Riess} et~al.,}{{Riess}
  et~al.}{2021a}]{Riess2021b}
{Riess} A.~G.,  et~al., 2021a, arXiv e-prints, \href
  {https://ui.adsabs.harvard.edu/abs/2021arXiv211204510R} {p. arXiv:2112.04510}

\bibitem[\protect\citeauthoryear{{Riess}, {Casertano}, {Yuan}, {Bowers},
  {Macri}, {Zinn}  \& {Scolnic}}{{Riess} et~al.}{2021b}]{Riess2021}
{Riess} A.~G.,  {Casertano} S.,  {Yuan} W.,  {Bowers} J.~B.,  {Macri} L.,
  {Zinn} J.~C.,   {Scolnic} D.,  2021b, \mn@doi [\apjl]
  {10.3847/2041-8213/abdbaf}, \href
  {https://ui.adsabs.harvard.edu/abs/2021ApJ...908L...6R} {908, L6}

\bibitem[\protect\citeauthoryear{{Ripepi}, {Molinaro}, {Musella}, {Marconi},
  {Leccia}  \& {Eyer}}{{Ripepi} et~al.}{2019}]{Ripepi2019}
{Ripepi} V.,  {Molinaro} R.,  {Musella} I.,  {Marconi} M.,  {Leccia} S.,
  {Eyer} L.,  2019, \mn@doi [\aap] {10.1051/0004-6361/201834506}, \href
  {https://ui.adsabs.harvard.edu/abs/2019A&A...625A..14R} {625, A14}

\bibitem[\protect\citeauthoryear{{Ripepi} et~al.,}{{Ripepi}
  et~al.}{2020}]{Ripepi2020}
{Ripepi} V.,  et~al., 2020, \mn@doi [\aap] {10.1051/0004-6361/202038714}, \href
  {https://ui.adsabs.harvard.edu/abs/2020A&A...642A.230R} {642, A230}

\bibitem[\protect\citeauthoryear{{Ripepi} et~al.,}{{Ripepi}
  et~al.}{2021a}]{Ripepi2021a}
{Ripepi} V.,  et~al., 2021a, \mn@doi [\mnras] {10.1093/mnras/stab2460}, \href
  {https://ui.adsabs.harvard.edu/abs/2021MNRAS.508.4047R} {508, 4047}

\bibitem[\protect\citeauthoryear{{Ripepi} et~al.,}{{Ripepi}
  et~al.}{2021b}]{Ripepi2021b}
{Ripepi} V.,  et~al., 2021b, \mn@doi [\aap] {10.1051/0004-6361/202040123},
  \href {https://ui.adsabs.harvard.edu/abs/2021A&A...647A.111R} {647, A111}

\bibitem[\protect\citeauthoryear{{Ripepi} et~al.,}{{Ripepi}
  et~al.}{2022}]{Ripepi2022}
{Ripepi} V.,  et~al., 2022, \mn@doi [\aap] {10.1051/0004-6361/202142649}, \href
  {https://ui.adsabs.harvard.edu/abs/2022A&A...659A.167R} {659, A167}

\bibitem[\protect\citeauthoryear{Romaniello et~al.,}{Romaniello
  et~al.}{2008a}]{romaniello2008influence}
Romaniello M.,  et~al., 2008a, Astronomy \& Astrophysics, 488, 731

\bibitem[\protect\citeauthoryear{{Romaniello} et~al.,}{{Romaniello}
  et~al.}{2008b}]{Romaniello2008}
{Romaniello} M.,  et~al., 2008b, \mn@doi [\aap] {10.1051/0004-6361:20065661},
  \href {https://ui.adsabs.harvard.edu/abs/2008A&A...488..731R} {488, 731}

\bibitem[\protect\citeauthoryear{{Romano}, {Karakas}, {Tosi}  \&
  {Matteucci}}{{Romano} et~al.}{2010}]{romano2010sc}
{Romano} D.,  {Karakas} A.~I.,  {Tosi} M.,   {Matteucci} F.,  2010, \mn@doi
  [\aap] {10.1051/0004-6361/201014483}, \href
  {https://ui.adsabs.harvard.edu/abs/2010A&A...522A..32R} {522, A32}

\bibitem[\protect\citeauthoryear{{Sandage} \& {Tammann}}{{Sandage} \&
  {Tammann}}{2006}]{Sandage2006}
{Sandage} A.,  {Tammann} G.~A.,  2006, \mn@doi [\araa]
  {10.1146/annurev.astro.43.072103.150612}, \href
  {https://ui.adsabs.harvard.edu/abs/2006ARA&A..44...93S} {44, 93}

\bibitem[\protect\citeauthoryear{Sasselov}{Sasselov}{1986}]{sasselov1986normal}
Sasselov D.~D.,  1986, Publications of the Astronomical Society of the Pacific,
  98, 561

\bibitem[\protect\citeauthoryear{{Silva Aguirre} et~al.,}{{Silva Aguirre}
  et~al.}{2018}]{silvaguirre2018}
{Silva Aguirre} V.,  et~al., 2018, \mn@doi [\mnras] {10.1093/mnras/sty150},
  \href {https://ui.adsabs.harvard.edu/abs/2018MNRAS.475.5487S} {475, 5487}

\bibitem[\protect\citeauthoryear{{Simmerer}, {Sneden}, {Cowan}, {Collier},
  {Woolf}  \& {Lawler}}{{Simmerer} et~al.}{2004}]{simmerer2004}
{Simmerer} J.,  {Sneden} C.,  {Cowan} J.~J.,  {Collier} J.,  {Woolf} V.~M.,
  {Lawler} J.~E.,  2004, \mn@doi [\apj] {10.1086/424504}, \href
  {https://ui.adsabs.harvard.edu/abs/2004ApJ...617.1091S} {617, 1091}

\bibitem[\protect\citeauthoryear{{Skowron} et~al.,}{{Skowron}
  et~al.}{2019}]{Skowron2019}
{Skowron} D.~M.,  et~al., 2019, \mn@doi [Science] {10.1126/science.aau3181},
  \href {https://ui.adsabs.harvard.edu/abs/2019Sci...365..478S} {365, 478}

\bibitem[\protect\citeauthoryear{Spina et~al.,}{Spina
  et~al.}{2021}]{spina2021galah}
Spina L.,  et~al., 2021, Monthly Notices of the Royal Astronomical Society,
  503, 3279

\bibitem[\protect\citeauthoryear{{Spitoni}, {Silva Aguirre}, {Matteucci},
  {Calura}  \& {Grisoni}}{{Spitoni} et~al.}{2019}]{spitoni2019}
{Spitoni} E.,  {Silva Aguirre} V.,  {Matteucci} F.,  {Calura} F.,   {Grisoni}
  V.,  2019, \mn@doi [\aap] {10.1051/0004-6361/201834188}, \href
  {https://ui.adsabs.harvard.edu/abs/2019A&A...623A..60S} {623, A60}

\bibitem[\protect\citeauthoryear{Surman, Mclaughlin, Ruffert, Janka  \&
  Hix}{Surman et~al.}{2008}]{surman2008r}
Surman R.,  Mclaughlin G.~C.,  Ruffert M.,  Janka H.-T.,   Hix W.~R.,  2008,
  The Astrophysical Journal, 679, L117

\bibitem[\protect\citeauthoryear{{Verde}, {Treu}  \& {Riess}}{{Verde}
  et~al.}{2019}]{verde2019}
{Verde} L.,  {Treu} T.,   {Riess} A.~G.,  2019, \mn@doi [Nature Astronomy]
  {10.1038/s41550-019-0902-0}, \href
  {https://ui.adsabs.harvard.edu/abs/2019NatAs...3..891V} {3, 891}

\bibitem[\protect\citeauthoryear{Yong, Carney  \& Friel}{Yong
  et~al.}{2012}]{yong2012elemental}
Yong D.,  Carney B.~W.,   Friel E.~D.,  2012, The Astronomical Journal, 144, 95

\bibitem[\protect\citeauthoryear{da Silva et~al.,}{da~Silva
  et~al.}{2022}]{da2022new}
da Silva R.,  et~al., 2022, arXiv preprint arXiv:2202.07945

\makeatother
\end{thebibliography}
